\documentclass[10pt,prl,aps,twocolumn,superscriptaddress,preprintnumbers]{revtex4-1}

\usepackage[utf8]{inputenc}
\usepackage{epigraph}
\usepackage{amsmath}
\usepackage{wrapfig}
\usepackage{epsfig}
\usepackage{amssymb}
\usepackage{graphicx}
\usepackage{slashed}
\usepackage[dvipsnames]{xcolor}
\usepackage{comment}
\usepackage{natbib}
\usepackage{enumitem}

\usepackage[hidelinks, colorlinks=true]{hyperref}

\begin{document}
\title{Regge spectroscopy of higher twist states in $\mathcal{N}=4$ supersymmetric Yang-Mills theory}

\author{Rob Klabbers}
%\email{rob.klabbers@hu-berlin.de}
\affiliation{Humboldt-Universität zu Berlin,~Zum Gro{\ss}en Windkanal 2, 12489 Berlin, Germany}
\author{Michelangelo Preti}
%\email{michelangelo.preti@kcl.ac.uk}
\affiliation{Department of Mathematics, King’s College London
Strand WC2R 2LS, London, UK}
\author{Istv\'{a}n M. Sz\'{e}cs\'{e}nyi}
%\email{istvan.szecsenyi@su.se}
\affiliation{Nordita, KTH Royal Institute of Technology and Stockholm University, Hannes Alfvéns väg 12, SE-106 91 Stockholm, Sweden}

\begin{abstract}
    We study a family of higher-twist Regge trajectories in $\mathcal{N}=4$ supersymmetric Yang-Mills theory using the Quantum Spectral Curve. 
    We explore the many-sheeted Riemann surface connecting the different trajectories and show the interplay between the degenerate non-local operators known as horizontal trajectories. We resolve their degeneracy analytically by computing the first non-trivial order of the Regge intercept at weak coupling, which exhibits new behaviour: it depends linearly on the coupling. This is consistent with our numerics, which interpolate all the way to strong coupling. 
\end{abstract}

\preprint{$\substack{\text{HU-EP-23/41} \\ \text{HU-Mathematik-2023-2}}$}

\maketitle

\textit{Introduction.---} Regge theory has been instrumental for our understanding of (strongly interacting) quantum field theory (QFT), especially of elastic high-energy forward scattering. Its greatest success is arguably in providing an explanation for the rising cross sections $\sigma_{\text{tot}} \sim ~s^{\alpha_P-1}$ in quantum chromodynamics (QCD), which depends on the intercept $\alpha_P$ of an exchanged Pomeron trajectory. Such Regge trajectories are the analytic continuation of local operators in complex spin. The intercept $\alpha_P$, the minimum of this trajectory, was computed by exploiting the emergent integrability at weak coupling, leading to the celebrated Balitsky--Fadin--Kuraev--Lipatov (BFKL) equation that resums all leading logarithmic contributions \cite{kuraev1976multiregge,kuraev1977pomeranchuk,balitsky1978pomeranchuk}.  

For conformal field theories (CFTs), Regge theory takes a special form \cite{Costa:2012cb}, making it possible to study Regge behaviour by analysing the analytic continuation of CFT data such as scaling dimensions. It was understood only recently that to describe this continuation properly, one should use --generically non-local-- light-ray operators \cite{Balitsky:1987bk,kravchuk2018light}. This is because the usual operator product expansion of local operators fails when considered in Lorentzian signature, the natural signature for Regge behaviour. Light-ray operators can be constructed explicitly \cite{kravchuk2018light,balitsky2023two}, and serve as building blocks for horizontal trajectories (HTs), which --at least for gauge theories-- are qualitatively similar to null Wilson lines \cite{Kologlu:2019mfz,Chang:2020qpj,caron2023detectors}. In the free theory, these HTs constitute a family of operators with constant spin but arbitrary scaling dimensions.  On general grounds, one can see that such trajectories exist in virtually any CFT, play an essential role in the computation of Regge intercepts through mixing at finite coupling \cite{caron2015does,caron2023detectors,caron2023leading}, and can even form the dominant contribution. Nevertheless, only a few explicit studies \cite{caron2023detectors,Korchemsky:2003rc} into these phenomena exist. 

One of those explicit cases is the minimal twist sector of $\mathcal{N}=4$ supersymmetric Yang-Mills (SYM), where the leading BFKL Pomeron intercept receives contributions from a single HT mixing with the trajectories of the local twist-two operators and their shadows. 
Its value can be computed very efficiently using the Quantum Spectral Curve (QSC), based on the integrability of (planar) $\mathcal{N}=4$ SYM \cite{Alfimov:2014bwa,gromov2015pomeron,gromov2015nnlo}. However, the minimal-twist sector enjoys many simplifications compared to the general theory, allowing for computational shortcuts that are absent when studying a generic scattering process.

\begin{figure}
{    \centering
\includegraphics[width=\columnwidth]{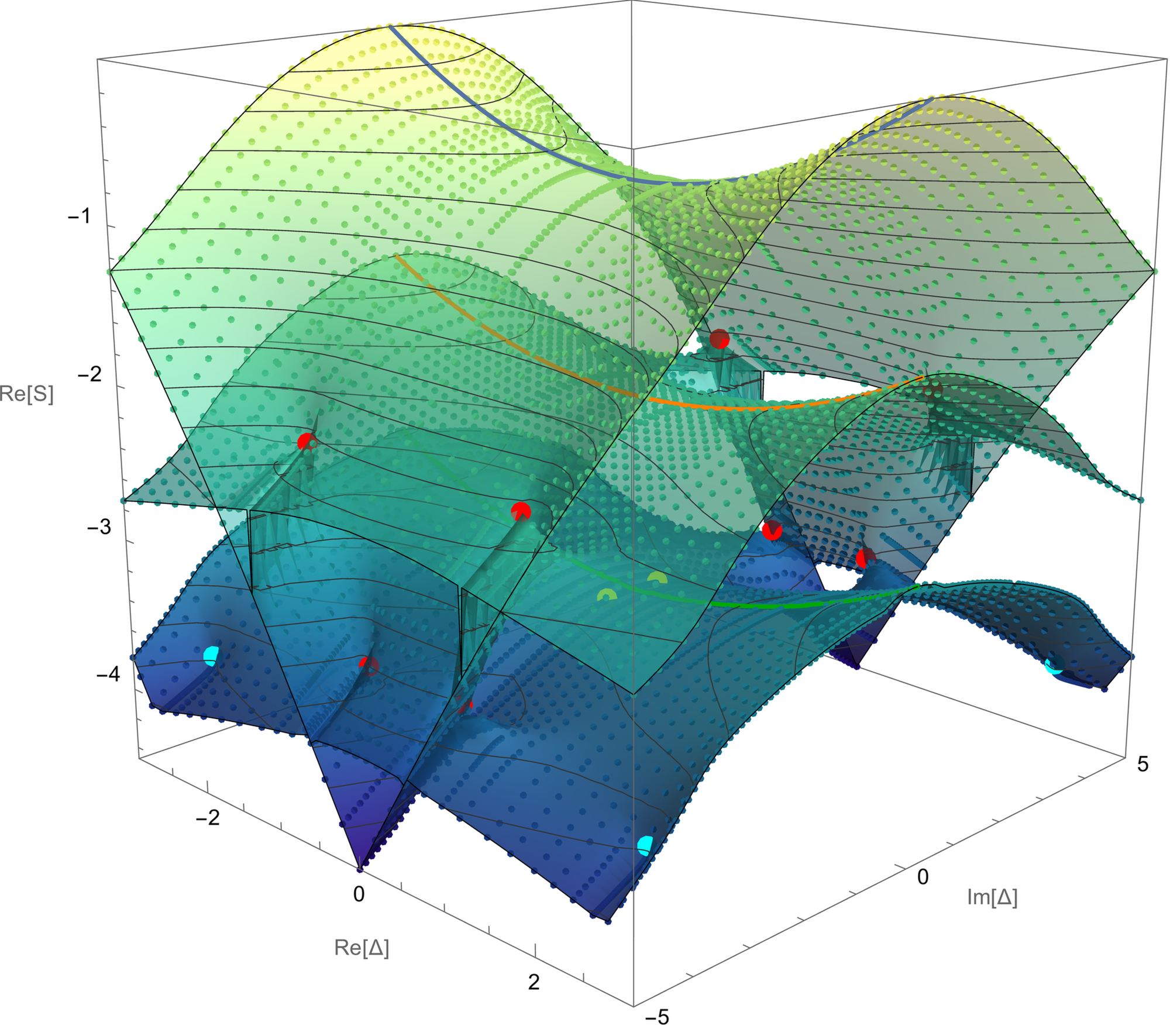}
}
    \caption{ Three sheets of the Riemann surface connecting the different twist-trajectories at $g=1/2$. Dots stand for the branch points. The surface is generated with over 9000 points. The blue, orange and green curves are twist-$3$ and -$5$ trajectories also appearing in FIG.~\ref{fig:spectrumandtraj} for different values of $g$. 
    \label{fig:RiemannReal}}
\end{figure}
In this Letter, we move beyond twist two and study a family of higher-twist states and their mixing with HTs, both perturbatively and non-perturbatively. Our numerics demonstrate how at finite coupling their Regge trajectories are connected to each other, forming a Riemann surface by complex spin continuation (see FIG.~ \ref{fig:RiemannReal}). Moreover, we show the degeneracy of two HTs, as $g\to 0$, explicitly for the first time, by computing the first order of the weak-coupling expansion of the intercept analytically. We resolve the mixing, caused by the contributions of the degenerate HTs, by modifying the QSC ansatz, and observe novel behaviour: the whole BFKL regime, and accordingly the intercept, depends \textit{linearly} on the coupling $g$. This is consistent with our numerical intercept computation, which also interpolates all the way to the strong coupling.

This Letter also forms a stepping stone for the non-perturbative study of other relevant Regge trajectories, such as the Odderon \cite{lukaszuk1973possible,bartels1980high,jaroszewicz1980infra,KWIECINSKI1980413}, the leading charge-conjugacy-odd contribution, which explains the discrepancy between hadron-hadron and hadron-antihadron scattering at high energies. Recent measurements \cite{adloff2002search,antchev2020elastic,abazov2021odderon, MARTYNOV2018414} have confirmed the existence of this discrepancy. However, despite existing weak- \cite{janik1999solution} and strong-coupling approaches \cite{brower2009odderon,brower2015strong}, our fundamental understanding of the Odderon remains lacking, and calls for a non-perturbative study into the mechanisms underlying it. With the established connections \cite{kotikov2004dglap,Kotikov:2010nd,gromov2015nnlo} between $\mathcal{N}=4$ SYM and QCD in mind, this work forms a first step in that direction. 

\vskip 0.1in
\textit{Twist-three states.---}
Our starting point are the superprimary states in planar $\mathcal{N}=4$ SYM with classical twist $\Delta_0 - S = 3$, i.e. $\mathfrak{psu}(2,2,|4)$-highest weight states (in so-called ABA-grading) built up from (single-trace) operators containing three of the fundamental fields. Using data from \cite{marboe2018full}, we select the lowest-lying family of such states, forming the trajectory with the largest intercept $\alpha = S(\Delta=0)$ \footnote[99]{This definition is dependent on the precise choice of conformal primaries out of a supermultiplet, with simple shifts of quantum numbers relating them. For example, the definition of the intercept in  \cite{Costa:2012cb,Brower:2014wha} is $\tilde{\alpha}=\tilde{S}(\Delta=2)$, that is related to our choice as $\tilde{\alpha}= \alpha+2$.}. The local operators $\mathcal{O}_S$ ($S=0,2,4,\ldots$) on this trajectory have quantum numbers $(J_1, J_2,J_3,\Delta,S,S_2) = (3,0,0,3+S+\gamma,S,0)$ with $\gamma$ their anomalous dimension, and are parity singlets \cite{kristjansen2012review}. The Regge behaviour of the three-point couplings of this trajectory was studied recently in \cite{Homrich:2022mmd}. At tree-level, they can be represented in the commonly used form $\mathcal{O}_S = \text{Tr}(D^S Z Z Z)+ \text{permutations}$ (see SM for more details). 

\begin{figure}
{    \centering
\includegraphics[width=\columnwidth]{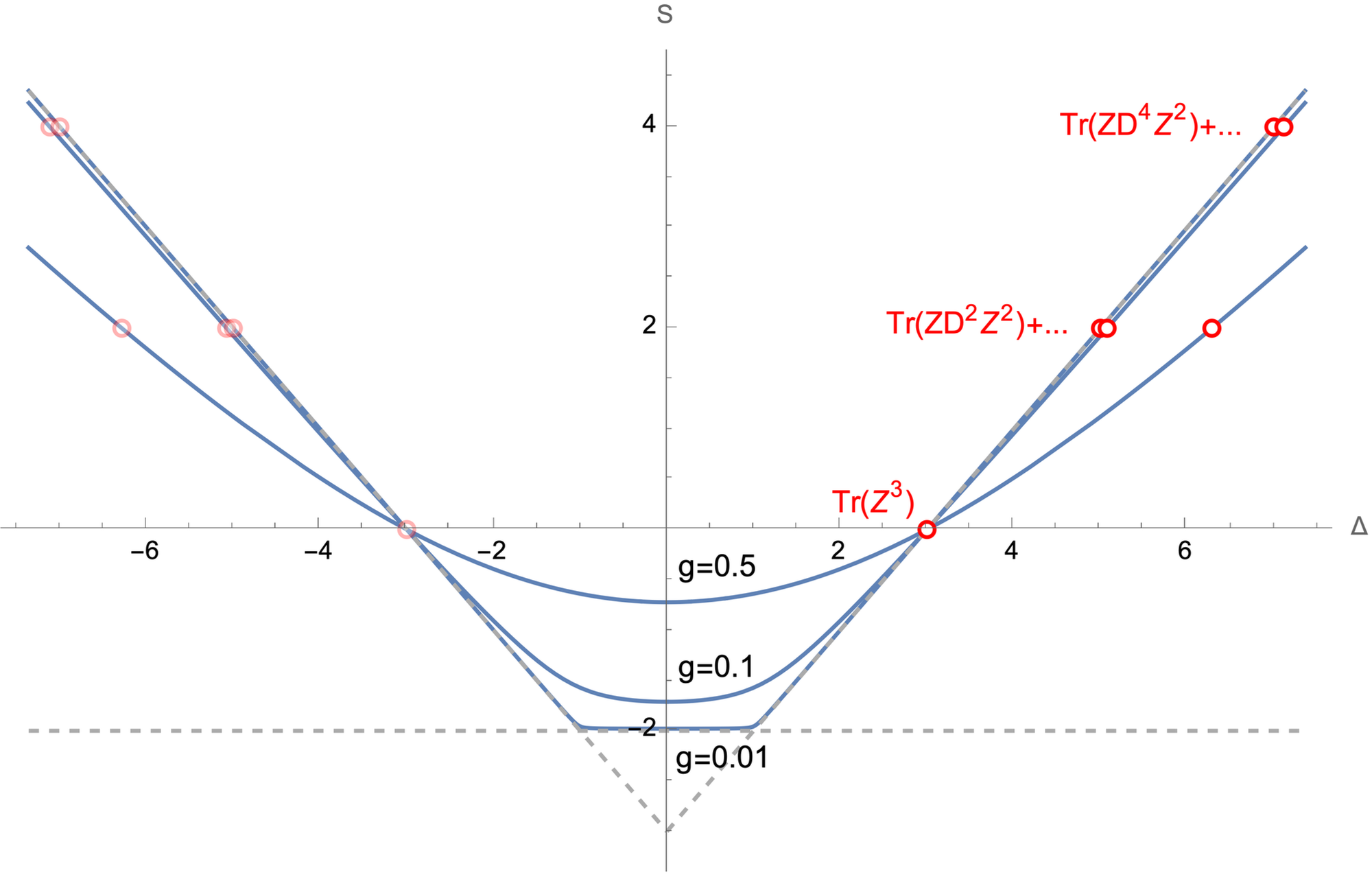}
}
    \caption{ Regge trajectories $S(\Delta)$ corresponding to the twist-3 operators $\mathcal{O}_0,\mathcal{O}_2$, and $\mathcal{O}_4$ (indicated by red circles, and transparent red circles for their shadows) for different values of $g$. Each trajectory is produced by about 400 points. 
    \label{fig:leadingtraj}}
\end{figure}

\vskip 0.1in
\textit{Quantum Spectral Curve.---} The spectrum of single trace operators in planar $\mathcal{N}=4$ SYM is encoded in the QSC \cite{Gromov:2013pga,Gromov:2014caa,Gromov:2015wca} (see also \cite{Gromov:2014bva,Hegedus:2016eop,Gromov:2017blm,Alfimov:2018cms,Marboe:2018ugv,marboe2018full,Gromov:2023hzc} and references therein). In this formulation, a supermultiplet is associated to a unique set of functions $\mathbf{P}_{a}(u)$ and $\mathbf{Q}_{i}(u)$ (with $a,i=1,...,4$), that have fixed analytic structure and satisfy a set of coupled difference equations called the QQ-system, the manifestation of the integrable structure. The quantum numbers are carried by their large-$u$ asymptotics
\begin{equation}\label{asymptotics}
\mathbf{P}_{a}(u) \sim A_a u^{-\tilde{M}_{a}}\qquad\text{and}\qquad \mathbf{Q}_{i}(u) \sim B_i u^{\hat{M}_{i}-1}\,,
\end{equation}
where $\tilde{M}_{a}$ are certain linear combinations of the $J_k$, $\hat{M}_{i}$ are combinations of $\Delta, S$ and $S_2$. The constants $A_a$, $B_i$ are also fixed. 

\begin{figure*}
{    \centering
\includegraphics[trim={0 22cm 0 0},clip,width=\textwidth]{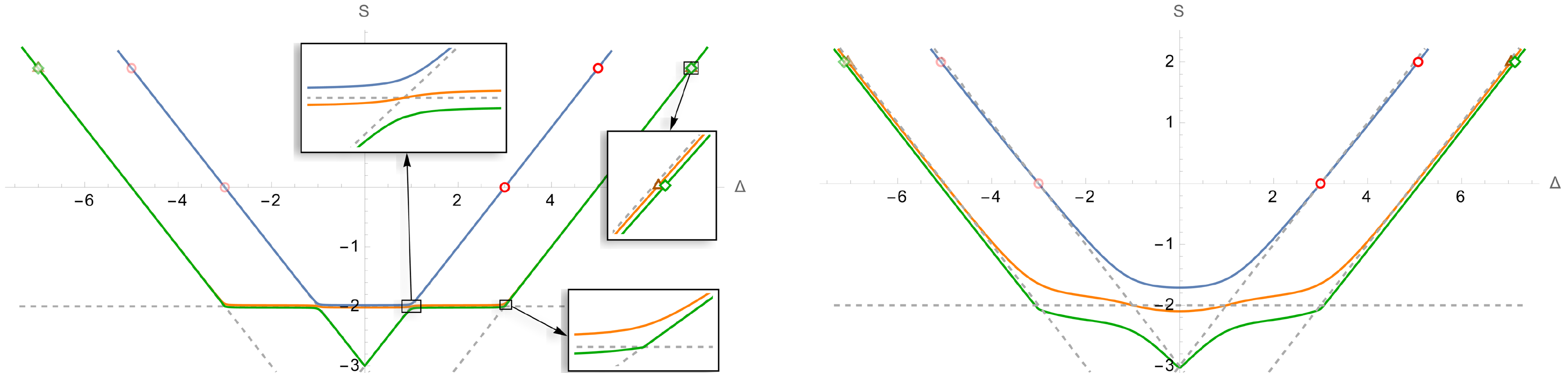}
}
    \caption{The intersection of the Riemann surface $S(\Delta)$ with the real $(\Delta,S)$ plane for $g=1/100$ (left) and $g=1/10$ (right). The blue curve is the Regge trajectory appearing in FIG. \ref{fig:leadingtraj}, the orange and green curves belong to the twist-$5$ trajectories accessed by continuation on the Riemann surface. The dashed lines show the zero-coupling limit of the curves. Local operators are indicated by markers, and shadows by their transparent counterparts.
    \label{fig:spectrumandtraj}}
\end{figure*}

One can implement the analyticity requirements on the $\mathbf{P}_a$ functions explicitly to simplify the analysis.  We use the rescaling symmetry $\mathbf{P}_a=(gx)^{-\Lambda}\mathbf{p}_a$, where $g=\sqrt{\lambda}/4\pi$ is the 't Hooft coupling, and $x(u)= \frac{u+\sqrt{u-2g}\sqrt{u+2g}}{2g}$ is the Zhukovsky variable. With an appropriate choice of $\Lambda$, the $\mathbf{p}_a$ have only square-root short branch cuts on the region $[-2g,2g]$. In our situation, $\Lambda=3/2$, and $\mathbf{p}_a$ have well-defined parity in $x$, leading to the ansatz
\begin{equation}\label{Pansatz}
\!\mathbf{p}_{a}\!=\!A_a(gx)^{-\tilde{M}_{a}+\Lambda}\!+\!\sum_{n=1}^{\infty}\!\left\{\frac{c_{1,n}}{x^{2n+1}},
\frac{c_{2,n}}{x^{2n}},
\frac{c_{3,n}}{x^{2n}},
\frac{c_{4,n}}{x^{2n-3}}\right\}\!,
\end{equation}
where we can gauge-fix $c_{4,2}~=~0$. This parametrisation converges in a neighbourhood of the cut on the second sheet, where the continued function $\tilde{\mathbf{P}}_a$ is obtained by the map $x\mapsto 1/x$.

To solve the QSC, we use various equivalent reformulations depending on the particular problem at hand (see S M for more details): for our analytic computations, we use the $\mathbf{P}\mu$-formulation \cite{Gromov:2013pga}, and compute either $\Delta$ perturbatively in the coupling $g$ as a function of $S$ or vice versa (while keeping the other quantum numbers fixed). For our numerics, on the other hand, we use some of the relations from the QQ-system. In both cases, we close the system by imposing the \emph{gluing conditions} \cite{alfimov2018bfkl}. These conditions follow from analyticity properties of the $\mathbf{Q}_i$ and relate the continued function $\tilde{\mathbf{Q}}^i(u)$ to the first-sheet $\mathbf{Q}_j(-u)$ through the \emph{gluing matrix} $M$. For our chosen quantum numbers they look as follows: 
\begin{equation}\begin{split}\label{gluing}
\tilde{\mathbf{Q}}^i(u)&=M^{ij}(u)\mathbf{Q}_j(-u)\,,\\
\tilde{\mathbf{Q}}_i(u)&=-\left(M^{-1}\right)_{ji}(u)\mathbf{Q}^j(-u)\,.
\end{split}\end{equation}
The minus sign in the lower equation is a consequence of the choice of $\Lambda$ to be half-integer rather than integer and therefore differs from the Pomeron case \cite{alfimov2018bfkl}. 

Generally, the gluing matrix $M$ is hermitian, analytic in the whole complex plane and $i$-periodic. For local operators with integer $S$ it is given by the constant matrix $M^{ij}=\ell_{1}\delta_{i,3-j}+\ell_{2}\delta_{i,7-j}$, where $\ell_{k}$ are complex numbers. To analytically continue in $S$, we need to include $u$-dependence in the gluing matrix \cite{gromov2015nnlo,alfimov2018bfkl}. The minimal choice to keep all its required properties is the following
\begin{equation}
\!\!M\!=\!\!
\setlength\arraycolsep{1pt}
\footnotesize
\begin{pmatrix}
\ell_{1} & \ell_{2} & \ell_{3} & \color{gray!80}{0}\\
\ell_{2} & \color{gray!80}{0} & \color{gray!80}{0} & \color{gray!80}{0}\\
\ell_{3} & \color{gray!80}{0} & \ell_{4} & \ell_{5}\\
\color{gray!80}{0} & \color{gray!80}{0} & \ell_{5} & \color{gray!80}{0}
\end{pmatrix}\normalsize
\!+\!
\footnotesize\begin{pmatrix}
\color{gray!80}{0} & \color{gray!80}{0} & \ell_{6} & \color{gray!80}{0}\\
\color{gray!80}{0} & \color{gray!80}{0} & \color{gray!80}{0} & \color{gray!80}{0}\\
\ell_{7} & \color{gray!80}{0} & \color{gray!80}{0} & \color{gray!80}{0}\\
\color{gray!80}{0} & \color{gray!80}{0} & \color{gray!80}{0} & \color{gray!80}{0}
\end{pmatrix}\normalsize
\! e^{2\pi u}
\!+\!
\footnotesize\begin{pmatrix}
\color{gray!80}{0} & \color{gray!80}{0} & \ell_{7} & \color{gray!80}{0}\\
\color{gray!80}{0} & \color{gray!80}{0} & \color{gray!80}{0} & \color{gray!80}{0}\\
\ell_{6} & \color{gray!80}{0} & \color{gray!80}{0} & \color{gray!80}{0}\\
\color{gray!80}{0} & \color{gray!80}{0} & \color{gray!80}{0} & \color{gray!80}{0}
\end{pmatrix}\normalsize
\! e^{-2\pi u}\,.
\end{equation}
This is strongly motivated by our analytic calculation for weak coupling close to the $\mathcal{O}_S$ local operators. (cf. SM).

\vskip 0.1in
\textit{Regge trajectories and Riemann surface.---} 
The QSC equations, together with the gluing conditions, allow for an efficient numerical algorithm to calculate the $S(\Delta)$ function (cf. SM).
Our strategy is to start from the weak coupling data of the local operators $\mathcal{O}_2$ and $\mathcal{O}_4$ computed using \cite{marboe2018full} and then move along the Regge trajectory varying $\Delta$ to obtain $S$. Moreover, we can alter the value of the coupling $g$ moving to the non-perturbative regime. This leads to the trajectories presented in the Chew-Frautschi plot in FIG.~\ref{fig:leadingtraj} where the points correspond to local operators at various values of $S$ (and their shadows related by $\Delta\!\rightarrow \!-\Delta$). These trajectories all pass through the BPS-point $(3,0)$, belonging to the protected local operator $\mathcal{O}_0 = \text{Tr}(Z^3)$, and $(-3,0)$, belonging to its shadow \footnote{Around these points, one can study the near-BPS limit of the expansion, i.e. the Taylor series of $\Delta$ with respect to $S$: $\Delta= 3+(1+f_1) S + f_2 S^2 +\ldots$, with $f_1$ the slope \cite{Basso:2011rs} and $f_2$ the curvature \cite{Gromov:2014bva} functions, both known exactly. Fitting our data in FIG.~\ref{fig:leadingtraj}, we reproduce $f_1$ and $f_2$ for $g=1/100,1/10,1/2$ with 8 and 7 digits of precision, respectively.}. 

For $g\ll 1$, FIG.~\ref{fig:leadingtraj} shows that the trajectory develops a plateau at $S=-2$ for $|\Delta|<1$, resembling the Pomeron trajectory \cite{Alfimov:2014bwa}, and signals the presence of HTs \cite{caron2023detectors}. To probe the mixing between the HTs and the twist-$3$ trajectory, we map out the neighbourhood of the plateau for \emph{complex} $\Delta$, finding several branch points (BPs) that connect sheets of a non-trivial Riemann surface. The first three sheets of the surface are presented in FIG.~\ref{fig:RiemannReal} for $g=1/2$, and the coloured dots indicate the location of the different BPs \footnote{We choose the branch cuts to run perpendicular to and away from the real $(\Delta,S)$ plane.}.

The analysis shows second-order behaviour for the monodromy around the BPs at finite $g$. Reducing $g$, the BPs approach the real $(\Delta,S)$ plane at certain integer $(\Delta, S)$ points from both sides and "cut" the Riemann surface into disconnected planes at $g=0$. On the real $(\Delta,S)$ plane, the asymptotic lines for the trajectories are the HTs and the fixed twist lines. For finite $g$, these asymptotic lines mix due to the presence of the BPs and form the trajectories. This transition is presented in FIG.~\ref{fig:spectrumandtraj} with the three Regge trajectories highlighted in colour blue, orange and green. 

In particular, the red BPs limit to $(\pm 1, -2)$, the endpoints of the plateau. This mixing resolves \emph{two} degenerate HTs and the twist-$3$ line, being the first explicit occurrence of this phenomenon, forming the blue, orange, and green trajectories. The degeneracy of HTs plays an important role in the computation of the weak-coupling intercept and its novel behaviour.

Moreover, the orange and green trajectories undergo further mixing at $(\pm 3,-2)$ including the cyan BPs \footnote{In FIG.~\ref{fig:RiemannReal} the only visible cyan BPs are on the third sheet. However, they appear also in the second sheet but outside the region of our plot.}, before asymptoting to the twist-$5$ line. 
The mixing happens between at least four trajectories: the two HTs, and the two twist-$5$ trajectories. 
We can identify the local operators $\mathcal{O}_\pm$ belonging to these trajectories at $S=2$ as being the only two length-$5$ parity singlet superprimaries with $g=0$ charges $(3,0,0,7,2,0)$ \footnote{We note that at tree-level $\mathcal{O}_\pm$ are long linear combinations of length-$5$ elementary fields and their derivatives and are not simply a combination of Tr$D^2 Z ZZZZ$- and Tr$DZDZZZZ$-type states (cf. SM). The protected operator Tr$(Z^5)$ is also not on either trajectory.}. Using the QSC we are able to compute their non-perturbative scaling dimension $\Delta_{\pm}$ (framed plot in FIG.~\ref{fig:spectrumandtraj}) matching the following weak coupling expansion
\begingroup\makeatletter\def\f@size{8}\check@mathfonts
\begin{align}\label{Deltapm}
\Delta_{\pm}\!=&7+\left(10\pm 2 \sqrt{5}\right) \!g^2\!-\!\left(\frac{63}{2}\pm\frac{27 \sqrt{5}}{2}\right) \!g^4\!+\!\left(\frac{1869}{8}\pm\frac{4287}{8 \sqrt{5}}\right) g^6\nonumber\\
&-\left(\!\left(140\pm68 \sqrt{5}\right)\zeta_3+\frac{69071}{32}\pm\frac{155773}{32 \sqrt{5}}\right)g^8\!+\mathcal{O}(g^{10}) , 
\end{align}\endgroup
obtained with the solver \cite{marboe2018full}. 

We expect similar behaviour for mixing between the two HTs and odd twist-$k$ trajectories around $(\pm k,-2)$. We see signs of mixing around $(0,-3)$ as well (green BPs), although preliminary calculations show that there are no HTs at $S=-3$ (and, in fact, not for any odd $S$). However, the understanding of further sheets of the Riemann surface and the possible roles of other HTs, e.g. with constant spin $S=-4$, is out of the scope of this Letter. 

\begin{figure}
{    \centering
\includegraphics[width=\columnwidth]{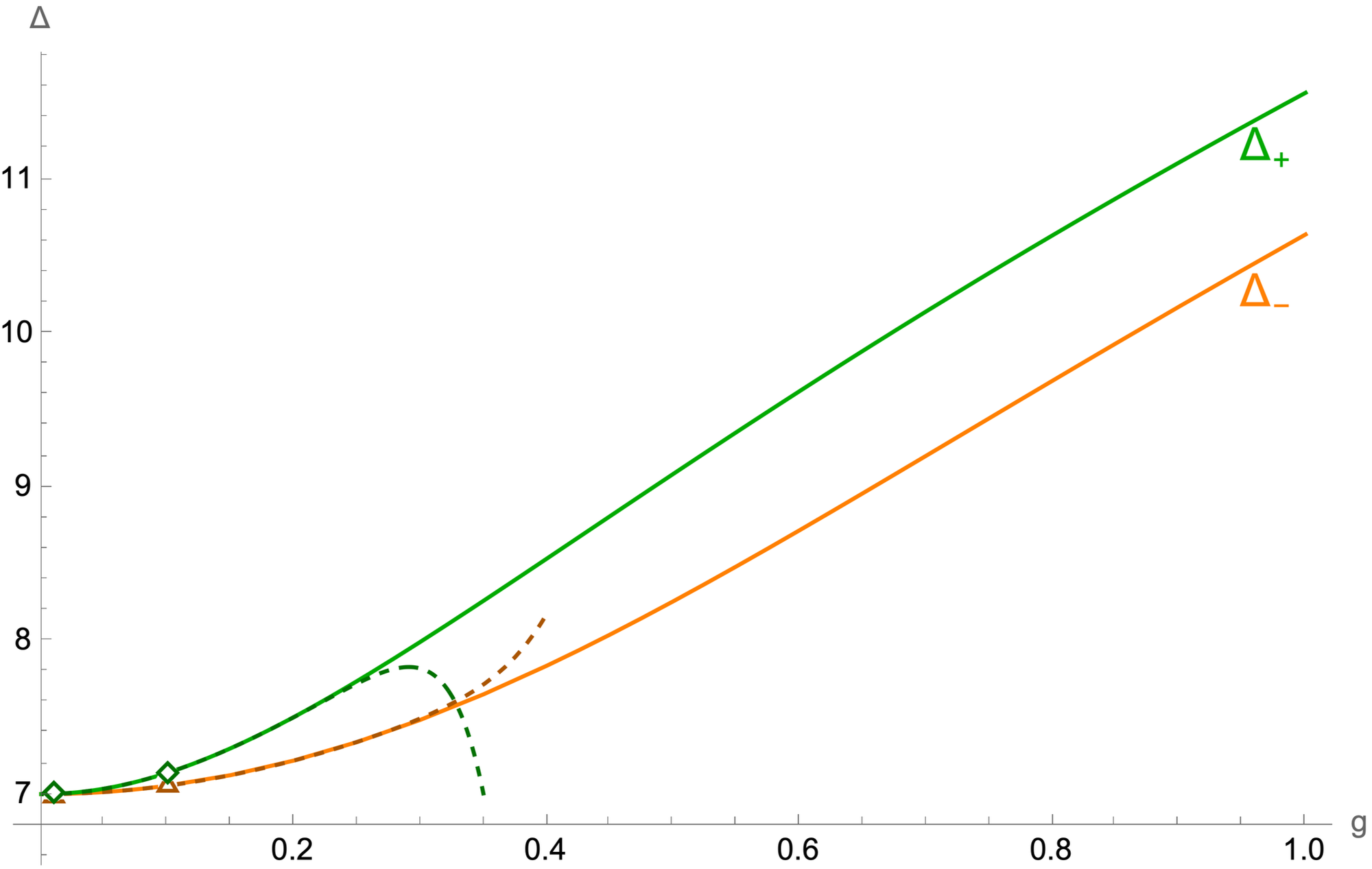}
}
    \caption{The non-perturbative scaling dimension of the twist-$5$ operators $\mathcal{O}_\pm$appearing on the orange and green trajectories of FIG. \ref{fig:spectrumandtraj} (with markers here representing the same value of $\Delta_\pm$ in FIG. \ref{fig:spectrumandtraj}). Dashed lines are the weak coupling predictions \eqref{Deltapm}.
    \label{fig:spectrum_twist_5}}
\end{figure}

\vskip 0.1in
\textit{Resolution of the HT-degeneracy.---} The weak-coupling intercept $\alpha=S(0)$ of the blue trajectory can be computed using QSC while considering $S$ as a function of $\Delta$ and focusing on finding solutions close to $(\Delta,S)\sim (0,-2)$. However, we have to be careful, since for $g=0$ also the orange trajectory passes through $(0,-2)$, and we have to account for this degeneracy. For the perturbative calculation in $g$, we take the ansatz 
\begin{equation}\label{eq:S_ansatz}
    S=-2+\sum_{i=1}^\infty I_i(\Delta) g^i
\end{equation}
and fix the remaining quantum numbers as $(J_1,J_2,J_3,S_2) = (3,0,0,0)$. We stress the inclusion of odd powers of $g$ to the intercept, which is a novelty compared to all known cases. In \cite{marboe2018full}, degeneracy observed in the $\mathbf{P}_a$ functions was resolved by including odd powers of $g$ in the $\mathbf{P}_a$-ansatz, but this did not result in odd powers in the resulting $S(\Delta)$. In our case, the trajectories (and consequently the intercept) are degenerate at weak coupling for $|\Delta|<1$ as shown in FIG.~\ref{fig:spectrumandtraj}, and we allow for odd $g$ dependence in all QSC quantities to give room for the description of new behaviour. 

Let's focus on the leading order of \eqref{eq:S_ansatz}. To compute $I_1$, we solve the QSC along similar steps as in the calculation of the Pomeron intercept \cite{Alfimov:2014bwa} (c.f. SM for the detailed calculation). First, given the ansatz \eqref{Pansatz} and the QSC equations in the $\mathbf{P}\mu$ formulation, we fix the $\mathbf{P}_a$ solutions up to next-to-leading order (NLO) in $g$. Then, from these solutions, we derive a fourth-order difference equation for the $\mathbf{Q}_i$ functions, which factorises to give a second-order one for $\mathbf{Q}_1$ and $\mathbf{Q}_3$ \footnote{Eq. \eqref{2ndorder} was first derived by M. Alfimov \cite{misha} and independently rederived by the authors. We thank him for sharing his result.} as follows
\begin{eqnarray}\label{2ndorder}
 && \frac{2 (u+i)+i g I_1 }{ (u+i)^{3/2}}\mathbf{Q}_i(u+i)+\frac{2 (u-i)-i g I_1}{ (u-i)^{3/2}}\mathbf{Q}_i(u-i)  \nonumber \\ 
  && - \frac{1 + 8 u^2 - \Delta^2 + 2 g I_1}{2 u^{5/2}}\mathbf{Q}_i (u)=0\,.
\end{eqnarray}
We solve \eqref{2ndorder} through a Mellin transformation \cite{Faddeev:1994zg,korchemsky1995bethe} and with the solution of the generalised hypergeometric differential equation \cite{smith1939}. The solutions up to NLO for $\mathbf{Q}_{1,3}$ are analytic in the upper half plane, have the desired asymptotic \eqref{asymptotics} for large $u$, and have only  $I_1$ as a free parameter. To fix $I_1$, we impose the gluing conditions \eqref{gluing}, which simplify to 
\begin{equation}\label{gluingexplicit}
\tilde{\mathbf{Q}}_1(u)=-\mathbf{Q}_3(-u)/\ell_2 \quad\text{and}\quad\tilde{\mathbf{Q}}_3(u)=\mathbf{Q}_1(-u)\, \ell_2\, , 
\end{equation}
 as well as require a regularity condition on the real axis. Due to the quantum numbers, the $\mathbf{Q}_{1,3}$ functions have an extra $\sqrt{u}$ branch cut at small coupling in addition to the Zhukovsky cut. Hence, rather than the usual combination of $\mathbf{Q}_i + \tilde{\mathbf{Q}}_i$, we must demand that 
 \begin{equation}\label{Qregularity}
\mathbf{Q}_i \sqrt{x}^{-1}+\tilde{\mathbf{Q}}_i \sqrt{x}
\end{equation}
is analytic on the real axis or equivalently that it is regular around $u\sim 0$ at small coupling 
\footnote{We thank N. Gromov for discussions on this point}. This, combined with \eqref{gluingexplicit}, leads to the consistency equation $I_1(\Delta)^2=4$, yielding two QSC solutions with $I_1(\Delta)= \pm 2$ that, interestingly, are independent of the value of $\Delta$. This matches our numerical observations: each solution belongs to one of the degenerate HTs and, as anticipated in \eqref{eq:S_ansatz}, the degeneracy of the intercept is resolved at linear order in $g$.

\begin{figure}
{    \centering
\includegraphics[width=\columnwidth]{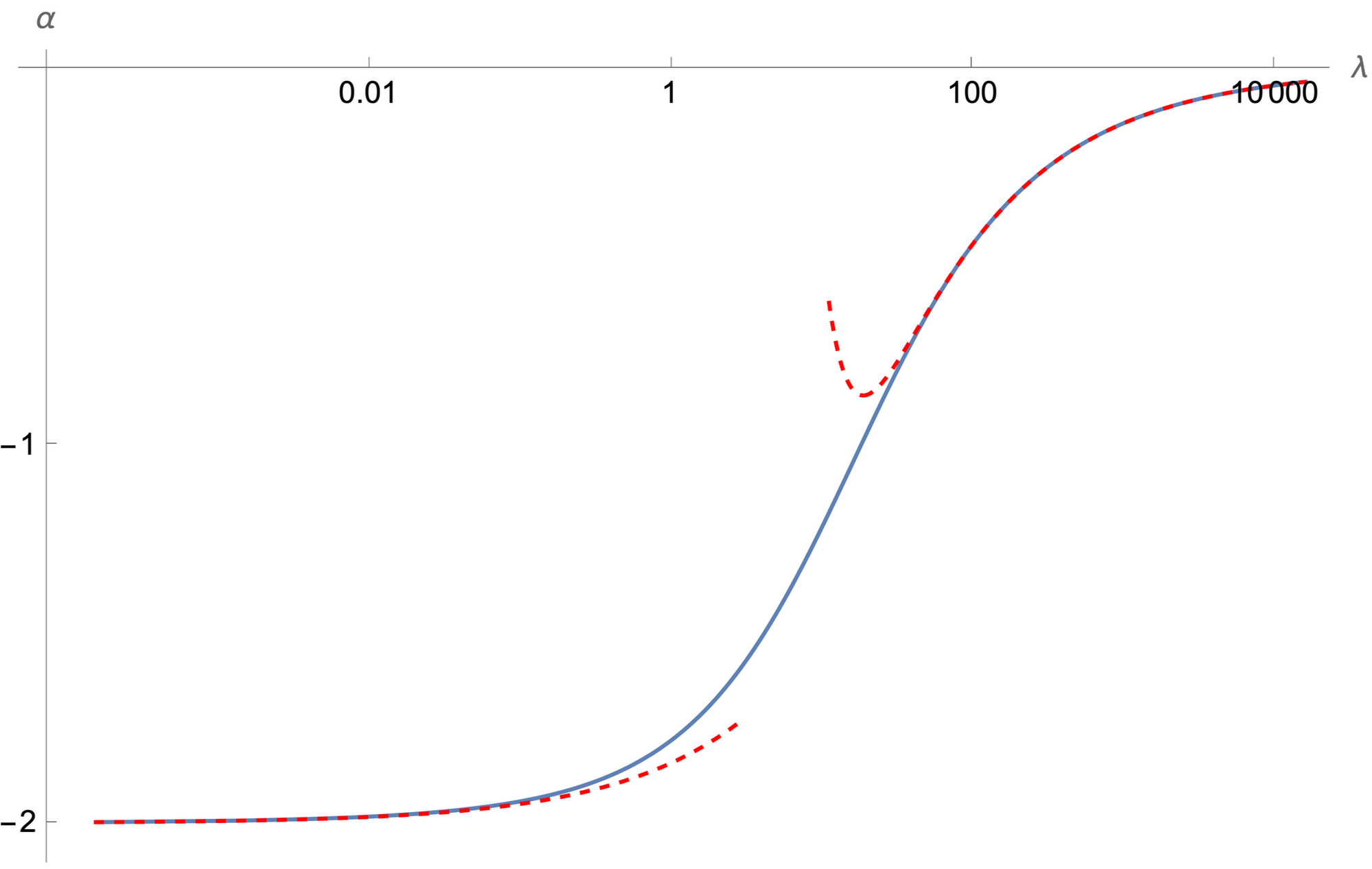}
}
    \caption{The non-perturbative intercept as a function of the coupling $\lambda=16\pi^2g^2$ (we consider the range $g \in [10^{-3},10]$). The solid line is obtained by our numerical procedure (around 250 points),  while the dashed lines are the perturbative predictions.
    \label{fig:intercept}}
\end{figure}

\vskip 0.1in
\textit{Numerical intercept.---} Finally, using our numerical algorithm, we test our prediction computing the non-perturbative intercept $\alpha(g)$ for a wide range of values of the coupling as presented in FIG.~\ref{fig:intercept} \footnote{To help the convergence of our algorithm, we compute the intercept at $\Delta=10^{-7}$.}. The curve nicely interpolates between our weak coupling 1-loop computation and the 6-loop strong coupling expansion derived in \cite{Gromov:2014bva,brower2015strong} for twist $\tau=3$. Considering that at weak coupling our data points reach 60-80 digits of precision, we fit the curve and obtain the following expansion 
\begin{equation}\begin{split}\label{interceptweak}
\alpha&=-2 +2g+16\log 2 \;g^2-\!\frac{2 \pi ^2}{3} g^3\!\\
&-204.77377158292661 g^4 \!+\!136.29333638813 g^5 \\
&+\! 4733.39078974 g^6 \!- 6116.79585 g^7\!\!+..., 
\end{split}\end{equation}
where the transcendental coefficients of the $g^2$ and $g^3$ terms fit with at least $17$ significant digits to the found numerical values while the other terms are presented with their significant digits. 

Notice that the $g^2$ term coincides with the $\Delta=0$ value of the well-known Pomeron eigenvalue \cite{jaroszewicz1982gluonic,lipatov1986bare,kotikov2004dglap}. Indeed, probing the expansion of $S(\Delta)$ for several values of the scaling dimension in the region $|\Delta|<1$, we confirm the familiar form for the $g^2$ coefficient 
\begin{equation}-4 \left(\psi\!\left(\frac{1\!-\!\Delta }{2}\right)\!+\!\psi\!\left(\frac{1\!+\!\Delta }{2}\right)\!-\!2 \psi(1)\right)\!\,,
\end{equation}
where $\psi$ is the digamma function. 

Furthermore, in addition to the linear term, the $g^3$ coefficient is also independent of $\Delta$. Hence, it is natural to ask if all the odd powers of $g$ have this feature, but our precision is not enough to corroborate it.

\vskip 0.1in
\textit{Discussion.---}
In this Letter, we showed that contrary to all other known cases for $\mathcal{N}=4$ SYM theory, the intercept \eqref{interceptweak} can have a linear dependence on the coupling $g$, which has important consequences. 

First, it facilitates a novel, fully non-perturbative relationship between the two trajectories that are degenerate at $g=0$: starting from a point on the blue trajectory in the $|\Delta|<1$ region, the analytic continuation $g\mapsto-g$, and the monodromy around the BP at $(\pm1,-2)$ at fixed $g$, lead to the same point on the orange trajectory. This means these two trajectories and their intercepts are related by the $g\mapsto-g$ exchange as long as $|\Delta|<1$. 

Second, in conformal Regge-theory \cite{Costa:2012cb}, the defining equation for the intercept is  $\Delta(S)=0$ \footnotemark[99]. Inverting this relation, one can hope to relate the perturbative Laurent expansions of the anomalous dimension and the intercept. This approach was successful for the Pomeron trajectory (see, e.g. \cite{kotikov2007dressing}); however, it breaks down in our case. The perturbative anomalous dimensions in $\mathcal{N}=4$ SYM are a function of $g^2$; hence the order by order inversion does not reproduce the linear $g$-dependence of the intercept. This shows the importance of using non-perturbative tools for Regge spectroscopy. The QSC formalism is a viable candidate for this, and this Letter also presents, through numerical optimisations and the revised regularity condition \eqref{Qregularity}, a step forward in making it applicable to more general trajectories.  

Although the full Riemann surface is much larger than we have explored in this Letter, we observe several of its features that seem to be true more generally: trajectories do not seem to cross or diverge, and in the $g\to 0$ limit, all trajectories asymptote to horizontal or fixed-twist lines. We also note that all the local operators we discussed on the Riemann surface have the same $J_1,J_2,J_3,S_2$ quantum numbers, have even spin, and are all parity singlets, corroborating the conjecture from \cite{caron2023detectors} that, in a generic CFT, \emph{all} states with the same global symmetries (including the discrete ones) are part of the same Riemann surface. 

\vskip 0.1in
\begin{acknowledgments}
\textit{Acknowledgments.---}
We thank M. Alfimov for his contributions at the early stage of this project and useful comments during its realisation. We thank A. Cavagli\`a, B. Eden, S. Ekhammar, A. Georgoudis,  J. Julius,  G. Korchemsky, P. Kravchuk, F. Levkovich-Maslyuk, J. Penedones, D. Volin, K. Zarembo, and especially N. Gromov for insightful discussions.

The work of MP is supported by European Research Council (ERC) under the European Union’s Horizon 2020 research and innovation programme (grant agreement No. 865075) EXACTC.

The work of IMSZ was supported by the grant “Exact Results in Gauge and String Theories” from the Knut and Alice Wallenberg foundation. IMSZ  received support from Nordita that is supported in part by NordForsk. 
\end{acknowledgments}

\newpage

\onecolumngrid

\setcounter{secnumdepth}{2} 
\numberwithin{equation}{section}

\setcounter{equation}{0}
\setcounter{section}{0}
\setcounter{figure}{0}
\setcounter{table}{0}
\setcounter{page}{1}
\renewcommand\theequation{S\Roman{section}.\arabic{equation}}

\renewcommand{\thesection}{\Roman{section}}
\renewcommand{\thefigure}{S\arabic{figure}}
\renewcommand{\thetable}{S\arabic{table}}
\renewcommand{\tablename}{TABLE}

\renewcommand{\bibnumfmt}[1]{[SM#1]}
\renewcommand{\citenumfont}[1]{SM#1}

\section*{Supplemental material}
\setcounter{section}{0}
\section{ Conformal primaries on twist-three trajectories}\label{sec:conf_prim}

Describing superconformal multiplets of planar $\mathcal{N}=4$ super Yang-Mills theory requires more than just the six quantum numbers (or Cartan charges) $(J_1, J_2, J_3, \Delta_0, S, S_2 )$ at tree-level (i.e. in the free theory), being the three $R$-symmetry charges $J_i$, the Lorenz spins $S$ and $S_2$ and the classical scaling dimension $\Delta_0$. This latter charge receives quantum corrections and must in practice be computed separately \emph{after} specifying which multiplet one wishes to study. Following \cite{SMmarboe2018full} one can label multiplets by oscillator numbers $\{n_{\mathbf{b}_1},n_{\mathbf{b}_2} | n_{\mathbf{f}_1}, n_{\mathbf{f}_2},n_{\mathbf{f}_3},n_{\mathbf{f}_4} | n_{\mathbf{a}_1}, n_{\mathbf{a}_2} \}$, which count the number of each of the appearing oscillators in the oscillator representation of $\mathfrak{psu}(2,2|4)$ at tree level. In addition, one must choose a grading for the algebra, with two common choices being the ABA or beauty grading (see, e.g. \cite{SMBeisert:2010kp}). Generically, there are multiple super highest-weight states with the same oscillator numbers, hence additional information is needed to fully specify them. 

In order to determine which states to study, we start by considering all multiplets with super-highest weights with $\Delta_0 \leq 10$, twist $\Delta_0 -S=3$ and length three (see \cite{SMmarboe2018full} and its ancillary files) and plot all the one-loop anomalous dimensions as a function of spin $S$. 
\begin{figure}[!h]
{    \centering
\includegraphics[width=0.8\textwidth]{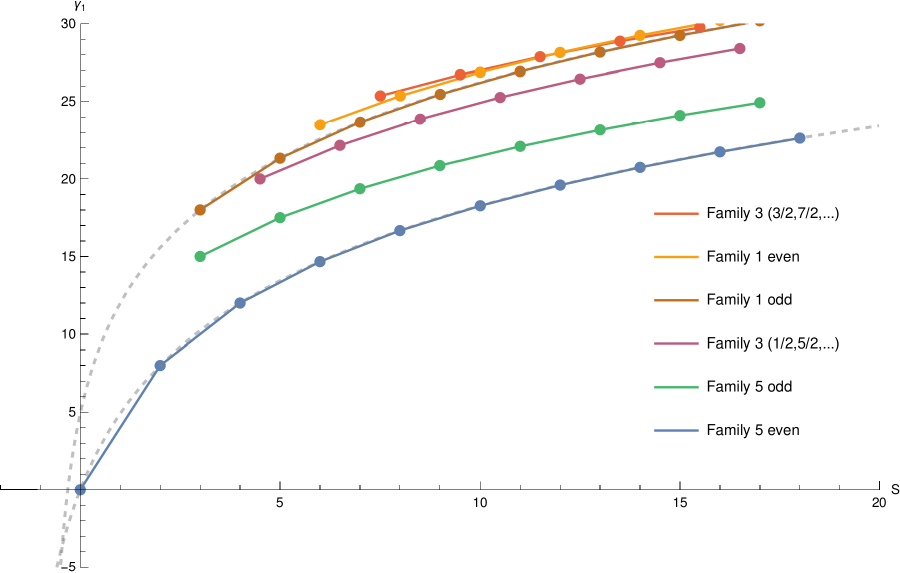}
}
    \caption{The one-loop anomalous dimension $\gamma_1$ of the first six lowest-lying twist-three parity-singlet trajectories. The dots indicate the exact one-loop scaling of local operators, connected by straight lines to guide the eye. The dashed lines indicate the known analytic expressions for the analytic continuation of $\gamma_1$. 
    \label{fig:traj_all}}
\end{figure}
It is not difficult to recognise trajectories in this picture (see FIG.~\ref{fig:traj_all}, and this yields five families of oscillators: 
\begin{center}
\begin{tabular}{c|c|c|c}
family & oscillator numbers & Cartan numbers & local operators at\\
\hline
1 & $\{0,S-1 | 2,2,0,0| S+1,0\}$ & $(2,0,0,3+S,S,-1)$ & $S=3,5,6,7,\ldots$ \\
2 & $\{0,S+1| 3,3,1,1 | S-1,0\}$ & $(2,0,0,3+S,S,1)$ & $S=3,5,6,7,\ldots$ \\
3 & \quad$\{0,S-1/2 | 3,2,0,0| S+1/2,0\}$\quad &\quad $(5/2,1/2,-1/2,3+S,S,-1/2)$\quad & $S=7/2,9/2,\ldots$ \\
4 & $\{0,S+1/2 | 3,3,1,0 | S-1/2,0\}$ & $(5/2,1/2,-1/2,3+S,S,1/2)$ & $S=7/2,9/2,\ldots$ \\
5 & $\{0,S | 3,3,0,0 | S,0\}$ & $(3,0,0,3+S,S,0)$ & $S=0,2,3,4,\ldots$ \\
\end{tabular}
\end{center}
Families 1 and 2, as well as families 3 and 4, are conjugate and have the same scaling dimensions. For increasing $S$, the number of multiplets associated with a single choice of oscillator numbers increases as well, and as such, there are more than five trajectories associated with these five families. In particular, generically, states with even and odd spin $S$ are on different trajectories. We focus on family $5$, the lowest-lying trajectory in FIG.~\ref{fig:traj_all} containing the protected operator Tr$(Z^3)$ with oscillator numbers $\{0,0 | 3,3,0,0| 0,0\} $. This is the only family with $S_2 = 0$, simplifying some of our analysis. The local super-highest weights on this trajectory are those with oscillator numbers $\{0,S|3,3,0,0|S,0\}$ for $S\geq 0$ and even spin $S\in 2\mathbb{Z}$ and are parity singlets \cite{SMkristjansen2012review}. For $S\geq 2$, these states are all super highest-weight states in ABA-grading. They are linear combinations of single-trace operators with the same oscillator numbers (after a consistent choice of relating fields and oscillators, we follow \cite{SMmarboe2018full}) consisting of derivatives and the scalar $Z$-field, and give the first few local operators at tree level in the following table: 
\begin{center}
\begin{tabular}{c|l}
S &  single-trace form \\
\hline
0 & $\text{Tr}(Z^3)$ \\
2 & $2\text{Tr}(DZ DZ Z)-\text{Tr}(D^2Z Z Z)$ \\
4 &  $18\text{Tr}(D^2Z D^2Z Z)-\text{Tr}(D^3Z Z DZ)-8 \text{Tr}(D^3Z DZ Z)+\text{Tr}(D^4Z Z Z)$ \\
6 & $400 \text{Tr}(D^3Z DZ D^2Z)-400 \text{Tr}(D^3Z D^2Z DZ)-25 \text{Tr}(D^4Z D^2Z Z)+4 \text{Tr}(D^5Z Z DZ)-4 \text{Tr}(D^5Z DZ Z)$
\end{tabular}
\end{center}
where the normalisation is chosen for convenience. For $S=6$ we selected the parity singlet out of the three-dimensional space of super-highest weights. This specifies all needed input to consider the behaviour of the trajectory in the BFKL regime.

\section{Quantum Spectral Curve}\label{SM:QSC}
\subsection{General formalism}
The spectrum of single trace operators in planar $\mathcal{N}=4$ SYM is encoded in the Quantum Spectral Curve (QSC) \cite{SMGromov:2013pga,Gromov:2015wca}. In this formulation, a supermultiplet is associated to a unique set of functions $\mathbf{P}_{a}(u)$ and $\mathbf{Q}_{i}(u)$ with $a,i=1,...,4$, which depend on the spectral parameter $u$. Global charges are carried by their large-$u$ asymptotics
\begin{equation}\label{SMasymptotics}
\mathbf{P}_{a}(u) \sim A_a u^{-\tilde{M}_{a}},\quad \mathbf{Q}_{i}(u) \sim B_iu^{\hat{M}_{i}-1}\qquad\text{and}\qquad
\mathbf{P}^{a}(u) \sim A^a u^{\tilde{M}_{a}-1},\quad \mathbf{Q}^{i}(u) \sim B^iu^{-\hat{M}_{i}}\,,
\end{equation}
which in general are given by 
\begin{equation}\begin{split}
\tilde{M}_a&=\left\{\frac{J_1+J_2-J_2+2}{2} ,\frac{J_1-J_2+J_2}{2} ,\frac{-J_1+J_2+J_2+2}{2},\frac{-J_1-J_2-J_2}{2} \right\}\\
\hat{M}_i&=\left\{\frac{\Delta -S_2-S+2}{2} ,\frac{\Delta +S_2+S}{2},\frac{-\Delta +S_2-S+2}{2} ,\frac{-\Delta -S_2+S}{2} \right\}
\end{split}\end{equation}
with coefficients fixed by the following relations (without summation over $a$ and $j$)
\begin{equation}
\label{eq:A_and_B}
A^aA_a=i \frac{\prod_j(\tilde{M}_a-\hat{M}_j)}{\prod_{b\neq a}(\tilde{M}_a-\tilde{M}_b)}\quad\text{and}\quad
B^jB_j=i \frac{\prod_a(\hat{M}_j-\tilde{M}_a)}{\prod_{k\neq j}(\hat{M}_j-\hat{M}_k)}. 
\end{equation}
In the case in which operators belong to the so-called left-right symmetric subsector (as in our case), $\mathbf{P}^a$ and $\mathbf{Q}^i$ are related to their lower-index counterparts through the constant matrix $\chi^{ab} =(-1)^{a}\delta_{a,5-b}$.

In order to solve the QSC, in our case meaning determining the dimension $\Delta(S)$ or the spin $S(\Delta)$ of a given family of operators, we exploit the fact that $\mathbf{P}_a$ is analytic apart from a square-root short cut $[-2g,2g]$ where $g=\sqrt{\lambda}/4\pi$ is the 't Hooft coupling. Notice that in the whole QSC system, the only information about the coupling is encoded in the branch point locations!
We also use the rescaling symmetry $\mathbf{P}_a=(gx)^{-\Lambda}\mathbf{p}_a$ that preserves its analyticity property and asymptotic behaviour with $\Lambda$ an arbitrary constant. Those two properties allow to parametrise $\mathbf{p}_a$ as follows
\begin{equation}\label{SMPansatz}
\mathbf{p}_{a}=A_a(gx)^{-\tilde{M}_{a}+\Lambda}+\sum_{n=1}^{\infty}\left\{\frac{c_{1,n}}{x^{2n+1}},
\frac{c_{2,n}}{x^{2n}},
\frac{c_{3,n}}{x^{2n}},
\frac{c_{4,n}}{x^{2n-3}}\right\},
\end{equation}
where $x(u)= \frac{u+\sqrt{u-2g}\sqrt{u+2g}}{2g}$ is the Zhukovsky variable and $\Lambda=3/2$ fixes the parity of the expansion.
The $H$-symmetry of the QSC equations allows us to gauge-fix $c_{4,2}~=~0$.
This parametrisation converges in a neighbourhood of the cut on the second sheet, where the continued function $\tilde{\mathbf{P}}^a$ is obtained by the map $x\mapsto 1/x$.

\subsection{Numerical algorithm}
There are several different ways to solve the QSC, as there are typically many more equations and relations than there are unknowns, but whether one is able to solve these equations depends very much on the particular state under consideration. It is therefore important to choose a subset of equations that are efficient, i.e. not overdetermined, to solve yet still form yield a solution, i.e. they are closed. We will now first describe which choice we make for our numerical algorithm. 

Given the ansatz for $\mathbf{P}_a$, $\mathbf{Q}_i$ is obtained by solving the P-Q system
\begin{equation}\label{PQsystem}
Q_{a|i}^+ - Q_{a|i}^- =\mathbf{P}_a\,\mathbf{Q}_i \quad\text{with}\quad
\mathbf{Q}_{i} = -\mathbf{P}^{a} Q_{a|i}^+
\end{equation}
with the auxiliary function $Q_{a|i}^{\pm}(u)=Q_{a|i}(u\pm\tfrac{i}{2})$ bounded at large $u$ as
\begin{equation}
Q_{a|i}\sim i \frac{A_aB_i}{\tilde{M}_{a}-\hat{M}_i}u^{-\tilde{M}_{a}+\hat{M}_i}\;.
\end{equation}
Substituting the second equation of \eqref{PQsystem} into the first one, we first solve the homogeneous finite difference equation to determine $Q_{a|i}(u)$, then 
we evaluate $\mathbf{Q}_i(u)$ for $u\in [-2g,2g]$ along the real axis.
The function $\mathbf{Q}_i$ has an infinite series of short cuts in the lower half-plane.
Going through the first short cut at $[-2g,2g]$, one would find a function
$\tilde{\mathbf{Q}}_i(u)$ that is analytic in the lower-half-plane and obtained by the 2nd equation of \eqref{PQsystem} substituting $\mathbf{P}$ with $\tilde{\mathbf{P}}$.

In order to close the system of equations, the crucial observation is that $\tilde{\mathbf{Q}}$ has the same analytic structure as the initial $\mathbf{Q}_i$ or $\mathbf{Q}^i$ with $u\mapsto -u$. In general, this is true for the complex conjugated $\bar{\mathbf{Q}}$, but given the definite parity of the $\mathbf{P}$-functions under $u\leftrightarrow -u$, we can use the following weaker condition
\begin{equation}\begin{split}\label{SMgluing}
\tilde{\mathbf{Q}}^i(u)&=M^{ij}(u)\mathbf{Q}_j(-u)\,,\\
\tilde{\mathbf{Q}}_i(u)&=-\left(M^{-1}\right)_{ji}(u)\mathbf{Q}^j(-u)\,.
\end{split}
\end{equation}
In general, the gluing matrix $M$ is hermitian, analytic in the whole complex plane and $i$-periodic. For physical states (with integer $S$) it is given by the constant matrix $M^{ij}=\ell_{1}\delta_{i,3-j}+\ell_{2}\delta_{i,7-j}$, where $\ell_{k}$ are complex numbers. To move away from physical states along the Regge trajectories allowing $S$ to be non-integer, the gluing matrix has to depend on $u$. The minimal choice to keep all its required properties is the following
\begin{equation}
\!\!M\!=\!\!
\setlength\arraycolsep{1pt}
\footnotesize
\begin{pmatrix}
\ell_{1} & \ell_{2} & \ell_{3} & \color{gray!80}{0}\\
\ell_{2} & \color{gray!80}{0} & \color{gray!80}{0} & \color{gray!80}{0}\\
\ell_{3} & \color{gray!80}{0} & \ell_{4} & \ell_{5}\\
\color{gray!80}{0} & \color{gray!80}{0} & \ell_{5} & \color{gray!80}{0}
\end{pmatrix}\normalsize
\!+\!
\footnotesize\begin{pmatrix}
\color{gray!80}{0} & \color{gray!80}{0} & \ell_{6} & \color{gray!80}{0}\\
\color{gray!80}{0} & \color{gray!80}{0} & \color{gray!80}{0} & \color{gray!80}{0}\\
\ell_{7} & \color{gray!80}{0} & \color{gray!80}{0} & \color{gray!80}{0}\\
\color{gray!80}{0} & \color{gray!80}{0} & \color{gray!80}{0} & \color{gray!80}{0}
\end{pmatrix}\normalsize
\! e^{2\pi u}
\!+\!
\footnotesize\begin{pmatrix}
\color{gray!80}{0} & \color{gray!80}{0} & \ell_{7} & \color{gray!80}{0}\\
\color{gray!80}{0} & \color{gray!80}{0} & \color{gray!80}{0} & \color{gray!80}{0}\\
\ell_{6} & \color{gray!80}{0} & \color{gray!80}{0} & \color{gray!80}{0}\\
\color{gray!80}{0} & \color{gray!80}{0} & \color{gray!80}{0} & \color{gray!80}{0}
\end{pmatrix}\normalsize
\! e^{-2\pi u}\,. \label{eq:SMgluingmatrix}
\end{equation}

The QSC equations, together with the gluing conditions, allow for an efficient numerical algorithm (see also \cite{SMGromov:2017blm}).
Given the coupling $g$ and the spin $S$, the aim is to determine with high-precision the parameters $c_{a,n}$ in \eqref{SMPansatz}, from which one reads off the scaling dimension from
the exponents in \eqref{SMasymptotics}. The algorithm can be summarized as follows
\begin{itemize}[leftmargin=*]
\item Truncate the series \eqref{SMPansatz} at order $n=N_{P}$ and guess the initial set of values $\{c_{a,n}\}_{seed}$ for the coefficients. The value of $N_P$ is proportional to the precision wanted for the output, namely higher $N_P$, higher the number of significant digits. In our case, we set it in the range $N_P=[6,25]$ depending on the needed precision. The algorithm is very sensible to the choice $\{c_{a,n}\}_{seed}$ for a given $g$ and $S$. In order to select good starting values for those parameters, our strategy is to solve the QSC system analytically for a few orders at weak coupling and then read off the $\{c_{a,n}\}_{seed}$ from there. 
\item Given the $\mathbf{P}$-functions in terms of $\{c_{a,n}\}_{seed}$, find the series $Q_{a|i}=u^{-\tilde{M}_{a}+\hat{M}_i}\sum_{n=0}^{N_Q}\tfrac{B_{a,i,n}}{u^{2n}}$ up to the cut $N_Q$ at large $u$ solving \eqref{PQsystem}. From this equation, we obtain a linear system of equations on the coefficients $B_{a,i,n}$ at large $u$. The value of $N_Q$ contributes to the final precision of the output but to a lesser extent than $N_P$. In our case, we set it in the range $N_Q=[10,20]$. 
\item Starting from $Q_{a|i}$ at large $\text{Im}(u)$, move down to the real
axis iterating \eqref{PQsystem} to find $Q_{a|i}(u+i/2)$.
\item Having $Q_{a|i}$ computed, reconstruct $\mathbf{Q_i}$ and $\tilde{\mathbf{Q}}^i$ in terms of $\{c_{a,n}\}_{seed}$ using the second equation of \eqref{PQsystem} and its version for the analytically continued functions.
\item Constrain parameters using the gluing conditions \eqref{SMgluing}. Due to the truncation at $N_P$, gluing condition cannot be satisfied exactly. Hence, the strategy is to minimize the discrepancy in \eqref{SMgluing} by adjusting $\Delta$ and $\{c_{a,n}\}_{seed}$ by a small amount $\epsilon$ at some set of probe points on the cut $u_k\in (-2g,2g)$. The number of sample points on the cut has to be sufficiently big to have enough equations to close the system. In our case we set this number to be $4N_P$. Update the parameters using the Newton method and iterate until the target precision is reached.
\end{itemize}
Once the algorithm converges to the desired precision, we store $g$, $S$ and $\Delta$ together with the final values of the parameters $\{c_{a,n}\}$. Then, we vary the desired parameter by a small amount $\delta$ (for instance, we can keep $g$ fixed and vary the spin such as $S'=S+\delta$) and interpolate the $\{c_{a,n}\}$ accordingly to generate the new set of starting parameters $\{c_{a,n}\}_{seed}$. Now we use $g$, $S'$ and the new $\{c_{a,n}\}_{seed}$ to restart the algorithm. It is understood that the more runs we complete, the more points we have for efficient interpolation and to generate new starting values for the parameters, thus allowing us to increase the value of $\delta$. However, it often happens that due to some divergences appearing in the QSC equations when $S$ or $\Delta$ are close to integer values, one has to drastically reduce the value of the step $\delta$ in order to allow the algorithm to converge. In those cases, sometimes it is also useful to allow some of the parameters to move in the imaginary direction to go around the divergence.

Finally, since $S$ and $\Delta$ enter the QSC system on equal footing, we can easily switch from computing $\Delta(S)$ to finding $S$ for a given $\Delta$. This is very useful to determine the quantities of interest of this work such as the Regge trajectories and the intercept. 

\subsection{The $\mathbf{P}\mu$-system}
For our analytic computation, we follow a slightly different approach to our numerical algorithm described above. Rather than using \eqref{PQsystem} directly to pass the ansatz \eqref{SMPansatz} to the $\mathbf{Q}_i$ and perform all equation solving there, we first use the $\mathbf{P}\mu$-system to constrain the ansatz directly \cite{SMGromov:2013pga}. 

The $\mathbf{P}\mu$-system relates the $\mathbf{P}_a$ functions to their second-sheet evaluations $\tilde{\mathbf{P}}_a$ through the equations
\begin{equation}
\label{eq:Pmu1}
    \tilde{\mathbf{P}}_a = \mu_{ab} P^b, 
\end{equation}
where $\mu$ is a totally antisymmetric $4\times 4$ matrix depending on $u$ with infinitely many branch points at $\pm 2g +i\mathbb{Z}$. Denoting its continuation through the real-axis branch cut by $\tilde{\mu}_{ab}$, it satisfies
\begin{equation}
\label{eq:Pmu2}
    \mu_{ab}-\tilde{\mu}_{ab} = \tilde{\mathbf{P}}_a \mathbf{P}_b - \mathbf{P}_a \tilde{\mathbf{P}}_b\, ,\quad \mu_{ab}^{[2]} = \tilde{\mu}_{ab}
\end{equation}
in the strip $0<\text{Im}(u)<1$. Together with the prescribed asymptotics \eqref{SMasymptotics} and the requirements that all $\mathbf{P}_a$ and $\mu_{ab}$ have no poles and are bounded at branch points, this forms the $\mathbf{P}\mu$-system.

\section{Analytic continuation close to the local operators at weak coupling}
\label{sec:TQ_weak_coupling}
\subsection{Derivation of the TQ equation}
The twist-three trajectory we selected in Section~\ref{sec:conf_prim} can be analysed using the QSC. As a first step, we look for a solution to the $\mathbf{P}\mu$-system  at the quantum numbers $(J_1,J_2,J_3,S_2) = (3,0,0,0)$ of a local operator with $\Delta=3+S+\mathcal{O}(g^2)$ and arbitrary $S$. Such a solution must have $\mathbf{P}_a$ asymptotics \eqref{SMasymptotics} parametrised by 
\begin{equation}
    \tilde{M}_a = \frac{1}{2} \{ 5, 3, -1 ,-3\}\,. 
\end{equation}
We can use the $H$-symmetry of the QSC to choose a convenient gauge for the $A$ coefficients in \eqref{SMasymptotics}, as only their product is constrained by \eqref{eq:A_and_B}. Writing $A_aA^a$ for the right-hand side in \eqref{eq:A_and_B} we set
\begin{equation}
    A_a = \{-A_1A^1,A_2A^2, 1, 1\}\, , \quad A^a = \{-1,1,A_3 A^3, A_4A^4\}\,. 
\end{equation}
The convenience of this choice is evident when we first analyse the QSC at $g=0$: expanding $A_1 A^1$ it is easy to see that $A_1$ vanishes, meaning that $\mathbf{P}_1$ does as well, simplifying the equations significantly. As originally described in \cite{SMMarboe:2014gma}, the $\mathbf{P}\mu$-system splits and yields an equation involving only $\mu_{12}$ and $\mathbf{P}_2$ and $\mathbf{P}_3$. Introducing $Q(u)= \alpha \, \mu_{12}(u+i/2)$ with  some unimportant normalisation factor $\alpha$, one can derive an equation that takes the form of a $TQ$-equation
\begin{equation}
    T Q(u) + \frac{1}{(\mathbf{P}_2^+)^2} Q(u+i) + \frac{1}{(\mathbf{P}_2^-)^2} Q(u-i) = 0\,,
\end{equation}
where $T=\frac{\mathbf{P}_3^-}{\mathbf{P}_2^-} - \frac{\mathbf{P}_3^+}{\mathbf{P}_2^+} +\frac{1}{(\mathbf{P}_2^-)^2} +\frac{1}{(\mathbf{P}_2^+)^2}$. Approximating all $\mathbf{P}_a$ by their lowest order solution and plugging them in the result then yields the equation
\begin{equation}
\label{eq:TQ}
    \left[2 u^3 +\left(-\frac{3}{2} -S(S+2)\right)u\right] Q(u)=\left(u+\frac{i}{2}\right)^3 Q(u+i) +\left(u-\frac{i}{2}\right)^3 Q(u-i) \, .
\end{equation}

By solving the $TQ$ equation for arbitrary, complex, $S$ we ultimately gain information about the asymptotic behaviour of $\mu_{12}$, which we can use to postulate the correct behaviour for the gluing matrix \eqref{eq:SMgluingmatrix}. For this calculation, we follow the idea worked out in \cite{SMJanik:2013nqa}.

\subsection{Solution of the TQ equation}

We solve the difference equation \eqref{eq:TQ} along the steps explained in detail in Section \ref{sec:solving_diff_eq}.  First, we relate the difference equation to a differential equation through Mellin transformation \eqref{eq:Q_ansatz}. By searching for the solution in the form $\tilde{Q}(z)=\sqrt{z(1-z)}f(4z(1-z))$, we find that $f(y)$ satisfies exactly the generalised hypergeometric differential equation with parameters $\alpha_1=-S/2, \alpha_2=1+S/2,\alpha_3=1/2$, and $\beta_1=\beta_2=1$. Using the solutions \eqref{eq:log_solution_n3_b} and the relations \eqref{eq:int_identity}, \eqref{eq:int_identity_2}, and \eqref{eq:polygamm_multiplications_theorem}, leads to three solutions
\begin{equation}
\begin{aligned}
Q^{(1)}(u)=&\sum_{n=0}^\infty \frac{(-S/2)_n (1+S/2)_n (1/2-i u)_n (1/2+i u)_n}{(n!)^4}={}_{4}F_{3}
    \left( \genfrac{}{}{0pt}{0}{-S/2,1+S/2, 1/2+i u,1/2-i u}{1,1, 1};1\right) \,, \\
Q^{(2)}(u)=&\sum_{n=0}^\infty \frac{(-S/2)_n (1+S/2)_n (1/2-i u)_n (1/2+i u)_n}{(n!)^4} \\
&\times \left\{\psi_0\left(-S/2+n\right)+ \psi_0\left(1+S/2+n\right)+ \psi_0\left(1/2-iu+n\right)+\psi_0\left(1/2+iu+n\right) - 4\psi_0\left(1+n\right) \right\} \,, \\
Q^{(3)}(u)=&\sum_{n=0}^\infty \frac{(-S/2)_n (1+S/2)_n (1/2-i u)_n (1/2+i u)_n}{(n!)^4} \\
&\times \Bigg\{\left(\psi_0\left(-S/2+n\right)+ \psi_0\left(1+S/2+n\right)+ \psi_0\left(1/2-iu+n\right)+\psi_0\left(1/2+iu+n\right) - 4\psi_0\left(1+n\right)\right)^2 \\
&\,\,\,+\psi_1\left(-S/2+n\right)+ \psi_1\left(1+S/2+n\right)+ \psi_1\left(1/2-iu+n\right)+\psi_1\left(1/2+iu+n\right) - 4\psi_1\left(1+n\right)\Bigg\} \,, \label{eq:sol_Q3}
\end{aligned}
\end{equation}
where $(x)_n$ is the Pochhammer symbol and $\psi_k(x)$ are the polygamma functions.

The first function $Q^{(1)}(u)$ is exactly the known solution for even integer spin $S$ \cite{SMBeccaria:2007cn}. It properly reproduces the 1-loop anomalous dimension
\begin{equation}
\gamma_{1} = 2i \partial_u \log \frac{Q(u+i/2)}{Q(u-i/2)} \bigg|_{u=0} = 8 S_1\left(\frac{S}{2}\right) \, ,
\end{equation}
where 
\begin{equation}
    S_a(M)=\sum_{l=1}^M \frac{(\mathrm{sign}(a))^l}{l^{|a|}}
\end{equation}
are the harmonic numbers.

$Q^{(1)}(u)$ is not a solution of \eqref{eq:TQ} for general $S$ that we can easily see from numerical evaluation. $Q^{(2)}(u)$ and $Q^{(3)}(u)$ satisfy \eqref{eq:TQ} for general $S$, hence the full solution should have the form 
\begin{equation}
    Q(u)=f_1(u)Q^{(2)}(u) +f_2(u)Q^{(3)}(u) \,,
    \label{eq:Q_full_ansatz}
\end{equation}
where $f_{j}(u)$ are $i$-periodic functions. 

Following the argument in \cite{SMJanik:2013nqa}, we can extract the derivatives of our desired solution around $u=\pm i/2$ by doing a subtle analysis of $Q^{(1)}(u)$, namely
\begin{equation}
\begin{aligned}
Q(\pm i/2)&=1 \,,\\
Q'(\pm i/2)&=\mp2i S_1(M) \,,\\
Q''(\pm i/2)&=-4 S_1(M)^2 \,,\\
Q'''(\pm i/2)&=\pm 8i \left[S_1(M)^3-S_3(M)\right]\,,
\label{eq:Q_derivatives}
\end{aligned}    
\end{equation}
where $M=S/2$ .
To analytically continue the harmonic sums we can use the representation \cite{SMKotikov:2005gr}
\begin{equation}
    S_k(x)=\frac{(-1)^{k-1}}{(k-1)!}\left(\psi_{k-1}(x+1)-\psi_{k-1}(1)\right)\,. 
 \label{eq:harmonic_sum_continuation}
\end{equation}

With all these ingredients, \eqref{eq:Q_derivatives} restricts the functional forms of the $f_j(u)$ in \eqref{eq:Q_full_ansatz}, and we can postulate the full physical solution of the TQ equation \eqref{eq:TQ} for general $S$
\begin{equation}
    Q(u)=\frac{1+\cosh (2\pi u)}{2\pi}\left( -\cot (\pi S/2)Q^{(2)}(u) +\frac{1}{2\pi} Q^{(3)}(u)\right) \,.
    \label{eq:Q_full_ansatz_solution}
\end{equation}

We perform consistency checks for this $Q(u)$ in the next Section.

\subsection{Consistency checks of the solution}

We perform two consistency checks for the solution \eqref{eq:Q_full_ansatz_solution}. The first is to take the $S\to 2M$ limit of the formula with $M$ being an integer. This should give the local operator solutions  $Q^{(1)}(u)$. It is straightforward to  expand the solutions $Q^{(2)}(u)$ and $Q^{(3)}(u)$ around $S=2M+\epsilon$
\begin{equation}
\begin{aligned}
Q^{(2)}(u)&=\sum_{k=-1}^\infty \hat{q}_k^{(2)}(u) \epsilon^k \,,\\ 
Q^{(3)}(u)&=\sum_{k=-2}^\infty \hat{q}_k^{(3)}(u) \epsilon^k \,,
\end{aligned}
\end{equation}
where the coefficients $\hat{q}_k^{(j)}$ are typically  infinite sums. 

\begin{figure}
    \centering
    \includegraphics[width=0.49\columnwidth]{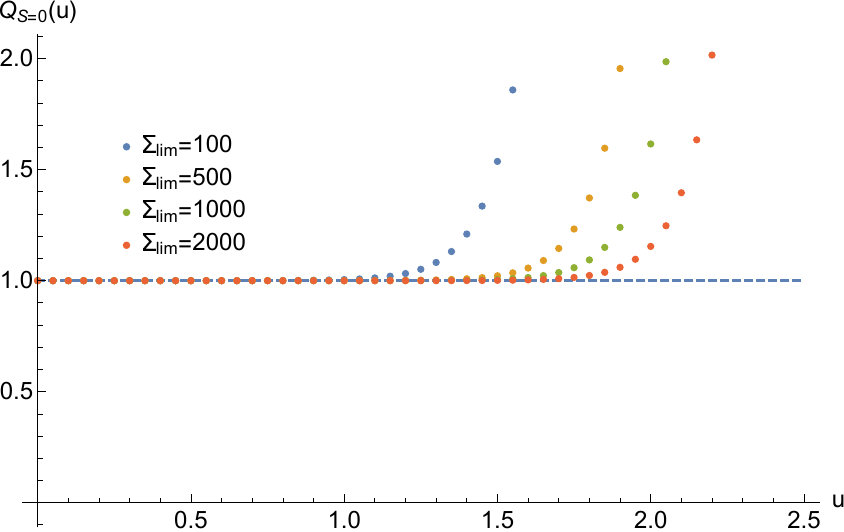}\hfill
    \includegraphics[width=0.49\columnwidth]{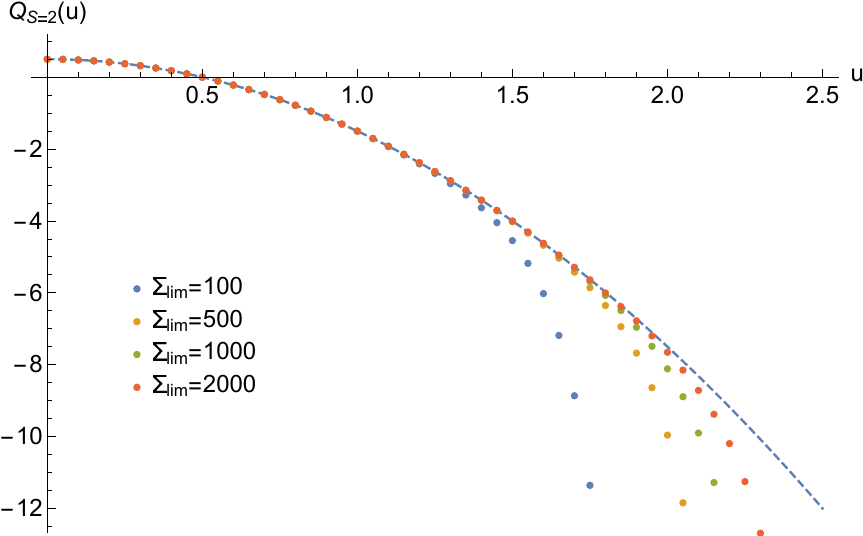}
   
    \vspace{0.3 cm}
    \includegraphics[width=0.49\columnwidth]{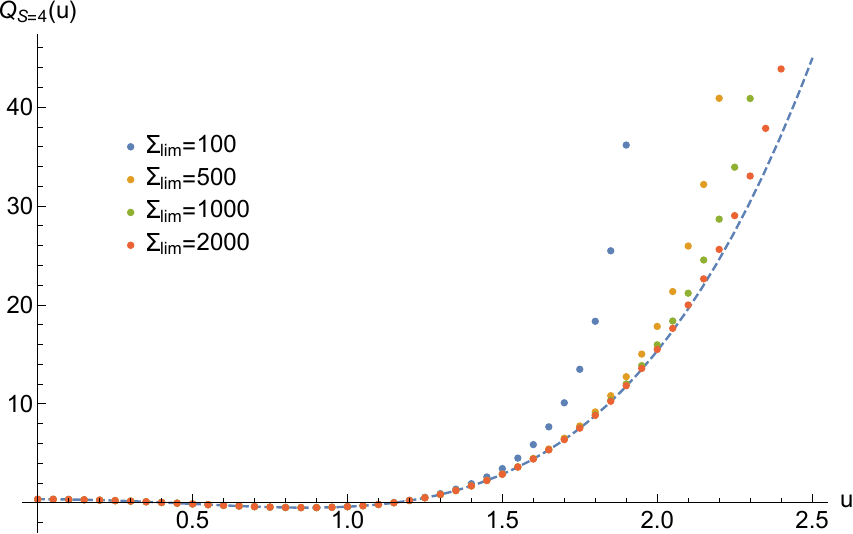} \hfill \includegraphics[width=0.49\columnwidth]{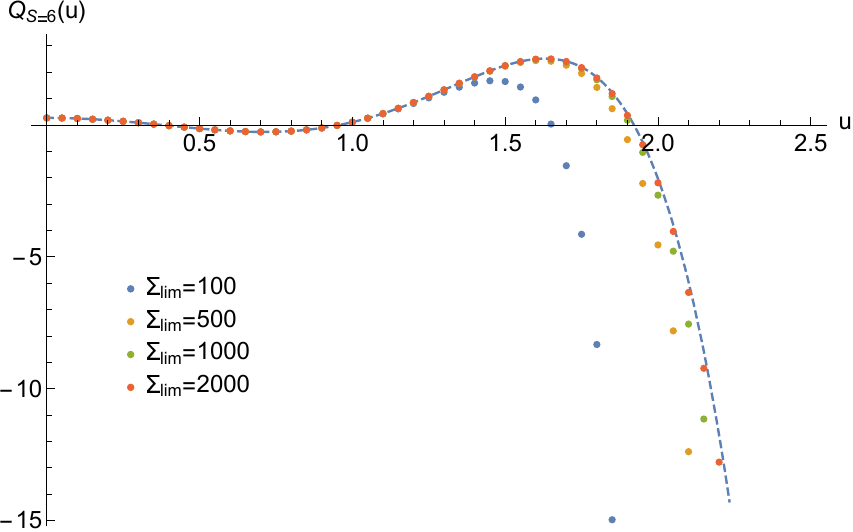} 
    \caption{Numerical evaluation of the even integer ansatz \eqref{eq:Q_ansatz_even_spin}  (dots) for $S=0,2,4,6$ against the known polynomial solution $Q^{(1)}(u)$ (dashed line). By increasing the summation limit ($\Sigma_{\mathrm{lim}}$), the numerical data clearly gets closer to the expected solution.  }
    \label{fig:even_integer_spin_limit}
\end{figure}

Combining the coefficients into the expansion of the ansatz \eqref{eq:Q_full_ansatz_solution}, we immediately see that all negative powers of $\epsilon$ cancel out, and the finite part is
\begin{equation}
    \lim_{S\to2M}Q(u)=\frac{1+\cosh(2\pi u)}{2\pi}\left(\frac{\pi}{3}\hat{q}_{-1}^{(2)}(u)-\frac{1}{\pi}\hat{q}_{1}^{(2)}(u)+\frac{1}{2\pi}\hat{q}_{0}^{(3)}(u)\right)\,.\label{eq:Q_ansatz_even_spin}
\end{equation}
$\hat{q}_{-1}^{(2)}(u)$ and $\hat{q}_{0}^{(3)}(u)$ are finite sums (interestingly $\hat{q}_{-1}^{(2)}(u)$ is exactly the expected polynomial solution), but $\hat{q}_{0}^{(2)}(u)$ contains an infinite sum, hence we cannot evaluate the solutions in closed form. However, as FIG.~\ref{fig:even_integer_spin_limit} shows, the numerical evaluation shows increasingly better agreement with $Q^{(1)}(u)$ for higher upper summation limits. 

As a second check of  \eqref{eq:Q_full_ansatz_solution}, we extract the small spin limit (i.e. $S\to 0$) of our solution. According to  \cite{SMBasso:2012ex}, for small spin 
\begin{equation}
    R(u)=\frac{Q(u+i/2)}{Q(u-i/2)}=1+r(u) S+\mathcal{O}(S^2) \,.
\end{equation}
The function $r(u)$ has a mode expansion
\begin{equation}
r(u)=\sum_{|m|<J-1} \kappa_m r^{(m)}(u)\,, \label{eq:small_spin_Basso_ansatz}
\end{equation}
with 
\begin{equation}
    r^{(m)}(u)=-i\frac{(2\pi m)^{1-J}}{ J }\frac{e^{2m\pi u}}{u^J} \big[\Gamma(1+j)-\Gamma(1+J,2 m\pi u)\big]\,,
\end{equation}
where $\Gamma(a,x)$ is the incomplete Gamma function. $J=J_1=3$ in our case and the mode coefficients satisfy the constraints
\begin{equation}
\begin{aligned}
\sum_{|m|<J-1} \kappa_m
&=1 \,,\\
\sum_{|m|<J-1}   \kappa_m m
&=0 \,. \label{eq:kappa_constraints}
\end{aligned}
\end{equation}

\begin{figure}
    \centering
    \includegraphics[scale=0.8]{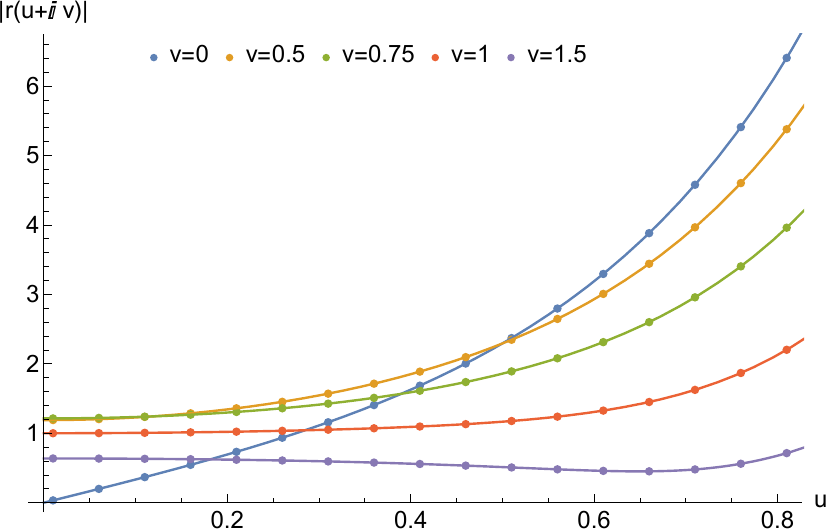}
    \caption{The comparison of $|r(u+i v)|$ calculated from  our solution $r(u)=Q_1(u+i/2)-Q_1(u-i/2)$ numerically with maximum summation index $1000$ (dots) and the theoretical prediction \eqref{eq:small_spin_Basso_ansatz} with coefficients $\kappa_1=\kappa_{-1}=1/2$, and $\kappa_2=\kappa_{-2}=0$, i.e. \eqref{eq:small_spin_solution} (solid line).}
    \label{fig:small_spin_r}
\end{figure}

The small spin expansion on our solutions $Q^{(2)}(u)$ and $Q^{(2)}(u)$ takes the form
\begin{equation}
\begin{aligned}
Q^{(2)}(u)&=\sum_{k=-1}^\infty \tilde{q}_k^{(2)}(u) S^k \,,\\ 
Q^{(3)}(u)&=\sum_{k=-2}^\infty \tilde{q}_k^{(3)}(u) S^k \,, 
\end{aligned}
\end{equation}
that leads to the expansion of our conjecture \eqref{eq:Q_full_ansatz_solution}
\begin{equation}
Q(u)=\tilde{Q}_0(u)+\tilde{Q}_1(u)S +\mathcal{O}(S^2)\,,
\end{equation}
with
\begin{equation}
\begin{aligned}
Q_0(u)&=\left(1+\cosh (2\pi u)\right)\frac{\pi^2 \tilde{q}_{-1}^{(2)}(u)-12\tilde{q}_{1}^{(2)}(u)+3\tilde{q}_{0}^{(3)}(u)}{12\pi^2} \,,\\
Q_1(u)&=\left(1+\cosh (2\pi u)\right)\frac{\pi^2 \tilde{q}_{0}^{(2)}(u)-12\tilde{q}_{2}^{(2)}(u)+3\tilde{q}_{1}^{(3)}(u)}{12\pi^2} \,.
\end{aligned}
\end{equation}
The coefficients $\tilde{q}_{1}^{(2)}(u)$, $\tilde{q}_{2}^{(2)}(u)$ and $\tilde{q}_{1}^{(3)}(u)$ contain infinite sums. With careful numerical evaluation,  one can convince themselves that $Q_0(u)=1$, and $r(u)=\left(Q_1(u+i/2)-Q_1(u-i/2)\right)$ is nothing else, but the  \eqref{eq:small_spin_Basso_ansatz} with mode coefficients $\kappa_1=\kappa_{-1}=1/2$, and $\kappa_2=\kappa_{-2}=0$( see FIG.~\ref{fig:small_spin_r} ). This also means, the functional form of $r(u)$ for $J=3$ is 
\begin{equation}
    r(u)=\left(\frac{i}{u}+\frac{i}{ 2\pi^2 u^3}-\frac{i\cosh(2
    \pi u)}{2\pi^2 u^3}\right)\,.\label{eq:small_spin_solution}
\end{equation}

With this result, the postulated form \eqref{eq:Q_full_ansatz_solution} passes all necessary crosschecks, and we conclude it is indeed the proper analytically continued solution. Since $Q\sim \mu_{12}^+$, this result is connected to the behaviour of the gluing matrix \cite{SMalfimov2018bfkl}, and it fully supports our claim  \eqref{eq:SMgluingmatrix}.

\section{Solution close to the intercept at weak couplings}
\label{sec:intercept_weak coupling}
\subsection{Ansatz}
Our aim in this Section is to solve the QSC directly at the degenerate point $(\Delta,S) = (0,-2)$ (while keeping $(J_1,J_2,J_3,S_2) = (3,0,0,0)$), where we expect multiple solutions to the QSC at weak coupling. This Section mostly follows the same steps as those in \cite{SMAlfimov:2014bwa}, where the weak-coupling Pomeron intercept was computed analytically. However, rather than formulating everything as a double-scaling limit as done in \cite{SMAlfimov:2014bwa}, we find it more convenient to directly look for solutions for which $\Delta$ is arbitrary and $S$ is parametrised by 
\begin{equation}
 S=-2+\sum_{i=1}^\infty I_i(\Delta )g^i\, , 
\end{equation}
where we have stressed that the coefficients $I_i$ depend on $\Delta$. The novel feature of this expansion is that we will allow for odd powers of the coupling $g$. It has been observed before that in degenerate cases such as ours, the $\mathbf{P}_a$ and $Q_i$ functions require odd-power corrections in $g$ in order to resolve the degeneracy \cite{SMmarboe2018full}. However, for local operators, it never happens that the relationship between $\Delta$ and $S$ also receives such corrections. Nevertheless, motivated by our numerics, we include odd powers and will see that these are indeed relevant! 

\subsection{Fixing NLO for $\mathbf{P}_a$}
The first step is to solve the $\mathbf{P}\mu$-system until order $g$ (next-to-leading order (NLO)). To do this, we plug the ansatz \eqref{SMPansatz} into the $\mathbf{P}\mu$-equations \eqref{eq:Pmu1}--\eqref{eq:Pmu2} and look for a solution with the correct analyticity. This fixes the first three $\mathbf{P}_a$ 
\begin{equation}
    \mathbf{P}_1 = -i \frac{(\Delta^2-1)(\Delta^2-7)}{192 u^{5/2}} -i g \frac{(\Delta^2+7)I_1}{16 u^{5/2}}\, , \quad  \mathbf{P}_2 = -i \frac{(\Delta^2-1)(\Delta^2-5)}{96 u^{3/2}} -i g \frac{(\Delta^2-5)I_1}{8 u^{5/2}}\, ,\quad \mathbf{P}_3 = u^{1/2}\, , 
\end{equation}
while the final $\mathbf{P}_4$ is given by
\begin{equation}
\label{eq:P4LO}
    \mathbf{P}_4 = \frac{u^2 + c_{4,-1,4}}{u^{1/2}} + g \frac{c_{4,-1,5}}{\sqrt{u}}\, ,
\end{equation}
containing two free parameters. Next is to try to find the $\mu_{ab}$ at LO (i.e. at order $g^{-5}$) by postulating the most general ansatz consistent with the required analyticity of all functions in the $\mathbf{P}\mu$-system. To ensure $\tilde{P}_a$ is regular at $u=0$,  $\mu_{ab}$ must be polynomial in $u$ up to an $i$-periodic function, which must be of the form $C_1 +C_2 \cosh^2 \pi u$ according to the asymptotics of $\mu_{ab}$. Plugging this ansatz into the equation
\begin{equation}
    \mu_{ab}^{[+2]} = \tilde{\mu}_{ab}  = \mu_{ab} + \mathbf{P}_a \mu_{bc}\mathbf{P}^c - \mathbf{P}_b \mu_{ac} \mathbf{P}^c, 
\end{equation}
which follows from \eqref{eq:Pmu1}--\eqref{eq:Pmu2}, indeed constrains the $\mu_{ab}$ ansatz. To fully fix the $\mu_{ab}$ solution we continue the $\mathbf{P}_a$ to give us $\tilde{\mathbf{P}}_a$ and impose \eqref{eq:Pmu1}, yielding fully fixed expressions for the $\mu_{ab}$. It is noteworthy that $\mathbf{P}_4$ remains unfixed, as the procedure above only yields the expression $c_{4,-1,4} = (\Delta^2-1)/(\Delta^2-49)$ for one of the free parameters in \eqref{eq:P4LO}, mimicking what happens in the Pomeron case. 

To fix the free parameters $c_{4,-1,5}$, we need to include the NLO contributions for the $\mu_{ab}$ (i.e. at order $g^{-4}$)  and repeat the same steps as above.  The NLO ansatz for $\mu_{ab}$ are no longer polynomial and contain rational factors, such as a digamma function. Following the Appendix B in \cite{SMAlfimov:2014bwa} closely yields, after lengthy but straightforward computations, a solution that fixes the $\mu_{ab}$ functions fully up to NLO and the final free NLO coefficient $c_{4,-1,5} = 24 I_1 (\Delta^2-21)/(\Delta^2-49)$ for $\mathbf{P}_4$. This, therefore, fixes all $\mathbf{P}_a$ at NLO, allowing us to move on. 

\subsection{Fourth-order Baxter equation for $Q_i$}
As shown in \cite{SMAlfimov:2014bwa}, the $\mathbf{P}_a$ and $\mathbf{Q}_i$ functions can be related through a fourth-order difference equation for the $\mathbf{Q}_i$ with coefficients built up from the $\mathbf{P}_a$. Such equations are generally quite difficult to solve, but in our case (as it did in \cite{SMAlfimov:2014bwa}), it factorises into a product of second-order difference equations. The simplest factor of this product (which was first derived by M. Alfimov \cite{SMmisha} and independently rederived by the authors, who thank him for sharing his result) is (with $j=1,3$)
\begin{equation}\label{eq:sm_2ndorder}
 \frac{2 (u+i)+i g I_1 }{ (u+i)^{3/2}}\mathbf{Q}_j(u+i)+\frac{2 (u-i)-i g I_1}{ (u-i)^{3/2}}\mathbf{Q}_j(u-i)  - \frac{1 + 8 u^2 - \Delta^2 + 2 g I_1}{2 u^{5/2}}\mathbf{Q}_j (u)=0\,.
\end{equation}
It has two independent solutions at LO, $\mathbf{q}^{(1)}_{\text{LO}}(u)$ and $\mathbf{q}^{(2)}_{\text{LO}}(u)$, and two at NLO, $\mathbf{q}^{(1)}_{\text{NLO}}(u)$ and $\mathbf{q}^{(2)}_{\text{NLO}}(u)$. These can be found using Mellin transformation methods explained in detail in Section \ref{sec:solving_diff_eq}. 

After a $\mathbf{Q}_j(u)\mapsto u^{5/2}\mathbf{Q}_j(u)$ rescaling, the Mellin transformation relates the difference equation to a generalised hypergeometric differential equation with parameters $\alpha_1=(1-\Delta+g I_1)/2, \alpha_2=(1+\Delta+g I_1)/2$, and $\beta_1=1+g I_1/2$. The simpler solution of the differential equation is 
\begin{equation}
{}_{2}F_{1}
    \left( \genfrac{}{}{0pt}{0}{\alpha_1,\alpha_2,}{\beta_1};z\right)\,,
\end{equation}
that gives  $\tilde{\mathbf{q}}^{(1)}_{\text{LO}}(z)$ and $\tilde{\mathbf{q}}^{(1)}_{\text{NLO}}(z)$ solutions after expanding in $g$.  The second LO solution, $\tilde{\mathbf{q}}^{(2)}_{\text{LO}}(z)$ is given by \eqref{eq:log_solution_n2_b}, however, the second NLO solution is more complicated to construct. Assuming the \eqref{eq:diff_gen_sol} form for the series expansion of $\tilde{\mathbf{q}}^{(2)}_{\text{NLO}}(z)$ (the $\log^2(z)$ terms are crucial for this solution), one can solve a recursion for the coefficients, that will be an elaborate combination of polygamma functions. 

After the inverse transformations \eqref{eq:int_identity} and \eqref{eq:int_identity_2}, the solution of the difference equation have the form
\begin{equation}
\begin{aligned}
u^{-5/2}\mathbf{q}^{(1)}_{\text{LO}}(u)&=\sum_{n=0}^{\infty} \frac{((1+\Delta)/2)_n}{(2)_n}\frac{((1-\Delta)/2)_n}{(2)_n}\frac{(1+iu)_n}{n!}=\, {}_{3}F_{2} \left( \genfrac{}{}{0pt}{0}{1+i u,(1+\Delta)/2,(1-\Delta)/2}{2,2};1\right)\,, \\
u^{-5/2}\mathbf{q}^{(1)}_{\text{NLO}}(u)&=\frac{I_1}{2}\sum_{n=0}^{\infty} \frac{((1+\Delta)/2)_n}{(2)_n}\frac{((1-\Delta)/2)_n}{(2)_n}\frac{(1+iu)_n}{n!} \mathcal{A}_n \,, \\
u^{-5/2}\mathbf{q}^{(2)}_{\text{LO}}(u)&=\frac{4i}{u(1-\Delta^2)}+\sum_{n=0}^{\infty} \frac{((1+\Delta)/2)_n}{(2)_n}\frac{((1-\Delta)/2)_n}{(2)_n}\frac{(1+iu)_n}{(1)_n} \mathcal{B}_n \,, \\
u^{-5/2}\mathbf{q}^{(2)}_{\text{NLO}}(u)&=-\frac{2iI_1(\psi(iu)-\psi(1))}{u(1-\Delta^2)}+\sum_{n=0}^{\infty} \frac{((1+\Delta)/2)_n}{(2)_n}\frac{((1-\Delta)/2)_n}{(2)_n}\frac{(1+iu)_n}{n!} \mathcal{C}_n \,,
\end{aligned}
\end{equation}
where $\mathcal{A}_n, \mathcal{B}_n$ and $\mathcal{C}_n$ are certain combinations of $\psi_0$ and $\psi_1$ polygamma functions.

To find the two $\mathbf{Q}_i$ functions we need to select the right linear combination of these four solutions, where we note that their coefficients may be $i$-periodic functions. To find these, we write down the most general ansatz that solves the difference equation: let $Q(u) = f_1 \mathbf{q}_{\text{LO}}^{(1)} + f_2 \mathbf{q}_{\text{NLO}}^{(1)} + h_1 \mathbf{q}_{\text{LO}}^{(2)} + h_2 \mathbf{q}_{\text{NLO}}^{(2)} $, with prefactors
 \begin{equation}
 \label{eq:bigansatz}
 \begin{aligned}
 f_1(u) &= C_{10} + C_{11} g  + (A_{10} + A_{11}g) \coth^2(\pi u) \\
 f_2(u) &= C_{10} g  + A_{10} g \coth^2(\pi u) \\
 h_1(u) &= C_{30} + C_{31} g + (B_{30} + B_{31} g) \coth(\pi u) + (A_{30} + A_{31} g) \coth^2(\pi u)+(D_{30}+ D_{31} g) \tanh(\pi u) \\
 h_2(u) &= C_{41} g + B_{30} g \coth(\pi u) + A_{30} g \coth^2(\pi u) +D_{30} g \tanh( \pi u)\, , 
 \end{aligned}
 \end{equation}
 where all capital letters are coefficients. To fix these, we first require the solutions to be upper-half-plane analytic, i.e. there should be no poles in the upper half-plane. Next, for each $\mathbf{Q}_i$ separately we need to impose the relevant asymptotics \eqref{SMasymptotics}, which is somewhat more involved in our case. For example, the prescribed asymptotic of $\mathbf{Q}_1$  for large $u$ is
\begin{equation}
\begin{aligned}
        \mathbf{Q}_1 &\sim B_1 u^{(\Delta-S)/2} = \left( B_1^{(0)} + g B_1^{(1)} + \mathcal{O}(g^2)\right)u^{(\Delta+2 - I_1 g -\mathcal{O}(g^2))/2} \\
        &\approx B_1^{(0)} u^{(\Delta+2)/2} + \frac{g}{2} \left( B_1^{(1)} + I_1 B_1^{(0)} \log(u) \right) u^{(\Delta+2)/2} + \mathcal{O}(g^2)\, , 
    \end{aligned}
\end{equation}
where the $B_1^{(i)}$ are known explicitly from expanding \eqref{eq:A_and_B}. We see  from this that there should be a logarithmic correction to the asymptotic behaviour, which gives us extra equations to constrain the ansatz. Matching with the solutions to \eqref{eq:sm_2ndorder} indeed fixes all free coefficients -- some of which have very lengthy expressions -- except for $I_1$. 

\subsection{Revised regularity requirement}
Up until this point, we have managed to stay close to the Pomeron computation in \cite{SMAlfimov:2014bwa}, even though our expressions have become much more involved. At this point, however, we will have to deviate, as the strategy in \cite{SMAlfimov:2014bwa} crucially depends on the first non-trivial correction to $S$ to have order $g^2$. Instead, we turn to the gluing conditions \eqref{SMgluing} to help us constrain the final free parameter $I_1$. Conveniently, the equations for $\mathbf{Q}_1$ and $\mathbf{Q}_3$ factorise and take a very simple form 
\begin{equation}
\label{eq:sm_gluingexplicit}
\tilde{\mathbf{Q}}_1(u)=-\mathbf{Q}_3(-u)/\ell_2 \quad\text{and}\quad\tilde{\mathbf{Q}}_3(u)=\mathbf{Q}_1(-u)\, \ell_2\, , 
\end{equation}
with only one gluing matrix component, $\ell_2$. This allows us to express the continued $\tilde{\mathbf{Q}}_1$ and $\tilde{\mathbf{Q}}_3$ directly in terms of our found solutions. 

To use this knowledge, we look at the $\mathbf{Q}_i$ on the real line: for solutions associated with local operators, there are two square root branch points at $\pm 2g$, implying that $\mathbf{Q}_i+ \tilde{\mathbf{Q}}_i$ should be branch point free on the real line. In our current case, the monodromy around the two branch points is more complicated, and it is the combination $\mathbf{Q}_i/\sqrt{x}$ that has pure square root branch points \footnote{We confirmed this independently using numerics}. Hence rather than the usual regularity, we impose the revised condition that 
\begin{equation} 
\label{eq:sm_Qregularity}
\mathbf{Q}_i \sqrt{x}^{-1} + \tilde{\mathbf{Q}}_i \sqrt{x}
\end{equation}
must be regular on the real line. This implies that this combination must be pole-free at $u=0$ at small coupling. 

The small $u$ expansion of our $\mathbf{Q}_i$ is
\begin{equation}
\mathbf{Q}_i(u)\sim u^{3/2}\sum_{k=0}^\infty \mathbf{Q}_{i, k}^{(0)} u^k+g I_1 u^{1/2}\sum_{k=0}^\infty \mathbf{Q}_{i, k}^{(1)} u^k\,,
\end{equation}
with leading coefficients
\begin{equation}
\begin{aligned}
\mathbf{Q}_{1, 0}^{(0)}=\frac{4i\cos(\pi \Delta/2)}{\pi (1-\Delta^2)} \,, \quad \quad&
\mathbf{Q}_{1, 0}^{(1)}= -\frac{2\cos(\pi \Delta/2)}{\pi (1-\Delta^2)}\,,\\
\mathbf{Q}_{3, 0}^{(0)}= -\frac{4i}{\pi (1-\Delta^2)}\,,\quad \quad&\mathbf{Q}_{3, 0}^{(1)}=-\frac{2}{\pi (1-\Delta^2)} \,.
\end{aligned}
\end{equation}
Using the gluing conditions \eqref{eq:sm_gluingexplicit} to build the combinations \eqref{eq:sm_Qregularity}, we derive two consistency equations for the cancellation of the $u^{-1}$ singularity. Their combination leads to 
\begin{equation}
    I_1^2 =  \frac{\mathbf{Q}_{1, 0}^{(0)} \mathbf{Q}_{3, 0}^{(0)}}{\mathbf{Q}_{1, 0}^{(1)}\mathbf{Q}_{3, 0}^{(1)}} =4.
\end{equation}
The two solutions $I_1 = \pm 2$ are independent of $\Delta$, both parametrise a valid solution to the QSC, and correspond to the blue (plus sign, slightly above $S=-2$) and the orange trajectories (minus sign, slightly below $S=-2$) in the main text. 

This computation has shown us that there is indeed a degeneracy in the solutions of the QSC at the point $(\Delta, S) \approx (0,-2)$ for weak coupling. Moreover, we showed that there is a way to deal with the resolution of the degeneracy within the QSC.  At least in principle, this can be extended for more complicated cases of mixing in the Regge regime.

\section{Solving the difference equations}
\label{sec:solving_diff_eq}

\subsection{Solutions through Mellin transformation}

To solve the difference equations \eqref{eq:TQ} and \eqref{eq:sm_2ndorder}, we follow the strategy outlined in \cite{SMFaddeev:1994zg} and \cite{SMkorchemsky1995bethe}.  Let us denote the function in the difference equation by a generic $Q(u)$. We define the function $\tilde{Q}(z)$ through the relation (Mellin transformation) 
\begin{equation}
 Q(u) \sim \int_{0}^{1} \frac{dz}{2\pi i} z^{-i u-1} (z-1)^{i u-1} \tilde{Q}(z)\,. \label{eq:Q_ansatz}  
\end{equation}
By integration by parts, one finds the relation   
\begin{equation}
 u^{k} Q(u) \sim \int_{0}^{1} \frac{dz}{2\pi i} z^{-i u -1} (z-1)^{i u -1}\left(-i z(1-z) \frac{d}{dz}  \right)^k \tilde{Q}(z)\,,  
\end{equation}
which we can use to turn the difference equation for $Q(u)$ into a differential equation for $\tilde{Q}(z)$. The proportionality indicates the freedom to multiply by an $i$-periodic function, which is a freedom of the difference equation.

In both cases for  \eqref{eq:TQ} and \eqref{eq:sm_2ndorder}, the acquired differential equations are related to generalised hypergeometric equations, whose solutions can be written in the form of infinite series
\begin{equation}
\tilde{Q}(z)=\sum_{n=0}^{\infty} z^n\left[A_n+B_n \log(z)+C_n \log^2(z)\right]\,, \label{eq:diff_gen_sol}
\end{equation}
where the coefficients $A_n, B_n$ and $C_n$ depend on the parameters of the differential equations (see next Section for details).

Through the integral identity
\begin{equation}
\int_0^1 \mathrm{d}t \,t^a (1-t)^b = \frac{ \Gamma\left(1+a\right)\Gamma\left(1+b\right)}{\Gamma\left(2+a+b\right)} \quad \mathrm{for} \; \mathrm{Re}(a)>-1 \; \mathrm{and} \; \mathrm{Re}(b)>-1\,, \label{eq:int_identity}
\end{equation}
and its consequence
\begin{equation}
\int_0^1 \mathrm{d}t \,t^a (1-t)^b (\log(t))^c(\log(1-t))^d= \left[\frac{\partial^{c-1}}{\partial a^{c-1}}\frac{\partial^{d-1}}{\partial b^{d-1}}\frac{ \Gamma\left(1+a\right)\Gamma\left(1+b\right)}{\Gamma\left(2+a+b\right)}\right]\,, \label{eq:int_identity_2}
\end{equation}
the solutions of the form \eqref{eq:diff_gen_sol} can be turned into solutions for the difference equations following \eqref{eq:Q_ansatz}.

The multiplication theorem for the polygamma functions
\begin{equation}
k^{j+1} \psi_j (k y) =\delta_{j,0} k \log(k) +\sum_{l=0}^{k-1} \psi_j\left(y+\frac{l}{k}\right)\,,\label{eq:polygamm_multiplications_theorem}
\end{equation}
is also an important relation in simplifying the results. 

With the basis of solutions for the difference equations, we fix the full solutions by demanding physically motivated analytic and asymptotic properties (see Section \ref{sec:TQ_weak_coupling} and \ref{sec:intercept_weak coupling} for details )

\subsection{Solutions of the generalised hypergeometric differential equation}
\label{sec:hyp_gen_solutions}

The generalised hypergeometric differential equation has  the form 
\begin{equation}
    \theta \prod_{j=1}^q (\theta +\beta_j-1 ) - y \prod_{j=1}^p (\theta +\alpha_j ) Y(z) =0\,,
\end{equation}
where $ \theta = z\, \mathrm{d} / \mathrm{d} z$.
In case $p=q+1$, the equation has a more symmetric form
\begin{equation}
    \prod_{j=1}^{q+1} (\theta +\beta_j-1 ) - z \prod_{j=1}^{q+1} (\theta +\alpha_j )Y(z) =0\,, \label{eq:gen_hyp_diffeq}
\end{equation}
with $\beta_{q+1}=1$. 
\eqref{eq:gen_hyp_diffeq} has regular singular points at $z=0,1$ and $\infty$. If no two of the $\beta_j$ are equal or differ by an integer, we can write the $q+1$ independent solutions around $z=0$ as an infinite series
\begin{equation}
\begin{aligned}
    Y_j(z)&=z^{1-\beta_j}\prod_{t=1}^{q+1}\frac{\Gamma(1+\beta_t-\beta_j)}{\Gamma(1+\alpha_t-\beta_j)}\sum_{n=0}^{\infty} \prod_{t=1}^{q+1}\frac{\Gamma(1+\alpha_t-\beta_j+n)}{\Gamma(1+\beta_t-\beta_j+n)} z^n \\
    &= z^{1-\beta_j} \sum_{n=0}^{\infty}\prod_{t=1}^{q+1}\frac{(1+\alpha_t-\beta_j)_n}{(1+\beta_t-\beta_j)_n} z^n= z^{1-\beta_j} {}_{q+1}F_{q}
    \left( \genfrac{}{}{0pt}{0}{1+\alpha_1-\beta_j,\dots,1+\alpha_{q+1}-\beta_j  }{1+\beta_1-\beta_j,\widehat{\dots},1+\beta_{q+1}-\beta_j};z\right)\,,  \label{eq:gen_hyp_sol_basic}
\end{aligned}
\end{equation}
where $\widehat{\dots}$ denotes that the $1+\beta_j-\beta_j$ term is omitted, $(x)_n$ denotes the Pochhammer symbol, and ${}_{q+1}F_{q}$ denotes the generalised hypergeometric function. The series is convergent in the $|z|<1$ domain.
In case some of the $\beta_j$ are equal or differ by an integer, some of the  solutions \eqref{eq:gen_hyp_sol_basic} coincide or have no meaning; hence we need a different expression for the $q+1$ independent solutions.  \cite{SMScheidegger:2016ysn} calls this behaviour as $\beta_j$ being resonant, borrowing nomenclature from GKZ hypergeometric systems.
An integral representation of the new solutions was presented in \cite{SMnorlund1955}. The solutions given in \cite{SMsmith1939}, where also a proof on the line of the Frobenius theorem is given, are of the form of infinite series. This form is more suitable for our calculations, which we present in the following. 

Let us assume that $r$ number of $\beta_j$ coincide or differ by an integer. Without loss of generality, we take them to be the first $r$ parameters $\beta_1,\beta_2,\dots,\beta_r$ with ordering $\mathrm{Re}\beta_j\leq \mathrm{Re}\beta_{j+1}$. Furthermore, let us introduce the notation $l_{j}=\beta_{j+1}-\beta_{j}$ for $j=1, \dots, r-1$. 
If none of $\beta_1,\dots,\beta_r$ equal or differ from any $\alpha_j$ by an integer, then $Y_j(y)$ from \eqref{eq:gen_hyp_sol_basic} is a good solution for $j=1,r+1, \dots, q+1$, and the remaining independent solutions for $j=2,3,\dots,r $ are logarithmic in character and may be written as 
\begin{equation}
\begin{aligned}
    Y_j(z)&=\sum_{v=1}^{j} z^{1-\beta_v} \frac{(j-1)!}{(j-v)!}\left[\frac{\partial^{j-v}}{\partial w^{j-v}}z^{w} G_v(w,z)\right]_{w=0} \\
    &=(-1)^{j+1}\sum_{v=1}^{j-1}(-1)^v \binom{j-1}{v-1} (\log z)^{j-v} Y_v (z) + \sum_{v=1}^{j} z^{1-\beta_v} \frac{(j-1)!}{(j-v)!} G_v^{(j-v)}(0,z)\,, \label{eq:gen_hyp_sol_log}
\end{aligned}
\end{equation}
where $G_v^{(k)}(0,z)$ denotes the derivatives $\left[\partial_{w}^{k}G_v(w,z)\right]_{w=0}$ of the function 
\begin{equation}
 G_1(w,z)= 
 \sum_{n=0}^{\infty} \prod_{t=1}^{q+1}\frac{(1+\alpha_t-\beta_1+w)_n}{(1+\beta_t-\beta_1+w)_n} z^n\,,
\end{equation}
and 
\begin{equation}
\begin{aligned}
 G_v(w,z)=& (-1)^{v-1-\sum_{t=1}^{v-1} t l_t} \left(\frac{\pi w}{\sin \pi w}\right)^{1-v} \prod_{t=1}^{q+1}\frac{\Gamma(1+\beta_t-\beta_1+w)}{\Gamma(1+\alpha_t-\beta_1+w)} \\
 &\times\sum_{n=0}^{l_{v-1}-1}
 \left(\prod_{t=1}^{v-1}\Gamma(1+\alpha_t-\beta_v+w+n) \Gamma(\beta_v-\beta_t-w-n)\right) \\
 &\times \left(\prod_{t=v}^{q+1}\frac{\Gamma(1+\alpha_t-\beta_v+w+n)}{\Gamma(1+\beta_t-\beta_v+w+n)}\right)\left[(-1)^{v-1}z\right]^n  \,,
\end{aligned}
\end{equation}
for $v\geq2$. If $l_{v-1}=0$ then $G_v(w,z)=0$. We also see that for $Y_1(z)$ the two solutions \eqref{eq:gen_hyp_sol_basic} and \eqref{eq:gen_hyp_sol_log} coincide.

\subsection{Solutions relevant to our calculations}

Let us consider the case when $r=q+1$ and $l_j=0$, i.e. all $\beta_j$ coincide. In this situation \eqref{eq:gen_hyp_sol_log} describes all the independent solution to \eqref{eq:gen_hyp_diffeq} and it simplifies to 
\begin{equation}
        Y_j(z)= z^{1-\beta_1}\sum_{k=0}^{j-1}  \binom{j-1}{k}(\log z)^k  G_1^{(j-1-k)}(0,z)\,.\\
\end{equation}
In our calculation, we are interested in the solutions with $\beta_j=1$ for the case $q+1=2$ and $q+1=3$, that might be written according to  \eqref{eq:gen_hyp_sol_log} as 
\begin{equation}
\begin{aligned}
Y_1^{q+1=2}(z)&=\sum_{n=0}^{\infty} \frac{(\alpha_1)_n (\alpha_2)_n}{(n!)^2} z^n\,,\\
Y_2^{q+1=2}(z)&=\sum_{n=0}^{\infty} \frac{(\alpha_1)_n (\alpha_2)_n}{(n!)^2} z^n\left\{\log z +\sum_{t=1}^2\left(\psi_0(\alpha_t+n)-\psi_0(\alpha_t)\right) -2\left[\psi_0(1+n))-\psi_0(1)\right] \right\}\,, \label{eq:log_solution_n2}
\end{aligned}
\end{equation}
and 
\begin{equation}
\begin{aligned}
Y_1^{q+1=3}(z)=\sum_{n=0}^{\infty} \frac{(\alpha_1)_n (\alpha_2)_n (\alpha_3)_n}{(n!)^3} z^n&\,,\\
Y_2^{q+1=3}(z)=\sum_{n=0}^{\infty} \frac{(\alpha_1)_n (\alpha_2)_n (\alpha_3)_n}{(n!)^3} z^n&\left\{\log z +\sum_{t=1}^3\left(\psi_0(\alpha_t+n)-\psi_0(\alpha_t)\right) -3\left(\psi_0(1+n)-\psi_0(1)\right) \right\}\,, \\
Y_3^{q+1=3}(z)=\sum_{n=0}^{\infty} \frac{(\alpha_1)_n (\alpha_2)_n (\alpha_3)_n}{(n!)^3} z^n&\left\{(\log z)^2 +2 \log z \left(\sum_{t=1}^3\left(\psi_0(\alpha_t+n)-\psi_0(\alpha_t)\right) -3\left(\psi_0(1+n)-\psi_0(1)\right)\right) \right.\\
&+ \left(\sum_{t=1}^3\left(\psi_0(\alpha_t+n)-\psi_0(\alpha_t)\right) -3\left(\psi_0(1+n)-\psi_0(1)\right)\right)^2 \\
&\left.+\left(\sum_{t=1}^3\left(\psi_1(\alpha_t+n)-\psi_1(\alpha_t)\right) -3\left(\psi_1(1+n)-\psi_1(1)\right)\right)     \right\} \,, \label{eq:log_solution_n3}
\end{aligned}
\end{equation}
where $\psi_k(z)$ is the polygamma function of order $k$
\begin{equation}
\psi_k(z)=\frac{\mathrm{d}^{k+1}}{\mathrm{d} z^{k+1}}\log \Gamma(z)\,,
\end{equation}
with $\psi_0(z)$ being the digamma function.

The solutions are linearly independent, and we prefer to use the basis of solutions where the $n$ and $z$ independent terms in the curly brackets in \eqref{eq:log_solution_n2} and \eqref{eq:log_solution_n3} cancels out, namely
\begin{equation}
\begin{aligned}
\tilde{Y}_1^{q+1=2}(z)&=\sum_{n=0}^{\infty} \frac{(\alpha_1)_n (\alpha_2)_n}{(n!)^2} z^n\,,\\
\tilde{Y}_2^{q+1=2}(z)&=\sum_{n=0}^{\infty} \frac{(\alpha_1)_n (\alpha_2)_n}{(n!)^2} z^n\left\{\log z+\sum_{t=1}^2\psi_0(\alpha_t+n) -2\psi_0(1+n)) \right\}\,, \label{eq:log_solution_n2_b}
\end{aligned}
\end{equation}
and 
\begin{equation}
\begin{aligned}
\tilde{Y}_1^{q+1=3}(z)=\sum_{n=0}^{\infty} \frac{(\alpha_1)_n (\alpha_2)_n (\alpha_3)_n}{(n!)^3} z^n&\,,\\
\tilde{Y}_2^{q+1=3}(z)=\sum_{n=0}^{\infty} \frac{(\alpha_1)_n (\alpha_2)_n (\alpha_3)_n}{(n!)^3} z^n&\left\{\log z +\sum_{t=1}^3 \psi_0(\alpha_t+n) -3\psi_0(1+n) \right\}\,, \\
\tilde{Y}_3^{q+1=3}(z)=\sum_{n=0}^{\infty} \frac{(\alpha_1)_n (\alpha_2)_n (\alpha_3)_n}{(n!)^3} z^n&\left\{(\log z)^2 +2 \log z \left(\sum_{t=1}^3\psi_0(\alpha_t+n) -3\psi_0(1+n)\right) \right.\\
&+ \left(\sum_{t=1}^3\psi_0(\alpha_t+n) -3\psi_0(1+n)\right)^2 \\
&\left.+\left(\sum_{t=1}^3\psi_1(\alpha_t+n) -3\psi_1(1+n)\right)     \right\} \,. \label{eq:log_solution_n3_b}
\end{aligned}
\end{equation}

\section{Intercept data from QSC numerics}
In the following, we collect data for the intercept $\alpha$ computed with the numerical QSC for $\Delta=10^{-7}$. Our set of data ranges between 20 to 80 digits of precision. In Table \ref{tab:interceptdata}, we show the first 10 significant digits.
\begin{table}[!h]
    \centering
    \begin{tabular}{||cc|cc|cc|cc||}
    \hline
    $g$ & $\alpha$ & $g$ & $\alpha$ & $g$ & $\alpha$ & $g$ & $\alpha$ \\
    \hline\hline
0.001 & -1.9979889164 & 2.3 & -0.15817192359 & 4.9 & -0.073656778216 & 7.5 & -0.047995117936 \\ 0.005 & -1.9897236911 & 2.4 & -0.15149386850 & 5.0 & -0.072172779848 & 7.6 & -0.047360442858 \\ 0.01 & -1.9788995737 & 2.5 & -0.14535571366 & 5.1 & -0.070747376786 & 7.7 & -0.046742331932 \\ 0.05 & -1.8742687088 & 2.6 & -0.13969467466 & 5.2 & -0.069377167110 & 7.8 & -0.046140145154 \\ 0.1 & -1.7114508656 & 2.7 & -0.13445731552 & 5.3 & -0.068059007137 & 7.9 & -0.045553275066 \\ 0.2 & -1.3692071108 & 2.8 & -0.12959787553 & 5.4 & -0.066789987385 & 8.0 & -0.044981144714 \\ 0.3 & -1.0855411333 & 2.9 & -0.12507694210 & 5.5 & -0.065567411175 & 8.1 & -0.044423205758 \\ 0.4 & -0.87499456476 & 3.0 & -0.12086038919 & 5.6 & -0.064388775526 & 8.2 & -0.043878936720 \\ 0.5 & -0.72207523301 & 3.1 & -0.11691852130 & 5.7 & -0.063251754083 & 8.3 & -0.043347841360 \\ 0.6 & -0.60980939868 & 3.2 & -0.11322537832 & 5.8 & -0.062154181807 & 8.4 & -0.042829447164 \\ 0.7 & -0.52552731896 & 3.3 & -0.10975816702 & 5.9 & -0.061094041240 & 8.5 & -0.042323303949 \\ 0.8 & -0.46064821114 & 3.4 & -0.10649679312 & 6.0 & -0.060069450142 & 8.6 & -0.041828982553 \\ 0.9 & -0.40949271581 & 3.5 & -0.10342347385 & 6.1 & -0.059078650356 & 8.7 & -0.041346073627 \\ 1.0 & -0.36827947405 & 3.6 & -0.10052241527 & 6.2 & -0.058119997746 & 8.8 & -0.040874186504 \\ 1.1 & -0.33444482452 & 3.7 & -0.097779542107 & 6.3 & -0.057191953110 & 8.9 & -0.040412948143 \\ 1.2 & -0.30621100170 & 3.8 & -0.095182270419 & 6.4 & -0.056293073940 & 9.0 & -0.039962002149 \\ 1.3 & -0.28231615158 & 3.9 & -0.092719315381 & 6.5 & -0.055422006946 & 9.1 & -0.039521007852 \\ 1.4 & -0.26184439090 & 4.0 & -0.090380528006 & 6.6 & -0.054577481271 & 9.2 & -0.039089639453 \\ 1.5 & -0.24411707431 & 4.1 & -0.088156755878 & 6.7 & -0.053758302303 & 9.3 & -0.038667585218 \\ 1.6 & -0.22862179999 & 4.2 & -0.086039723856 & 6.8 & -0.052963346049 & 9.4 & -0.038254546729 \\ 1.7 & -0.21496505836 & 4.3 & -0.084021931505 & 6.9 & -0.052191553990 & 9.5 & -0.037850238180 \\ 1.8 & -0.20283998307 & 4.4 & -0.082096564579 & 7.0 & -0.051441928385 & 9.6 & -0.037454385717 \\ 1.9 & -0.19200394721 & 4.5 & -0.080257418354 & 7.1 & -0.050713527971 & 9.7 & -0.037066726823 \\ 2.0 & -0.18226270772 & 4.6 & -0.078498831014 & 7.2 & -0.050005464023 & 9.8 & -0.036687009732 \\ 2.1 & -0.17345899047 & 4.7 & -0.076815625575 & 7.3 & -0.049316896741 & 9.9 & -0.036314992890 \\ 2.2 & -0.16546414014 & 4.8 & -0.075203059111 & 7.4 & -0.048647031928 & 10 & -0.035950444437\\
    \hline
    \end{tabular}
    \caption{Non-perturbative data for the intercept $\alpha$. }
    \label{tab:interceptdata}
\end{table}

\section{Twist-$5$ local operators}
Even at tree-level, the first local operators on the twist-$5$ trajectories are very involved combinations of the fundamental fields of $\mathcal{N}=4$ SYM. We include a representation of these two super-highest weight vectors here.

\begingroup
\tiny
\thinmuskip=-2mu
\medmuskip=-2mu
\thickmuskip=-18mu
\noindent
$\mathcal{O}_{-}\quad=\quad 1.45 \text{Tr}\left(\,\mathcal{D}\,\Psi  \, Z\, Z\, Z\, \bar{\Psi } \,\right)+0.553 \text{Tr}\left(\,\mathcal{D}\,\Psi  \, Z\, Z\,
   \bar{\Psi } \, Z\,\right)+0.553 \text{Tr}\left(\,\mathcal{D}\,\Psi  \, Z\, \bar{\Psi } \, Z\, Z\,\right)+1.45
   \text{Tr}\left(\,\mathcal{D}\,\Psi  \, \bar{\Psi } \, Z\, Z\, Z\,\right)-0.106 \text{Tr}\left(\,\mathcal{D}\,Z\, \Psi  \, Z\,
   Z\, \bar{\Psi } \,\right)-0.106 \text{Tr}\left(\,\mathcal{D}\,Z\, \Psi  \, \bar{\Psi } \, Z\, Z\,\right)+0.106
   \text{Tr}\left(\,\mathcal{D}\,Z\, \Psi  \, Z\, Z\, \bar{\Psi } \,\right)+0.106 \text{Tr}\left(\,\mathcal{D}\,Z\, \Psi  \,
   \bar{\Psi } \, Z\, Z\,\right)-1.00 \text{Tr}\left(\,\mathcal{D}\,Z\, Z\, \Psi  \, Z\, \bar{\Psi } \,\right)-1.00
   \text{Tr}\left(\,\mathcal{D}\,Z\, Z\, \Psi  \, \bar{\Psi } \, Z\,\right)+1.00 \text{Tr}\left(\,\mathcal{D}\,Z\, Z\, \Psi  \,
   Z\, \bar{\Psi } \,\right)+1.00 \text{Tr}\left(\,\mathcal{D}\,Z\, Z\, \Psi  \, \bar{\Psi } \, Z\,\right)-1.79
   \text{Tr}\left(\,\mathcal{D}\,Z\, Z\, Z\, \Psi  \, \bar{\Psi } \,\right)+1.79 \text{Tr}\left(\,\mathcal{D}\,Z\, Z\, Z\, \Psi _{   
   }\, \bar{\Psi } \,\right)-0.894 \text{Tr}\left(\,\mathcal{D}\,Z\, Z\, Z\, X\, \mathcal{D}\,\bar{X}\,\right)+0.894 \text{Tr}\left(\,\mathcal{D}\,Z\, Z\, Z\, Y\,
   \mathcal{D}\,\bar{Y}\,\right)+0.894 \text{Tr}\left(\,\mathcal{D}\,Z\, Z\, Z\, \bar{Y}\, \mathcal{D}\,Y\,\right)-0.894 \text{Tr}\left(\,\mathcal{D}\,Z\, Z\, Z\, \bar{X}\,
   \mathcal{D}\,X\,\right)-3.58 \text{Tr}\left(\,\mathcal{D}\,Z\, Z\, Z\, \mathcal{D}\,Z\, \bar{Z}\,\right)-0.0528 \text{Tr}\left(\,\mathcal{D}\,Z\, Z\, Z\, \mathcal{D}\,X\,
   \bar{X}\,\right)+0.0528 \text{Tr}\left(\,\mathcal{D}\,Z\, Z\, Z\, \mathcal{D}\,Y\, \bar{Y}\,\right)+0.0528 \text{Tr}\left(\,\mathcal{D}\,Z\, Z\, Z\, \mathcal{D}\,\bar{Y}\,
   Y\,\right)-0.0528 \text{Tr}\left(\,\mathcal{D}\,Z\, Z\, Z\, \mathcal{D}\,\bar{X}\, X\,\right)-0.106 \text{Tr}\left(\,\mathcal{D}\,Z\, Z\, Z\, \bar{\Psi } \,
   \Psi  \,\right)+0.106 \text{Tr}\left(\,\mathcal{D}\,Z\, Z\, Z\, \bar{\Psi } \, \Psi  \,\right)-0.500
   \text{Tr}\left(\,\mathcal{D}\,Z\, Z\, X\, Z\, \mathcal{D}\,\bar{X}\,\right)+1.84 \text{Tr}\left(\,\mathcal{D}\,Z\, Z\, X\, \mathcal{D}\,Z\, \bar{X}\,\right)-0.500
   \text{Tr}\left(\,\mathcal{D}\,Z\, Z\, X\, \mathcal{D}\,\bar{X}\, Z\,\right)+0.500 \text{Tr}\left(\,\mathcal{D}\,Z\, Z\, Y\, Z\, \mathcal{D}\,\bar{Y}\,\right)-1.84
   \text{Tr}\left(\,\mathcal{D}\,Z\, Z\, Y\, \mathcal{D}\,Z\, \bar{Y}\,\right)+0.500 \text{Tr}\left(\,\mathcal{D}\,Z\, Z\, Y\, \mathcal{D}\,\bar{Y}\, Z\,\right)+0.500
   \text{Tr}\left(\,\mathcal{D}\,Z\, Z\, \bar{Y}\, Z\, \mathcal{D}\,Y\,\right)-1.84 \text{Tr}\left(\,\mathcal{D}\,Z\, Z\, \bar{Y}\, \mathcal{D}\,Z\, Y\,\right)+0.500
   \text{Tr}\left(\,\mathcal{D}\,Z\, Z\, \bar{Y}\, \mathcal{D}\,Y\, Z\,\right)-0.500 \text{Tr}\left(\,\mathcal{D}\,Z\, Z\, \bar{X}\, Z\, \mathcal{D}\,X\,\right)+1.84
   \text{Tr}\left(\,\mathcal{D}\,Z\, Z\, \bar{X}\, \mathcal{D}\,Z\, X\,\right)-0.500 \text{Tr}\left(\,\mathcal{D}\,Z\, Z\, \bar{X}\, \mathcal{D}\,X\, Z\,\right)-2.00
   \text{Tr}\left(\,\mathcal{D}\,Z\, Z\, \mathcal{D}\,Z\, Z\, \bar{Z}\,\right)+0.658 \text{Tr}\left(\,\mathcal{D}\,Z\, Z\, \mathcal{D}\,Z\, X\, \bar{X}\,\right)-0.658
   \text{Tr}\left(\,\mathcal{D}\,Z\, Z\, \mathcal{D}\,Z\, Y\, \bar{Y}\,\right)-0.658 \text{Tr}\left(\,\mathcal{D}\,Z\, Z\, \mathcal{D}\,Z\, \bar{Y}\, Y\,\right)+0.658
   \text{Tr}\left(\,\mathcal{D}\,Z\, Z\, \mathcal{D}\,Z\, \bar{X}\, X\,\right)-2.00 \text{Tr}\left(\,\mathcal{D}\,Z\, Z\, \mathcal{D}\,Z\, \bar{Z}\, Z\,\right)-0.500
   \text{Tr}\left(\,\mathcal{D}\,Z\, Z\, \mathcal{D}\,X\, \bar{X}\, Z\,\right)+0.500 \text{Tr}\left(\,\mathcal{D}\,Z\, Z\, \mathcal{D}\,Y\, \bar{Y}\, Z\,\right)+0.500
   \text{Tr}\left(\,\mathcal{D}\,Z\, Z\, \mathcal{D}\,\bar{Y}\, Y\, Z\,\right)-0.500 \text{Tr}\left(\,\mathcal{D}\,Z\, Z\, \mathcal{D}\,\bar{X}\, X\, Z\,\right)-1.00
   \text{Tr}\left(\,\mathcal{D}\,Z\, Z\, \bar{\Psi } \, \Psi  \, Z\,\right)+1.00 \text{Tr}\left(\,\mathcal{D}\,Z\, Z\, \bar{\Psi }_{   
   }\, \Psi  \, Z\,\right)-0.0528 \text{Tr}\left(\,\mathcal{D}\,Z\, X\, Z\, Z\, \mathcal{D}\,\bar{X}\,\right)+1.84 \text{Tr}\left(\,\mathcal{D}\,Z\, X\, Z\,
   \mathcal{D}\,Z\, \bar{X}\,\right)+1.84 \text{Tr}\left(\,\mathcal{D}\,Z\, X\, \mathcal{D}\,Z\, \bar{X}\, Z\,\right)-0.0528 \text{Tr}\left(\,\mathcal{D}\,Z\, X\,
   \mathcal{D}\,\bar{X}\, Z\, Z\,\right)+0.0528 \text{Tr}\left(\,\mathcal{D}\,Z\, Y\, Z\, Z\, \mathcal{D}\,\bar{Y}\,\right)-1.84 \text{Tr}\left(\,\mathcal{D}\,Z\, Y\, Z\,
   \mathcal{D}\,Z\, \bar{Y}\,\right)-1.84 \text{Tr}\left(\,\mathcal{D}\,Z\, Y\, \mathcal{D}\,Z\, \bar{Y}\, Z\,\right)+0.0528 \text{Tr}\left(\,\mathcal{D}\,Z\, Y\,
   \mathcal{D}\,\bar{Y}\, Z\, Z\,\right)+0.0528 \text{Tr}\left(\,\mathcal{D}\,Z\, \bar{Y}\, Z\, Z\, \mathcal{D}\,Y\,\right)+0.0528 \text{Tr}\left(\,\mathcal{D}\,Z\, \bar{Y}\,
   \mathcal{D}\,Y\, Z\, Z\,\right)-0.0528 \text{Tr}\left(\,\mathcal{D}\,Z\, \bar{X}\, Z\, Z\, \mathcal{D}\,X\,\right)-0.0528 \text{Tr}\left(\,\mathcal{D}\,Z\, \bar{X}\,
   \mathcal{D}\,X\, Z\, Z\,\right)-0.211 \text{Tr}\left(\,\mathcal{D}\,Z\, \mathcal{D}\,Z\, Z\, Z\, \bar{Z}\,\right)+0.500 \text{Tr}\left(\,\mathcal{D}\,Z\, \mathcal{D}\,Z\, Z\,
   X\, \bar{X}\,\right)-0.500 \text{Tr}\left(\,\mathcal{D}\,Z\, \mathcal{D}\,Z\, Z\, Y\, \bar{Y}\,\right)-0.500 \text{Tr}\left(\,\mathcal{D}\,Z\, \mathcal{D}\,Z\, Z\, \bar{Y}\,
   Y\,\right)+0.500 \text{Tr}\left(\,\mathcal{D}\,Z\, \mathcal{D}\,Z\, Z\, \bar{X}\, X\,\right)+0.658 \text{Tr}\left(\,\mathcal{D}\,Z\, \mathcal{D}\,Z\, X\, Z\,
   \bar{X}\,\right)+0.500 \text{Tr}\left(\,\mathcal{D}\,Z\, \mathcal{D}\,Z\, X\, \bar{X}\, Z\,\right)-0.658 \text{Tr}\left(\,\mathcal{D}\,Z\, \mathcal{D}\,Z\, Y\, Z\,
   \bar{Y}\,\right)-0.500 \text{Tr}\left(\,\mathcal{D}\,Z\, \mathcal{D}\,Z\, Y\, \bar{Y}\, Z\,\right)-0.658 \text{Tr}\left(\,\mathcal{D}\,Z\, \mathcal{D}\,Z\, \bar{Y}\, Z\,
   Y\,\right)-0.500 \text{Tr}\left(\,\mathcal{D}\,Z\, \mathcal{D}\,Z\, \bar{Y}\, Y\, Z\,\right)+0.658 \text{Tr}\left(\,\mathcal{D}\,Z\, \mathcal{D}\,Z\, \bar{X}\, Z\,
   X\,\right)+0.500 \text{Tr}\left(\,\mathcal{D}\,Z\, \mathcal{D}\,Z\, \bar{X}\, X\, Z\,\right)-0.211 \text{Tr}\left(\,\mathcal{D}\,Z\, \mathcal{D}\,Z\, \bar{Z}\, Z\,
   Z\,\right)-0.0528 \text{Tr}\left(\,\mathcal{D}\,Z\, \mathcal{D}\,X\, Z\, Z\, \bar{X}\,\right)-0.500 \text{Tr}\left(\,\mathcal{D}\,Z\, \mathcal{D}\,X\, Z\, \bar{X}\,
   Z\,\right)-0.894 \text{Tr}\left(\,\mathcal{D}\,Z\, \mathcal{D}\,X\, \bar{X}\, Z\, Z\,\right)+0.0528 \text{Tr}\left(\,\mathcal{D}\,Z\, \mathcal{D}\,Y\, Z\, Z\,
   \bar{Y}\,\right)+0.500 \text{Tr}\left(\,\mathcal{D}\,Z\, \mathcal{D}\,Y\, Z\, \bar{Y}\, Z\,\right)+0.894 \text{Tr}\left(\,\mathcal{D}\,Z\, \mathcal{D}\,Y\, \bar{Y}\, Z\,
   Z\,\right)+0.0528 \text{Tr}\left(\,\mathcal{D}\,Z\, \mathcal{D}\,\bar{Y}\, Z\, Z\, Y\,\right)+0.500 \text{Tr}\left(\,\mathcal{D}\,Z\, \mathcal{D}\,\bar{Y}\, Z\, Y\,
   Z\,\right)+0.894 \text{Tr}\left(\,\mathcal{D}\,Z\, \mathcal{D}\,\bar{Y}\, Y\, Z\, Z\,\right)-0.0528 \text{Tr}\left(\,\mathcal{D}\,Z\, \mathcal{D}\,\bar{X}\, Z\, Z\,
   X\,\right)-0.500 \text{Tr}\left(\,\mathcal{D}\,Z\, \mathcal{D}\,\bar{X}\, Z\, X\, Z\,\right)-0.894 \text{Tr}\left(\,\mathcal{D}\,Z\, \mathcal{D}\,\bar{X}\, X\, Z\,
   Z\,\right)-1.79 \text{Tr}\left(\,\mathcal{D}\,Z\, \bar{\Psi } \, \Psi  \, Z\, Z\,\right)-1.00 \text{Tr}\left(\,\mathcal{D}\,Z\, \bar{\Psi
   } \, Z\, \Psi  \, Z\,\right)-0.106 \text{Tr}\left(\,\mathcal{D}\,Z\, \bar{\Psi } \, Z\, Z\, \Psi _{   
   }\,\right)+1.79 \text{Tr}\left(\,\mathcal{D}\,Z\, \bar{\Psi } \, \Psi  \, Z\, Z\,\right)+1.00 \text{Tr}\left(\,\mathcal{D}\,Z\, \bar{\Psi
   } \, Z\, \Psi  \, Z\,\right)+0.106 \text{Tr}\left(\,\mathcal{D}\,Z\, \bar{\Psi } \, Z\, Z\, \Psi _{   
   }\,\right)+1.45 \text{Tr}\left(\,\mathcal{D}\,X\, Z\, Z\, Z\, \mathcal{D}\,\bar{X}\,\right)+0.553 \text{Tr}\left(\,\mathcal{D}\,X\, Z\, Z\, \mathcal{D}\,\bar{X}\,
   Z\,\right)+0.553 \text{Tr}\left(\,\mathcal{D}\,X\, Z\, \mathcal{D}\,\bar{X}\, Z\, Z\,\right)+1.45 \text{Tr}\left(\,\mathcal{D}\,X\, \mathcal{D}\,\bar{X}\, Z\, Z\,
   Z\,\right)-1.45 \text{Tr}\left(\,\mathcal{D}\,Y\, Z\, Z\, Z\, \mathcal{D}\,\bar{Y}\,\right)-0.553 \text{Tr}\left(\,\mathcal{D}\,Y\, Z\, Z\, \mathcal{D}\,\bar{Y}\,
   Z\,\right)-0.553 \text{Tr}\left(\,\mathcal{D}\,Y\, Z\, \mathcal{D}\,\bar{Y}\, Z\, Z\,\right)-1.45 \text{Tr}\left(\,\mathcal{D}\,Y\, \mathcal{D}\,\bar{Y}\, Z\, Z\,
   Z\,\right)-0.106 \text{Tr}\left(\,\bar{\Psi } \, \mathcal{F} \, Z\, Z\, \bar{\Psi } \,\right)-1.00 \text{Tr}\left(\,\bar{\Psi
   } \, \mathcal{F} \, Z\, \bar{\Psi } \, Z\,\right)-1.79 \text{Tr}\left(\,\bar{\Psi } \, \mathcal{F}_{ 
     }\, \bar{\Psi } \, Z\, Z\,\right)-0.500 \text{Tr}\left(\,\bar{\Psi } \, \Psi  \, \Psi  \, Z\,
   \bar{\Psi } \,\right)-0.658 \text{Tr}\left(\,\bar{\Psi } \, \Psi  \, \Psi  \, \bar{\Psi } \,
   Z\,\right)-0.658 \text{Tr}\left(\,\bar{\Psi } \, \Psi  \, Z\, \Psi  \, \bar{\Psi } \,\right)+0.0528
   \text{Tr}\left(\,\bar{\Psi } \, \Psi  \, Z\, Z\, \mathcal{D}\,\bar{X}\,\right)+1.00 \text{Tr}\left(\,\bar{\Psi } \, \Psi
    \, Z\, \bar{X}\, \mathcal{D}\,Z\,\right)+1.00 \text{Tr}\left(\,\bar{\Psi } \, \Psi  \, Z\, \mathcal{D}\,Z\,
   \bar{X}\,\right)+1.84 \text{Tr}\left(\,\bar{\Psi } \, \Psi  \, Z\, \bar{\Psi } \, \Psi  \,\right)+0.842
   \text{Tr}\left(\,\bar{\Psi } \, \Psi  \, \bar{X}\, Z\, \mathcal{D}\,Z\,\right)+1.00 \text{Tr}\left(\,\bar{\Psi } \, \Psi
    \, \bar{X}\, \mathcal{D}\,Z\, Z\,\right)+0.842 \text{Tr}\left(\,\bar{\Psi } \, \Psi  \, \mathcal{D}\,Z\, Z\,
   \bar{X}\,\right)+1.00 \text{Tr}\left(\,\bar{\Psi } \, \Psi  \, \mathcal{D}\,Z\, \bar{X}\, Z\,\right)+0.0528 \text{Tr}\left(\,\bar{\Psi
   } \, \Psi  \, \mathcal{D}\,\bar{X}\, Z\, Z\,\right)+1.84 \text{Tr}\left(\,\bar{\Psi } \, \Psi  \, \bar{\Psi
   } \, \Psi  \, Z\,\right)+1.84 \text{Tr}\left(\,\bar{\Psi } \, \Psi  \, \bar{\Psi } \, Z\,
   \Psi  \,\right)+0.500 \text{Tr}\left(\,\bar{\Psi } \, \Psi  \, \Psi  \, Z\, \bar{\Psi }_{   
   }\,\right)+0.658 \text{Tr}\left(\,\bar{\Psi } \, \Psi  \, \Psi  \, \bar{\Psi } \, Z\,\right)+0.658
   \text{Tr}\left(\,\bar{\Psi } \, \Psi  \, Z\, \Psi  \, \bar{\Psi } \,\right)-0.0528 \text{Tr}\left(\,\bar{\Psi
   } \, \Psi  \, Z\, Z\, \mathcal{D}\,Y\,\right)-1.00 \text{Tr}\left(\,\bar{\Psi } \, \Psi  \, Z\, Y\,
   \mathcal{D}\,Z\,\right)-1.00 \text{Tr}\left(\,\bar{\Psi } \, \Psi  \, Z\, \mathcal{D}\,Z\, Y\,\right)-1.84 \text{Tr}\left(\,\bar{\Psi }_{ 
     }\, \Psi  \, Z\, \bar{\Psi } \, \Psi  \,\right)-0.842 \text{Tr}\left(\,\bar{\Psi } \, \Psi _{ 
     }\, Y\, Z\, \mathcal{D}\,Z\,\right)-1.00 \text{Tr}\left(\,\bar{\Psi } \, \Psi  \, Y\, \mathcal{D}\,Z\, Z\,\right)-0.842
   \text{Tr}\left(\,\bar{\Psi } \, \Psi  \, \mathcal{D}\,Z\, Z\, Y\,\right)-1.00 \text{Tr}\left(\,\bar{\Psi } \, \Psi _{ 
     }\, \mathcal{D}\,Z\, Y\, Z\,\right)-0.0528 \text{Tr}\left(\,\bar{\Psi } \, \Psi  \, \mathcal{D}\,Y\, Z\, Z\,\right)-1.84
   \text{Tr}\left(\,\bar{\Psi } \, \Psi  \, \bar{\Psi } \, \Psi  \, Z\,\right)-1.84 \text{Tr}\left(\,\bar{\Psi
   } \, \Psi  \, \bar{\Psi } \, Z\, \Psi  \,\right)+0.106 \text{Tr}\left(\,\bar{\Psi } \, \Psi
    \, Z\, Z\, \mathcal{D}\,Z\,\right)+1.00 \text{Tr}\left(\,\bar{\Psi } \, \Psi  \, Z\, \mathcal{D}\,Z\, Z\,\right)+1.79
   \text{Tr}\left(\,\bar{\Psi } \, \Psi  \, \mathcal{D}\,Z\, Z\, Z\,\right)-1.00 \text{Tr}\left(\,\bar{\Psi } \, Z\,
   \mathcal{F} \, \bar{\Psi } \, Z\,\right)-0.500 \text{Tr}\left(\,\bar{\Psi } \, Z\, \Psi  \, \Psi _{ 
     }\, \bar{\Psi } \,\right)+0.500 \text{Tr}\left(\,\bar{\Psi } \, Z\, \Psi  \, Z\, \mathcal{D}\,\bar{X}\,\right)+0.842
   \text{Tr}\left(\,\bar{\Psi } \, Z\, \Psi  \, \bar{X}\, \mathcal{D}\,Z\,\right)+0.842 \text{Tr}\left(\,\bar{\Psi } \, Z\, \Psi
    \, \mathcal{D}\,Z\, \bar{X}\,\right)+0.500 \text{Tr}\left(\,\bar{\Psi } \, Z\, \Psi  \, \mathcal{D}\,\bar{X}\, Z\,\right)+1.84
   \text{Tr}\left(\,\bar{\Psi } \, Z\, \Psi  \, \bar{\Psi } \, \Psi  \,\right)+0.500 \text{Tr}\left(\,\bar{\Psi
   } \, Z\, \Psi  \, \Psi  \, \bar{\Psi } \,\right)-0.500 \text{Tr}\left(\,\bar{\Psi } \, Z\,
   \Psi  \, Z\, \mathcal{D}\,Y\,\right)-0.842 \text{Tr}\left(\,\bar{\Psi } \, Z\, \Psi  \, Y\, \mathcal{D}\,Z\,\right)-0.842
   \text{Tr}\left(\,\bar{\Psi } \, Z\, \Psi  \, \mathcal{D}\,Z\, Y\,\right)-0.500 \text{Tr}\left(\,\bar{\Psi } \, Z\, \Psi
    \, \mathcal{D}\,Y\, Z\,\right)-1.84 \text{Tr}\left(\,\bar{\Psi } \, Z\, \Psi  \, \bar{\Psi } \, \Psi _{ 
     }\,\right)+1.00 \text{Tr}\left(\,\bar{\Psi } \, Z\, \Psi  \, \mathcal{D}\,Z\, Z\,\right)-0.106 \text{Tr}\left(\,\bar{\Psi }_{ 
     }\, Z\, Z\, \mathcal{F} \, \bar{\Psi } \,\right)+0.894 \text{Tr}\left(\,\bar{\Psi } \, Z\, Z\, \Psi _{   
   }\, \mathcal{D}\,\bar{X}\,\right)-0.894 \text{Tr}\left(\,\bar{\Psi } \, Z\, Z\, \Psi  \, \mathcal{D}\,Y\,\right)+0.106
   \text{Tr}\left(\,\bar{\Psi } \, Z\, Z\, \Psi  \, \mathcal{D}\,Z\,\right)+0.894 \text{Tr}\left(\,\bar{\Psi } \, Z\, Z\, Y\,
   \mathcal{D}\,\Psi  \,\right)-0.894 \text{Tr}\left(\,\bar{\Psi } \, Z\, Z\, \bar{X}\, \mathcal{D}\,\Psi  \,\right)-0.0528
   \text{Tr}\left(\,\bar{\Psi } \, Z\, Z\, \mathcal{D}\,\Psi  \, \bar{X}\,\right)+0.0528 \text{Tr}\left(\,\bar{\Psi } \, Z\, Z\,
   \mathcal{D}\,\Psi  \, Y\,\right)+1.79 \text{Tr}\left(\,\bar{\Psi } \, Z\, Z\, \mathcal{D}\,Z\, \Psi  \,\right)-0.0528
   \text{Tr}\left(\,\bar{\Psi } \, Z\, Z\, \mathcal{D}\,Y\, \Psi  \,\right)+0.0528 \text{Tr}\left(\,\bar{\Psi } \, Z\, Z\,
   \mathcal{D}\,\bar{X}\, \Psi  \,\right)-1.79 \text{Tr}\left(\,\bar{\Psi } \, Z\, Z\, \bar{\Psi } \, \mathcal{F}_{   
   }\,\right)+0.342 \text{Tr}\left(\,\bar{\Psi } \, Z\, Y\, \Psi  \, \mathcal{D}\,Z\,\right)+0.500 \text{Tr}\left(\,\bar{\Psi } \,
   Z\, Y\, Z\, \mathcal{D}\,\Psi  \,\right)+0.500 \text{Tr}\left(\,\bar{\Psi } \, Z\, Y\, \mathcal{D}\,\Psi  \, Z\,\right)-1.00
   \text{Tr}\left(\,\bar{\Psi } \, Z\, Y\, \mathcal{D}\,Z\, \Psi  \,\right)-0.342 \text{Tr}\left(\,\bar{\Psi } \, Z\, \bar{X}\,
   \Psi  \, \mathcal{D}\,Z\,\right)-0.500 \text{Tr}\left(\,\bar{\Psi } \, Z\, \bar{X}\, Z\, \mathcal{D}\,\Psi  \,\right)-0.500
   \text{Tr}\left(\,\bar{\Psi } \, Z\, \bar{X}\, \mathcal{D}\,\Psi  \, Z\,\right)+1.00 \text{Tr}\left(\,\bar{\Psi } \, Z\,
   \bar{X}\, \mathcal{D}\,Z\, \Psi  \,\right)-0.500 \text{Tr}\left(\,\bar{\Psi } \, Z\, \mathcal{D}\,\Psi  \, \bar{X}\,
   Z\,\right)+0.500 \text{Tr}\left(\,\bar{\Psi } \, Z\, \mathcal{D}\,\Psi  \, Y\, Z\,\right)-0.342 \text{Tr}\left(\,\bar{\Psi } \,
   Z\, \mathcal{D}\,Z\, \Psi  \, \bar{X}\,\right)+0.342 \text{Tr}\left(\,\bar{\Psi } \, Z\, \mathcal{D}\,Z\, \Psi  \,
   Y\,\right)+1.00 \text{Tr}\left(\,\bar{\Psi } \, Z\, \mathcal{D}\,Z\, \Psi  \, Z\,\right)+1.00 \text{Tr}\left(\,\bar{\Psi } \,
   Z\, \mathcal{D}\,Z\, Z\, \Psi  \,\right)-1.00 \text{Tr}\left(\,\bar{\Psi } \, Z\, \mathcal{D}\,Z\, Y\, \Psi  \,\right)+1.00
   \text{Tr}\left(\,\bar{\Psi } \, Z\, \mathcal{D}\,Z\, \bar{X}\, \Psi  \,\right)-0.500 \text{Tr}\left(\,\bar{\Psi } \, Z\,
   \mathcal{D}\,Y\, \Psi  \, Z\,\right)+0.500 \text{Tr}\left(\,\bar{\Psi } \, Z\, \mathcal{D}\,\bar{X}\, \Psi  \, Z\,\right)-1.00
   \text{Tr}\left(\,\bar{\Psi } \, Z\, \bar{\Psi } \, \mathcal{F} \, Z\,\right)-0.658 \text{Tr}\left(\,\bar{\Psi }_{ 
     }\, Z\, \bar{\Psi } \, \Psi  \, \Psi  \,\right)+0.658 \text{Tr}\left(\,\bar{\Psi } \, Z\, \bar{\Psi
   } \, \Psi  \, \Psi  \,\right)-1.00 \text{Tr}\left(\,\bar{\Psi } \, Z\, \bar{\Psi } \, Z\,
   \mathcal{F} \,\right)+0.342 \text{Tr}\left(\,\bar{\Psi } \, Y\, \Psi  \, Z\, \mathcal{D}\,Z\,\right)+0.342
   \text{Tr}\left(\,\bar{\Psi } \, Y\, \Psi  \, \mathcal{D}\,Z\, Z\,\right)+0.342 \text{Tr}\left(\,\bar{\Psi } \, Y\, Z\, \Psi
    \, \mathcal{D}\,Z\,\right)+0.0528 \text{Tr}\left(\,\bar{\Psi } \, Y\, Z\, Z\, \mathcal{D}\,\Psi  \,\right)-0.842
   \text{Tr}\left(\,\bar{\Psi } \, Y\, Z\, \mathcal{D}\,Z\, \Psi  \,\right)+0.0528 \text{Tr}\left(\,\bar{\Psi } \, Y\,
   \mathcal{D}\,\Psi  \, Z\, Z\,\right)-0.842 \text{Tr}\left(\,\bar{\Psi } \, Y\, \mathcal{D}\,Z\, \Psi  \, Z\,\right)-1.00
   \text{Tr}\left(\,\bar{\Psi } \, Y\, \mathcal{D}\,Z\, Z\, \Psi  \,\right)-0.342 \text{Tr}\left(\,\bar{\Psi } \, \bar{X}\, \Psi
    \, Z\, \mathcal{D}\,Z\,\right)-0.342 \text{Tr}\left(\,\bar{\Psi } \, \bar{X}\, \Psi  \, \mathcal{D}\,Z\, Z\,\right)-0.342
   \text{Tr}\left(\,\bar{\Psi } \, \bar{X}\, Z\, \Psi  \, \mathcal{D}\,Z\,\right)-0.0528 \text{Tr}\left(\,\bar{\Psi } \,
   \bar{X}\, Z\, Z\, \mathcal{D}\,\Psi  \,\right)+0.842 \text{Tr}\left(\,\bar{\Psi } \, \bar{X}\, Z\, \mathcal{D}\,Z\, \Psi _{   
   }\,\right)-0.0528 \text{Tr}\left(\,\bar{\Psi } \, \bar{X}\, \mathcal{D}\,\Psi  \, Z\, Z\,\right)+0.842 \text{Tr}\left(\,\bar{\Psi }_{ 
     }\, \bar{X}\, \mathcal{D}\,Z\, \Psi  \, Z\,\right)+1.00 \text{Tr}\left(\,\bar{\Psi } \, \bar{X}\, \mathcal{D}\,Z\, Z\, \Psi _{ 
     }\,\right)-0.0528 \text{Tr}\left(\,\bar{\Psi } \, \mathcal{D}\,\Psi  \, Z\, Z\, \bar{X}\,\right)-0.500 \text{Tr}\left(\,\bar{\Psi
   } \, \mathcal{D}\,\Psi  \, Z\, \bar{X}\, Z\,\right)-0.894 \text{Tr}\left(\,\bar{\Psi } \, \mathcal{D}\,\Psi  \,
   \bar{X}\, Z\, Z\,\right)+0.0528 \text{Tr}\left(\,\bar{\Psi } \, \mathcal{D}\,\Psi  \, Z\, Z\, Y\,\right)+0.500 \text{Tr}\left(\,\bar{\Psi
   } \, \mathcal{D}\,\Psi  \, Z\, Y\, Z\,\right)+0.894 \text{Tr}\left(\,\bar{\Psi } \, \mathcal{D}\,\Psi  \, Y\,
   Z\, Z\,\right)-0.342 \text{Tr}\left(\,\bar{\Psi } \, \mathcal{D}\,Z\, \Psi  \, Z\, \bar{X}\,\right)-0.342 \text{Tr}\left(\,\bar{\Psi }_{ 
     }\, \mathcal{D}\,Z\, \Psi  \, \bar{X}\, Z\,\right)+0.342 \text{Tr}\left(\,\bar{\Psi } \, \mathcal{D}\,Z\, \Psi  \, Z\,
   Y\,\right)+0.342 \text{Tr}\left(\,\bar{\Psi } \, \mathcal{D}\,Z\, \Psi  \, Y\, Z\,\right)+0.106 \text{Tr}\left(\,\bar{\Psi } \,
   \mathcal{D}\,Z\, \Psi  \, Z\, Z\,\right)-0.342 \text{Tr}\left(\,\bar{\Psi } \, \mathcal{D}\,Z\, Z\, \Psi  \,
   \bar{X}\,\right)+0.342 \text{Tr}\left(\,\bar{\Psi } \, \mathcal{D}\,Z\, Z\, \Psi  \, Y\,\right)+0.106 \text{Tr}\left(\,\bar{\Psi }_{ 
     }\, \mathcal{D}\,Z\, Z\, Z\, \Psi  \,\right)-0.842 \text{Tr}\left(\,\bar{\Psi } \, \mathcal{D}\,Z\, Z\, Y\, \Psi _{   
   }\,\right)+0.842 \text{Tr}\left(\,\bar{\Psi } \, \mathcal{D}\,Z\, Z\, \bar{X}\, \Psi  \,\right)-0.842 \text{Tr}\left(\,\bar{\Psi }_{ 
     }\, \mathcal{D}\,Z\, Y\, \Psi  \, Z\,\right)-1.00 \text{Tr}\left(\,\bar{\Psi } \, \mathcal{D}\,Z\, Y\, Z\, \Psi _{   
   }\,\right)+0.842 \text{Tr}\left(\,\bar{\Psi } \, \mathcal{D}\,Z\, \bar{X}\, \Psi  \, Z\,\right)+1.00 \text{Tr}\left(\,\bar{\Psi }_{ 
     }\, \mathcal{D}\,Z\, \bar{X}\, Z\, \Psi  \,\right)-0.894 \text{Tr}\left(\,\bar{\Psi } \, \mathcal{D}\,Y\, \Psi  \, Z\,
   Z\,\right)-0.500 \text{Tr}\left(\,\bar{\Psi } \, \mathcal{D}\,Y\, Z\, \Psi  \, Z\,\right)-0.0528 \text{Tr}\left(\,\bar{\Psi }_{   
   }\, \mathcal{D}\,Y\, Z\, Z\, \Psi  \,\right)+0.894 \text{Tr}\left(\,\bar{\Psi } \, \mathcal{D}\,\bar{X}\, \Psi  \, Z\,
   Z\,\right)+0.500 \text{Tr}\left(\,\bar{\Psi } \, \mathcal{D}\,\bar{X}\, Z\, \Psi  \, Z\,\right)+0.0528 \text{Tr}\left(\,\bar{\Psi }_{ 
     }\, \mathcal{D}\,\bar{X}\, Z\, Z\, \Psi  \,\right)-0.106 \text{Tr}\left(\,\bar{\Psi } \, \bar{\Psi } \,
   \mathcal{F} \, Z\, Z\,\right)-0.500 \text{Tr}\left(\,\bar{\Psi } \, \bar{\Psi } \, \Psi  \, \Psi _{ 
     }\, Z\,\right)-0.658 \text{Tr}\left(\,\bar{\Psi } \, \bar{\Psi } \, \Psi  \, Z\, \Psi  \,\right)+0.500
   \text{Tr}\left(\,\bar{\Psi } \, \bar{\Psi } \, \Psi  \, \Psi  \, Z\,\right)+0.658 \text{Tr}\left(\,\bar{\Psi
   } \, \bar{\Psi } \, \Psi  \, Z\, \Psi  \,\right)-0.500 \text{Tr}\left(\,\bar{\Psi } \,
   \bar{\Psi } \, Z\, \Psi  \, \Psi  \,\right)+0.500 \text{Tr}\left(\,\bar{\Psi } \, \bar{\Psi }_{   
   }\, Z\, \Psi  \, \Psi  \,\right)-0.106 \text{Tr}\left(\,\bar{\Psi } \, \bar{\Psi } \, Z\, Z\,
   \mathcal{F} \,\right)-0.0528 \text{Tr}\left(\,\bar{\Psi } \, \Psi  \, Z\, Z\, \mathcal{D}\,\bar{Y}\,\right)-1.00
   \text{Tr}\left(\,\bar{\Psi } \, \Psi  \, Z\, \bar{Y}\, \mathcal{D}\,Z\,\right)-1.00 \text{Tr}\left(\,\bar{\Psi } \, \Psi
    \, Z\, \mathcal{D}\,Z\, \bar{Y}\,\right)-0.842 \text{Tr}\left(\,\bar{\Psi } \, \Psi  \, \bar{Y}\, Z\,
   \mathcal{D}\,Z\,\right)-1.00 \text{Tr}\left(\,\bar{\Psi } \, \Psi  \, \bar{Y}\, \mathcal{D}\,Z\, Z\,\right)-0.842 \text{Tr}\left(\,\bar{\Psi
   } \, \Psi  \, \mathcal{D}\,Z\, Z\, \bar{Y}\,\right)-1.00 \text{Tr}\left(\,\bar{\Psi } \, \Psi  \,
   \mathcal{D}\,Z\, \bar{Y}\, Z\,\right)-0.0528 \text{Tr}\left(\,\bar{\Psi } \, \Psi  \, \mathcal{D}\,\bar{Y}\, Z\, Z\,\right)+0.0528
   \text{Tr}\left(\,\bar{\Psi } \, \Psi  \, Z\, Z\, \mathcal{D}\,X\,\right)+1.00 \text{Tr}\left(\,\bar{\Psi } \, \Psi _{ 
     }\, Z\, X\, \mathcal{D}\,Z\,\right)+1.00 \text{Tr}\left(\,\bar{\Psi } \, \Psi  \, Z\, \mathcal{D}\,Z\, X\,\right)+0.842
   \text{Tr}\left(\,\bar{\Psi } \, \Psi  \, X\, Z\, \mathcal{D}\,Z\,\right)+1.00 \text{Tr}\left(\,\bar{\Psi } \, \Psi _{ 
     }\, X\, \mathcal{D}\,Z\, Z\,\right)+0.842 \text{Tr}\left(\,\bar{\Psi } \, \Psi  \, \mathcal{D}\,Z\, Z\, X\,\right)+1.00
   \text{Tr}\left(\,\bar{\Psi } \, \Psi  \, \mathcal{D}\,Z\, X\, Z\,\right)+0.0528 \text{Tr}\left(\,\bar{\Psi } \, \Psi _{ 
     }\, \mathcal{D}\,X\, Z\, Z\,\right)-0.106 \text{Tr}\left(\,\bar{\Psi } \, \Psi  \, Z\, Z\, \mathcal{D}\,Z\,\right)-1.00
   \text{Tr}\left(\,\bar{\Psi } \, \Psi  \, Z\, \mathcal{D}\,Z\, Z\,\right)-1.79 \text{Tr}\left(\,\bar{\Psi } \, \Psi _{ 
     }\, \mathcal{D}\,Z\, Z\, Z\,\right)-0.500 \text{Tr}\left(\,\bar{\Psi } \, Z\, \Psi  \, Z\, \mathcal{D}\,\bar{Y}\,\right)-0.842
   \text{Tr}\left(\,\bar{\Psi } \, Z\, \Psi  \, \bar{Y}\, \mathcal{D}\,Z\,\right)-0.842 \text{Tr}\left(\,\bar{\Psi } \, Z\, \Psi
    \, \mathcal{D}\,Z\, \bar{Y}\,\right)-0.500 \text{Tr}\left(\,\bar{\Psi } \, Z\, \Psi  \, \mathcal{D}\,\bar{Y}\,
   Z\,\right)+0.500 \text{Tr}\left(\,\bar{\Psi } \, Z\, \Psi  \, Z\, \mathcal{D}\,X\,\right)+0.842 \text{Tr}\left(\,\bar{\Psi } \,
   Z\, \Psi  \, X\, \mathcal{D}\,Z\,\right)+0.842 \text{Tr}\left(\,\bar{\Psi } \, Z\, \Psi  \, \mathcal{D}\,Z\, X\,\right)+0.500
   \text{Tr}\left(\,\bar{\Psi } \, Z\, \Psi  \, \mathcal{D}\,X\, Z\,\right)-1.00 \text{Tr}\left(\,\bar{\Psi } \, Z\, \Psi
    \, \mathcal{D}\,Z\, Z\,\right)-0.894 \text{Tr}\left(\,\bar{\Psi } \, Z\, Z\, \Psi  \, \mathcal{D}\,\bar{Y}\,\right)+0.894
   \text{Tr}\left(\,\bar{\Psi } \, Z\, Z\, \Psi  \, \mathcal{D}\,X\,\right)-0.106 \text{Tr}\left(\,\bar{\Psi } \, Z\, Z\, \Psi
    \, \mathcal{D}\,Z\,\right)-0.894 \text{Tr}\left(\,\bar{\Psi } \, Z\, Z\, X\, \mathcal{D}\,\Psi  \,\right)+0.894
   \text{Tr}\left(\,\bar{\Psi } \, Z\, Z\, \bar{Y}\, \mathcal{D}\,\Psi  \,\right)+0.0528 \text{Tr}\left(\,\bar{\Psi } \, Z\, Z\,
   \mathcal{D}\,\Psi  \, \bar{Y}\,\right)-0.0528 \text{Tr}\left(\,\bar{\Psi } \, Z\, Z\, \mathcal{D}\,\Psi  \, X\,\right)-1.79
   \text{Tr}\left(\,\bar{\Psi } \, Z\, Z\, \mathcal{D}\,Z\, \Psi  \,\right)+0.0528 \text{Tr}\left(\,\bar{\Psi } \, Z\, Z\,
   \mathcal{D}\,X\, \Psi  \,\right)-0.0528 \text{Tr}\left(\,\bar{\Psi } \, Z\, Z\, \mathcal{D}\,\bar{Y}\, \Psi  \,\right)-0.342
   \text{Tr}\left(\,\bar{\Psi } \, Z\, X\, \Psi  \, \mathcal{D}\,Z\,\right)-0.500 \text{Tr}\left(\,\bar{\Psi } \, Z\, X\, Z\,
   \mathcal{D}\,\Psi  \,\right)-0.500 \text{Tr}\left(\,\bar{\Psi } \, Z\, X\, \mathcal{D}\,\Psi  \, Z\,\right)+1.00
   \text{Tr}\left(\,\bar{\Psi } \, Z\, X\, \mathcal{D}\,Z\, \Psi  \,\right)+0.342 \text{Tr}\left(\,\bar{\Psi } \, Z\, \bar{Y}\,
   \Psi  \, \mathcal{D}\,Z\,\right)+0.500 \text{Tr}\left(\,\bar{\Psi } \, Z\, \bar{Y}\, Z\, \mathcal{D}\,\Psi  \,\right)+0.500
   \text{Tr}\left(\,\bar{\Psi } \, Z\, \bar{Y}\, \mathcal{D}\,\Psi  \, Z\,\right)-1.00 \text{Tr}\left(\,\bar{\Psi } \, Z\,
   \bar{Y}\, \mathcal{D}\,Z\, \Psi  \,\right)+0.500 \text{Tr}\left(\,\bar{\Psi } \, Z\, \mathcal{D}\,\Psi  \, \bar{Y}\,
   Z\,\right)-0.500 \text{Tr}\left(\,\bar{\Psi } \, Z\, \mathcal{D}\,\Psi  \, X\, Z\,\right)+0.342 \text{Tr}\left(\,\bar{\Psi } \,
   Z\, \mathcal{D}\,Z\, \Psi  \, \bar{Y}\,\right)-0.342 \text{Tr}\left(\,\bar{\Psi } \, Z\, \mathcal{D}\,Z\, \Psi  \,
   X\,\right)-1.00 \text{Tr}\left(\,\bar{\Psi } \, Z\, \mathcal{D}\,Z\, \Psi  \, Z\,\right)-1.00 \text{Tr}\left(\,\bar{\Psi } \,
   Z\, \mathcal{D}\,Z\, Z\, \Psi  \,\right)+1.00 \text{Tr}\left(\,\bar{\Psi } \, Z\, \mathcal{D}\,Z\, X\, \Psi  \,\right)-1.00
   \text{Tr}\left(\,\bar{\Psi } \, Z\, \mathcal{D}\,Z\, \bar{Y}\, \Psi  \,\right)+0.500 \text{Tr}\left(\,\bar{\Psi } \, Z\,
   \mathcal{D}\,X\, \Psi  \, Z\,\right)-0.500 \text{Tr}\left(\,\bar{\Psi } \, Z\, \mathcal{D}\,\bar{Y}\, \Psi  \, Z\,\right)-0.342
   \text{Tr}\left(\,\bar{\Psi } \, X\, \Psi  \, Z\, \mathcal{D}\,Z\,\right)-0.342 \text{Tr}\left(\,\bar{\Psi } \, X\, \Psi
    \, \mathcal{D}\,Z\, Z\,\right)-0.342 \text{Tr}\left(\,\bar{\Psi } \, X\, Z\, \Psi  \, \mathcal{D}\,Z\,\right)-0.0528
   \text{Tr}\left(\,\bar{\Psi } \, X\, Z\, Z\, \mathcal{D}\,\Psi  \,\right)+0.842 \text{Tr}\left(\,\bar{\Psi } \, X\, Z\,
   \mathcal{D}\,Z\, \Psi  \,\right)-0.0528 \text{Tr}\left(\,\bar{\Psi } \, X\, \mathcal{D}\,\Psi  \, Z\, Z\,\right)+0.842
   \text{Tr}\left(\,\bar{\Psi } \, X\, \mathcal{D}\,Z\, \Psi  \, Z\,\right)+1.00 \text{Tr}\left(\,\bar{\Psi } \, X\,
   \mathcal{D}\,Z\, Z\, \Psi  \,\right)+0.342 \text{Tr}\left(\,\bar{\Psi } \, \bar{Y}\, \Psi  \, Z\, \mathcal{D}\,Z\,\right)+0.342
   \text{Tr}\left(\,\bar{\Psi } \, \bar{Y}\, \Psi  \, \mathcal{D}\,Z\, Z\,\right)+0.342 \text{Tr}\left(\,\bar{\Psi } \,
   \bar{Y}\, Z\, \Psi  \, \mathcal{D}\,Z\,\right)+0.0528 \text{Tr}\left(\,\bar{\Psi } \, \bar{Y}\, Z\, Z\, \mathcal{D}\,\Psi _{   
   }\,\right)-0.842 \text{Tr}\left(\,\bar{\Psi } \, \bar{Y}\, Z\, \mathcal{D}\,Z\, \Psi  \,\right)+0.0528 \text{Tr}\left(\,\bar{\Psi }_{ 
     }\, \bar{Y}\, \mathcal{D}\,\Psi  \, Z\, Z\,\right)-0.842 \text{Tr}\left(\,\bar{\Psi } \, \bar{Y}\, \mathcal{D}\,Z\, \Psi _{ 
     }\, Z\,\right)-1.00 \text{Tr}\left(\,\bar{\Psi } \, \bar{Y}\, \mathcal{D}\,Z\, Z\, \Psi  \,\right)+0.0528 \text{Tr}\left(\,\bar{\Psi
   } \, \mathcal{D}\,\Psi  \, Z\, Z\, \bar{Y}\,\right)+0.500 \text{Tr}\left(\,\bar{\Psi } \, \mathcal{D}\,\Psi  \,
   Z\, \bar{Y}\, Z\,\right)+0.894 \text{Tr}\left(\,\bar{\Psi } \, \mathcal{D}\,\Psi  \, \bar{Y}\, Z\, Z\,\right)-0.0528
   \text{Tr}\left(\,\bar{\Psi } \, \mathcal{D}\,\Psi  \, Z\, Z\, X\,\right)-0.500 \text{Tr}\left(\,\bar{\Psi } \,
   \mathcal{D}\,\Psi  \, Z\, X\, Z\,\right)-0.894 \text{Tr}\left(\,\bar{\Psi } \, \mathcal{D}\,\Psi  \, X\, Z\, Z\,\right)+0.342
   \text{Tr}\left(\,\bar{\Psi } \, \mathcal{D}\,Z\, \Psi  \, Z\, \bar{Y}\,\right)+0.342 \text{Tr}\left(\,\bar{\Psi } \,
   \mathcal{D}\,Z\, \Psi  \, \bar{Y}\, Z\,\right)-0.342 \text{Tr}\left(\,\bar{\Psi } \, \mathcal{D}\,Z\, \Psi  \, Z\,
   X\,\right)-0.342 \text{Tr}\left(\,\bar{\Psi } \, \mathcal{D}\,Z\, \Psi  \, X\, Z\,\right)-0.106 \text{Tr}\left(\,\bar{\Psi } \,
   \mathcal{D}\,Z\, \Psi  \, Z\, Z\,\right)+0.342 \text{Tr}\left(\,\bar{\Psi } \, \mathcal{D}\,Z\, Z\, \Psi  \,
   \bar{Y}\,\right)-0.342 \text{Tr}\left(\,\bar{\Psi } \, \mathcal{D}\,Z\, Z\, \Psi  \, X\,\right)-0.106 \text{Tr}\left(\,\bar{\Psi }_{ 
     }\, \mathcal{D}\,Z\, Z\, Z\, \Psi  \,\right)+0.842 \text{Tr}\left(\,\bar{\Psi } \, \mathcal{D}\,Z\, Z\, X\, \Psi _{   
   }\,\right)-0.842 \text{Tr}\left(\,\bar{\Psi } \, \mathcal{D}\,Z\, Z\, \bar{Y}\, \Psi  \,\right)+0.842 \text{Tr}\left(\,\bar{\Psi }_{ 
     }\, \mathcal{D}\,Z\, X\, \Psi  \, Z\,\right)+1.00 \text{Tr}\left(\,\bar{\Psi } \, \mathcal{D}\,Z\, X\, Z\, \Psi _{   
   }\,\right)-0.842 \text{Tr}\left(\,\bar{\Psi } \, \mathcal{D}\,Z\, \bar{Y}\, \Psi  \, Z\,\right)-1.00 \text{Tr}\left(\,\bar{\Psi }_{ 
     }\, \mathcal{D}\,Z\, \bar{Y}\, Z\, \Psi  \,\right)+0.894 \text{Tr}\left(\,\bar{\Psi } \, \mathcal{D}\,X\, \Psi  \, Z\,
   Z\,\right)+0.500 \text{Tr}\left(\,\bar{\Psi } \, \mathcal{D}\,X\, Z\, \Psi  \, Z\,\right)+0.0528 \text{Tr}\left(\,\bar{\Psi }_{   
   }\, \mathcal{D}\,X\, Z\, Z\, \Psi  \,\right)-0.894 \text{Tr}\left(\,\bar{\Psi } \, \mathcal{D}\,\bar{Y}\, \Psi  \, Z\,
   Z\,\right)-0.500 \text{Tr}\left(\,\bar{\Psi } \, \mathcal{D}\,\bar{Y}\, Z\, \Psi  \, Z\,\right)-0.0528 \text{Tr}\left(\,\bar{\Psi }_{ 
     }\, \mathcal{D}\,\bar{Y}\, Z\, Z\, \Psi  \,\right)+1.45 \text{Tr}\left(\,\bar{\Psi } \, Z\, Z\, Z\, \mathcal{D}\,\Psi _{   
   }\,\right)+0.553 \text{Tr}\left(\,\bar{\Psi } \, Z\, Z\, \mathcal{D}\,\Psi  \, Z\,\right)+0.553 \text{Tr}\left(\,\bar{\Psi } \,
   Z\, \mathcal{D}\,\Psi  \, Z\, Z\,\right)+1.45 \text{Tr}\left(\,\bar{\Psi } \, \mathcal{D}\,\Psi  \, Z\, Z\, Z\,\right)+1.45
   \text{Tr}\left(\, \mathcal{D}^2\,Z\, Z\, Z\, Z\, \bar{Z}\,\right)-0.0528 \text{Tr}\left(\, \mathcal{D}^2\,Z\, Z\, Z\, X\, \bar{X}\,\right)+0.0528
   \text{Tr}\left(\, \mathcal{D}^2\,Z\, Z\, Z\, Y\, \bar{Y}\,\right)+0.0528 \text{Tr}\left(\, \mathcal{D}^2\,Z\, Z\, Z\, \bar{Y}\, Y\,\right)-0.0528
   \text{Tr}\left(\, \mathcal{D}^2\,Z\, Z\, Z\, \bar{X}\, X\,\right)+0.553 \text{Tr}\left(\, \mathcal{D}^2\,Z\, Z\, Z\, \bar{Z}\, Z\,\right)-0.500
   \text{Tr}\left(\, \mathcal{D}^2\,Z\, Z\, X\, Z\, \bar{X}\,\right)+0.500 \text{Tr}\left(\, \mathcal{D}^2\,Z\, Z\, Y\, Z\, \bar{Y}\,\right)+0.500
   \text{Tr}\left(\, \mathcal{D}^2\,Z\, Z\, \bar{Y}\, Z\, Y\,\right)-0.500 \text{Tr}\left(\, \mathcal{D}^2\,Z\, Z\, \bar{X}\, Z\, X\,\right)+0.553
   \text{Tr}\left(\, \mathcal{D}^2\,Z\, Z\, \bar{Z}\, Z\, Z\,\right)-0.894 \text{Tr}\left(\, \mathcal{D}^2\,Z\, X\, Z\, Z\, \bar{X}\,\right)-0.500
   \text{Tr}\left(\, \mathcal{D}^2\,Z\, X\, Z\, \bar{X}\, Z\,\right)-0.0528 \text{Tr}\left(\, \mathcal{D}^2\,Z\, X\, \bar{X}\, Z\, Z\,\right)+0.894
   \text{Tr}\left(\, \mathcal{D}^2\,Z\, Y\, Z\, Z\, \bar{Y}\,\right)+0.500 \text{Tr}\left(\, \mathcal{D}^2\,Z\, Y\, Z\, \bar{Y}\, Z\,\right)+0.0528
   \text{Tr}\left(\, \mathcal{D}^2\,Z\, Y\, \bar{Y}\, Z\, Z\,\right)+0.894 \text{Tr}\left(\, \mathcal{D}^2\,Z\, \bar{Y}\, Z\, Z\, Y\,\right)+0.500
   \text{Tr}\left(\, \mathcal{D}^2\,Z\, \bar{Y}\, Z\, Y\, Z\,\right)+0.0528 \text{Tr}\left(\, \mathcal{D}^2\,Z\, \bar{Y}\, Y\, Z\, Z\,\right)-0.894
   \text{Tr}\left(\, \mathcal{D}^2\,Z\, \bar{X}\, Z\, Z\, X\,\right)-0.500 \text{Tr}\left(\, \mathcal{D}^2\,Z\, \bar{X}\, Z\, X\, Z\,\right)-0.0528
   \text{Tr}\left(\, \mathcal{D}^2\,Z\, \bar{X}\, X\, Z\, Z\,\right)+1.45 \text{Tr}\left(\, \mathcal{D}^2\,Z\, \bar{Z}\, Z\, Z\, Z\,\right)-0.894
   \text{Tr}\left(\,\mathcal{D}\,\bar{\Psi } \, \Psi  \, Z\, Z\, \bar{X}\,\right)-0.500 \text{Tr}\left(\,\mathcal{D}\,\bar{\Psi }_{   
   }\, \Psi  \, Z\, \bar{X}\, Z\,\right)-0.0528 \text{Tr}\left(\,\mathcal{D}\,\bar{\Psi } \, \Psi  \, \bar{X}\, Z\,
   Z\,\right)+0.894 \text{Tr}\left(\,\mathcal{D}\,\bar{\Psi } \, \Psi  \, Z\, Z\, Y\,\right)+0.500 \text{Tr}\left(\,\mathcal{D}\,\bar{\Psi
   } \, \Psi  \, Z\, Y\, Z\,\right)+0.0528 \text{Tr}\left(\,\mathcal{D}\,\bar{\Psi } \, \Psi  \, Y\, Z\,
   Z\,\right)-1.45 \text{Tr}\left(\,\mathcal{D}\,\bar{\Psi } \, \Psi  \, Z\, Z\, Z\,\right)-0.500 \text{Tr}\left(\,\mathcal{D}\,\bar{\Psi }_{ 
     }\, Z\, \Psi  \, Z\, \bar{X}\,\right)+0.500 \text{Tr}\left(\,\mathcal{D}\,\bar{\Psi } \, Z\, \Psi  \, Z\,
   Y\,\right)-0.553 \text{Tr}\left(\,\mathcal{D}\,\bar{\Psi } \, Z\, \Psi  \, Z\, Z\,\right)-0.0528 \text{Tr}\left(\,\mathcal{D}\,\bar{\Psi
   } \, Z\, Z\, \Psi  \, \bar{X}\,\right)+0.0528 \text{Tr}\left(\,\mathcal{D}\,\bar{\Psi } \, Z\, Z\, \Psi  \,
   Y\,\right)-0.553 \text{Tr}\left(\,\mathcal{D}\,\bar{\Psi } \, Z\, Z\, \Psi  \, Z\,\right)-1.45 \text{Tr}\left(\,\mathcal{D}\,\bar{\Psi }_{ 
     }\, Z\, Z\, Z\, \Psi  \,\right)+0.0528 \text{Tr}\left(\,\mathcal{D}\,\bar{\Psi } \, Z\, Z\, Y\, \Psi  \,\right)-0.0528
   \text{Tr}\left(\,\mathcal{D}\,\bar{\Psi } \, Z\, Z\, \bar{X}\, \Psi  \,\right)+0.500 \text{Tr}\left(\,\mathcal{D}\,\bar{\Psi }_{   
   }\, Z\, Y\, Z\, \Psi  \,\right)-0.500 \text{Tr}\left(\,\mathcal{D}\,\bar{\Psi } \, Z\, \bar{X}\, Z\, \Psi  \,\right)+0.0528
   \text{Tr}\left(\,\mathcal{D}\,\bar{\Psi } \, Y\, \Psi  \, Z\, Z\,\right)+0.500 \text{Tr}\left(\,\mathcal{D}\,\bar{\Psi } \, Y\,
   Z\, \Psi  \, Z\,\right)+0.894 \text{Tr}\left(\,\mathcal{D}\,\bar{\Psi } \, Y\, Z\, Z\, \Psi  \,\right)-0.0528
   \text{Tr}\left(\,\mathcal{D}\,\bar{\Psi } \, \bar{X}\, \Psi  \, Z\, Z\,\right)-0.500 \text{Tr}\left(\,\mathcal{D}\,\bar{\Psi }_{   
   }\, \bar{X}\, Z\, \Psi  \, Z\,\right)-0.894 \text{Tr}\left(\,\mathcal{D}\,\bar{\Psi } \, \bar{X}\, Z\, Z\, \Psi _{   
   }\,\right)+0.894 \text{Tr}\left(\,\mathcal{D}\,\bar{\Psi } \, \Psi  \, Z\, Z\, \bar{Y}\,\right)+0.500 \text{Tr}\left(\,\mathcal{D}\,\bar{\Psi
   } \, \Psi  \, Z\, \bar{Y}\, Z\,\right)+0.0528 \text{Tr}\left(\,\mathcal{D}\,\bar{\Psi } \, \Psi  \, \bar{Y}\,
   Z\, Z\,\right)-0.894 \text{Tr}\left(\,\mathcal{D}\,\bar{\Psi } \, \Psi  \, Z\, Z\, X\,\right)-0.500 \text{Tr}\left(\,\mathcal{D}\,\bar{\Psi
   } \, \Psi  \, Z\, X\, Z\,\right)-0.0528 \text{Tr}\left(\,\mathcal{D}\,\bar{\Psi } \, \Psi  \, X\, Z\,
   Z\,\right)+1.45 \text{Tr}\left(\,\mathcal{D}\,\bar{\Psi } \, \Psi  \, Z\, Z\, Z\,\right)+0.500 \text{Tr}\left(\,\mathcal{D}\,\bar{\Psi }_{ 
     }\, Z\, \Psi  \, Z\, \bar{Y}\,\right)-0.500 \text{Tr}\left(\,\mathcal{D}\,\bar{\Psi } \, Z\, \Psi  \, Z\,
   X\,\right)+0.553 \text{Tr}\left(\,\mathcal{D}\,\bar{\Psi } \, Z\, \Psi  \, Z\, Z\,\right)+0.0528 \text{Tr}\left(\,\mathcal{D}\,\bar{\Psi
   } \, Z\, Z\, \Psi  \, \bar{Y}\,\right)-0.0528 \text{Tr}\left(\,\mathcal{D}\,\bar{\Psi } \, Z\, Z\, \Psi  \,
   X\,\right)+0.553 \text{Tr}\left(\,\mathcal{D}\,\bar{\Psi } \, Z\, Z\, \Psi  \, Z\,\right)+1.45 \text{Tr}\left(\,\mathcal{D}\,\bar{\Psi }_{ 
     }\, Z\, Z\, Z\, \Psi  \,\right)-0.0528 \text{Tr}\left(\,\mathcal{D}\,\bar{\Psi } \, Z\, Z\, X\, \Psi  \,\right)+0.0528
   \text{Tr}\left(\,\mathcal{D}\,\bar{\Psi } \, Z\, Z\, \bar{Y}\, \Psi  \,\right)-0.500 \text{Tr}\left(\,\mathcal{D}\,\bar{\Psi }_{   
   }\, Z\, X\, Z\, \Psi  \,\right)+0.500 \text{Tr}\left(\,\mathcal{D}\,\bar{\Psi } \, Z\, \bar{Y}\, Z\, \Psi  \,\right)-0.0528
   \text{Tr}\left(\,\mathcal{D}\,\bar{\Psi } \, X\, \Psi  \, Z\, Z\,\right)-0.500 \text{Tr}\left(\,\mathcal{D}\,\bar{\Psi } \, X\,
   Z\, \Psi  \, Z\,\right)-0.894 \text{Tr}\left(\,\mathcal{D}\,\bar{\Psi } \, X\, Z\, Z\, \Psi  \,\right)+0.0528
   \text{Tr}\left(\,\mathcal{D}\,\bar{\Psi } \, \bar{Y}\, \Psi  \, Z\, Z\,\right)+0.500 \text{Tr}\left(\,\mathcal{D}\,\bar{\Psi }_{   
   }\, \bar{Y}\, Z\, \Psi  \, Z\,\right)+0.894 \text{Tr}\left(\,\mathcal{D}\,\bar{\Psi } \, \bar{Y}\, Z\, Z\, \Psi _{   
   }\,\right)+1.45 \text{Tr}\left(\,\bar{\mathcal{F}} \, \mathcal{F} \, Z\, Z\, Z\,\right)+0.106 \text{Tr}\left(\,\bar{\mathcal{F}}_{ 
     }\, \Psi  \, \Psi  \, Z\, Z\,\right)+1.00 \text{Tr}\left(\,\bar{\mathcal{F}} \, \Psi  \, Z\, \Psi
    \, Z\,\right)+1.79 \text{Tr}\left(\,\bar{\mathcal{F}} \, \Psi  \, Z\, Z\, \Psi  \,\right)-0.106
   \text{Tr}\left(\,\bar{\mathcal{F}} \, \Psi  \, \Psi  \, Z\, Z\,\right)-1.00 \text{Tr}\left(\,\bar{\mathcal{F}}_{   
   }\, \Psi  \, Z\, \Psi  \, Z\,\right)-1.79 \text{Tr}\left(\,\bar{\mathcal{F}} \, \Psi  \, Z\, Z\, \Psi
    \,\right)+0.553 \text{Tr}\left(\,\bar{\mathcal{F}} \, Z\, \mathcal{F} \, Z\, Z\,\right)+1.00
   \text{Tr}\left(\,\bar{\mathcal{F}} \, Z\, \Psi  \, Z\, \Psi  \,\right)-1.00 \text{Tr}\left(\,\bar{\mathcal{F}}_{   
   }\, Z\, \Psi  \, Z\, \Psi  \,\right)+0.553 \text{Tr}\left(\,\bar{\mathcal{F}} \, Z\, Z\, \mathcal{F} \,
   Z\,\right)+0.106 \text{Tr}\left(\,\bar{\mathcal{F}} \, Z\, Z\, \Psi  \, \Psi  \,\right)-0.106
   \text{Tr}\left(\,\bar{\mathcal{F}} \, Z\, Z\, \Psi  \, \Psi  \,\right)+1.45 \text{Tr}\left(\,\bar{\mathcal{F}}_{   
   }\, Z\, Z\, Z\, \mathcal{F} \,\right)$
\endgroup

\begingroup
\tiny
\thinmuskip=-2mu
\medmuskip=-2mu
\thickmuskip=-18mu
\noindent
$\mathcal{O}_{+}\quad=\quad 0.553 \text{ Tr}\left(\mathcal{D}\,\Psi _{    }\, Z\, Z\, Z\, \bar{\Psi }_{    }\right)+1.45 \text{ Tr}\left(\mathcal{D}\,\Psi _{    }\, Z\, Z\,
   \bar{\Psi }_{    }\, Z\right)+1.45 \text{ Tr}\left(\mathcal{D}\,\Psi _{    }\, Z\, \bar{\Psi }_{    }\, Z\, Z\right)+0.553
   \text{ Tr}\left(\mathcal{D}\,\Psi _{    }\, \bar{\Psi }_{    }\, Z\, Z\, Z\right)-1.89 \text{ Tr}\left(\mathcal{D}\,Z\, \Psi _{    }\, Z\,
   Z\, \bar{\Psi }_{    }\right)-1.89 \text{ Tr}\left(\mathcal{D}\,Z\, \Psi _{    }\, \bar{\Psi }_{    }\, Z\, Z\right)+1.89
   \text{ Tr}\left(\mathcal{D}\,Z\, \Psi _{    }\, Z\, Z\, \bar{\Psi }_{    }\right)+1.89 \text{ Tr}\left(\mathcal{D}\,Z\, \Psi _{    }\,
   \bar{\Psi }_{    }\, Z\, Z\right)-1.00 \text{ Tr}\left(\mathcal{D}\,Z\, Z\, \Psi _{    }\, Z\, \bar{\Psi }_{    }\right)-1.00
   \text{ Tr}\left(\mathcal{D}\,Z\, Z\, \Psi _{    }\, \bar{\Psi }_{    }\, Z\right)+1.00 \text{ Tr}\left(\mathcal{D}\,Z\, Z\, \Psi _{    }\,
   Z\, \bar{\Psi }_{    }\right)+1.00 \text{ Tr}\left(\mathcal{D}\,Z\, Z\, \Psi _{    }\, \bar{\Psi }_{    }\, Z\right)+1.79
   \text{ Tr}\left(\mathcal{D}\,Z\, Z\, Z\, \Psi _{    }\, \bar{\Psi }_{    }\right)-1.79 \text{ Tr}\left(\mathcal{D}\,Z\, Z\, Z\, \Psi _{   
   }\, \bar{\Psi }_{    }\right)+0.894 \text{ Tr}\left(\mathcal{D}\,Z\, Z\, Z\, X\, \mathcal{D}\,\bar{X}\right)-0.894 \text{ Tr}\left(\mathcal{D}\,Z\, Z\, Z\, Y\,
   \mathcal{D}\,\bar{Y}\right)-0.894 \text{ Tr}\left(\mathcal{D}\,Z\, Z\, Z\, \bar{Y}\, \mathcal{D}\,Y\right)+0.894 \text{ Tr}\left(\mathcal{D}\,Z\, Z\, Z\, \bar{X}\,
   \mathcal{D}\,X\right)+3.58 \text{ Tr}\left(\mathcal{D}\,Z\, Z\, Z\, \mathcal{D}\,Z\, \bar{Z}\right)-0.947 \text{ Tr}\left(\mathcal{D}\,Z\, Z\, Z\, \mathcal{D}\,X\,
   \bar{X}\right)+0.947 \text{ Tr}\left(\mathcal{D}\,Z\, Z\, Z\, \mathcal{D}\,Y\, \bar{Y}\right)+0.947 \text{ Tr}\left(\mathcal{D}\,Z\, Z\, Z\, \mathcal{D}\,\bar{Y}\,
   Y\right)-0.947 \text{ Tr}\left(\mathcal{D}\,Z\, Z\, Z\, \mathcal{D}\,\bar{X}\, X\right)-1.89 \text{ Tr}\left(\mathcal{D}\,Z\, Z\, Z\, \bar{\Psi }_{    }\, \Psi
   _{    }\right)+1.89 \text{ Tr}\left(\mathcal{D}\,Z\, Z\, Z\, \bar{\Psi }_{    }\, \Psi _{    }\right)-0.500 \text{ Tr}\left(\mathcal{D}\,Z\,
   Z\, X\, Z\, \mathcal{D}\,\bar{X}\right)-0.842 \text{ Tr}\left(\mathcal{D}\,Z\, Z\, X\, \mathcal{D}\,Z\, \bar{X}\right)-0.500 \text{ Tr}\left(\mathcal{D}\,Z\, Z\, X\,
   \mathcal{D}\,\bar{X}\, Z\right)+0.500 \text{ Tr}\left(\mathcal{D}\,Z\, Z\, Y\, Z\, \mathcal{D}\,\bar{Y}\right)+0.842 \text{ Tr}\left(\mathcal{D}\,Z\, Z\, Y\,
   \mathcal{D}\,Z\, \bar{Y}\right)+0.500 \text{ Tr}\left(\mathcal{D}\,Z\, Z\, Y\, \mathcal{D}\,\bar{Y}\, Z\right)+0.500 \text{ Tr}\left(\mathcal{D}\,Z\, Z\, \bar{Y}\, Z\,
   \mathcal{D}\,Y\right)+0.842 \text{ Tr}\left(\mathcal{D}\,Z\, Z\, \bar{Y}\, \mathcal{D}\,Z\, Y\right)+0.500 \text{ Tr}\left(\mathcal{D}\,Z\, Z\, \bar{Y}\, \mathcal{D}\,Y\,
   Z\right)-0.500 \text{ Tr}\left(\mathcal{D}\,Z\, Z\, \bar{X}\, Z\, \mathcal{D}\,X\right)-0.842 \text{ Tr}\left(\mathcal{D}\,Z\, Z\, \bar{X}\, \mathcal{D}\,Z\,
   X\right)-0.500 \text{ Tr}\left(\mathcal{D}\,Z\, Z\, \bar{X}\, \mathcal{D}\,X\, Z\right)-2.00 \text{ Tr}\left(\mathcal{D}\,Z\, Z\, \mathcal{D}\,Z\, Z\,
   \bar{Z}\right)+3.34 \text{ Tr}\left(\mathcal{D}\,Z\, Z\, \mathcal{D}\,Z\, X\, \bar{X}\right)-3.34 \text{ Tr}\left(\mathcal{D}\,Z\, Z\, \mathcal{D}\,Z\, Y\,
   \bar{Y}\right)-3.34 \text{ Tr}\left(\mathcal{D}\,Z\, Z\, \mathcal{D}\,Z\, \bar{Y}\, Y\right)+3.34 \text{ Tr}\left(\mathcal{D}\,Z\, Z\, \mathcal{D}\,Z\, \bar{X}\,
   X\right)-2.00 \text{ Tr}\left(\mathcal{D}\,Z\, Z\, \mathcal{D}\,Z\, \bar{Z}\, Z\right)-0.500 \text{ Tr}\left(\mathcal{D}\,Z\, Z\, \mathcal{D}\,X\, \bar{X}\,
   Z\right)+0.500 \text{ Tr}\left(\mathcal{D}\,Z\, Z\, \mathcal{D}\,Y\, \bar{Y}\, Z\right)+0.500 \text{ Tr}\left(\mathcal{D}\,Z\, Z\, \mathcal{D}\,\bar{Y}\, Y\,
   Z\right)-0.500 \text{ Tr}\left(\mathcal{D}\,Z\, Z\, \mathcal{D}\,\bar{X}\, X\, Z\right)-1.00 \text{ Tr}\left(\mathcal{D}\,Z\, Z\, \bar{\Psi }_{    }\, \Psi
   _{    }\, Z\right)+1.00 \text{ Tr}\left(\mathcal{D}\,Z\, Z\, \bar{\Psi }_{    }\, \Psi _{    }\, Z\right)-0.947
   \text{ Tr}\left(\mathcal{D}\,Z\, X\, Z\, Z\, \mathcal{D}\,\bar{X}\right)-0.842 \text{ Tr}\left(\mathcal{D}\,Z\, X\, Z\, \mathcal{D}\,Z\, \bar{X}\right)-0.842
   \text{ Tr}\left(\mathcal{D}\,Z\, X\, \mathcal{D}\,Z\, \bar{X}\, Z\right)-0.947 \text{ Tr}\left(\mathcal{D}\,Z\, X\, \mathcal{D}\,\bar{X}\, Z\, Z\right)+0.947
   \text{ Tr}\left(\mathcal{D}\,Z\, Y\, Z\, Z\, \mathcal{D}\,\bar{Y}\right)+0.842 \text{ Tr}\left(\mathcal{D}\,Z\, Y\, Z\, \mathcal{D}\,Z\, \bar{Y}\right)+0.842
   \text{ Tr}\left(\mathcal{D}\,Z\, Y\, \mathcal{D}\,Z\, \bar{Y}\, Z\right)+0.947 \text{ Tr}\left(\mathcal{D}\,Z\, Y\, \mathcal{D}\,\bar{Y}\, Z\, Z\right)+0.947
   \text{ Tr}\left(\mathcal{D}\,Z\, \bar{Y}\, Z\, Z\, \mathcal{D}\,Y\right)+0.947 \text{ Tr}\left(\mathcal{D}\,Z\, \bar{Y}\, \mathcal{D}\,Y\, Z\, Z\right)-0.947
   \text{ Tr}\left(\mathcal{D}\,Z\, \bar{X}\, Z\, Z\, \mathcal{D}\,X\right)-0.947 \text{ Tr}\left(\mathcal{D}\,Z\, \bar{X}\, \mathcal{D}\,X\, Z\, Z\right)-3.79
   \text{ Tr}\left(\mathcal{D}\,Z\, \mathcal{D}\,Z\, Z\, Z\, \bar{Z}\right)+0.500 \text{ Tr}\left(\mathcal{D}\,Z\, \mathcal{D}\,Z\, Z\, X\, \bar{X}\right)-0.500
   \text{ Tr}\left(\mathcal{D}\,Z\, \mathcal{D}\,Z\, Z\, Y\, \bar{Y}\right)-0.500 \text{ Tr}\left(\mathcal{D}\,Z\, \mathcal{D}\,Z\, Z\, \bar{Y}\, Y\right)+0.500
   \text{ Tr}\left(\mathcal{D}\,Z\, \mathcal{D}\,Z\, Z\, \bar{X}\, X\right)+3.34 \text{ Tr}\left(\mathcal{D}\,Z\, \mathcal{D}\,Z\, X\, Z\, \bar{X}\right)+0.500
   \text{ Tr}\left(\mathcal{D}\,Z\, \mathcal{D}\,Z\, X\, \bar{X}\, Z\right)-3.34 \text{ Tr}\left(\mathcal{D}\,Z\, \mathcal{D}\,Z\, Y\, Z\, \bar{Y}\right)-0.500
   \text{ Tr}\left(\mathcal{D}\,Z\, \mathcal{D}\,Z\, Y\, \bar{Y}\, Z\right)-3.34 \text{ Tr}\left(\mathcal{D}\,Z\, \mathcal{D}\,Z\, \bar{Y}\, Z\, Y\right)-0.500
   \text{ Tr}\left(\mathcal{D}\,Z\, \mathcal{D}\,Z\, \bar{Y}\, Y\, Z\right)+3.34 \text{ Tr}\left(\mathcal{D}\,Z\, \mathcal{D}\,Z\, \bar{X}\, Z\, X\right)+0.500
   \text{ Tr}\left(\mathcal{D}\,Z\, \mathcal{D}\,Z\, \bar{X}\, X\, Z\right)-3.79 \text{ Tr}\left(\mathcal{D}\,Z\, \mathcal{D}\,Z\, \bar{Z}\, Z\, Z\right)-0.947
   \text{ Tr}\left(\mathcal{D}\,Z\, \mathcal{D}\,X\, Z\, Z\, \bar{X}\right)-0.500 \text{ Tr}\left(\mathcal{D}\,Z\, \mathcal{D}\,X\, Z\, \bar{X}\, Z\right)+0.894
   \text{ Tr}\left(\mathcal{D}\,Z\, \mathcal{D}\,X\, \bar{X}\, Z\, Z\right)+0.947 \text{ Tr}\left(\mathcal{D}\,Z\, \mathcal{D}\,Y\, Z\, Z\, \bar{Y}\right)+0.500
   \text{ Tr}\left(\mathcal{D}\,Z\, \mathcal{D}\,Y\, Z\, \bar{Y}\, Z\right)-0.894 \text{ Tr}\left(\mathcal{D}\,Z\, \mathcal{D}\,Y\, \bar{Y}\, Z\, Z\right)+0.947
   \text{ Tr}\left(\mathcal{D}\,Z\, \mathcal{D}\,\bar{Y}\, Z\, Z\, Y\right)+0.500 \text{ Tr}\left(\mathcal{D}\,Z\, \mathcal{D}\,\bar{Y}\, Z\, Y\, Z\right)-0.894
   \text{ Tr}\left(\mathcal{D}\,Z\, \mathcal{D}\,\bar{Y}\, Y\, Z\, Z\right)-0.947 \text{ Tr}\left(\mathcal{D}\,Z\, \mathcal{D}\,\bar{X}\, Z\, Z\, X\right)-0.500
   \text{ Tr}\left(\mathcal{D}\,Z\, \mathcal{D}\,\bar{X}\, Z\, X\, Z\right)+0.894 \text{ Tr}\left(\mathcal{D}\,Z\, \mathcal{D}\,\bar{X}\, X\, Z\, Z\right)+1.79
   \text{ Tr}\left(\mathcal{D}\,Z\, \bar{\Psi }_{    }\, \Psi _{    }\, Z\, Z\right)-1.00 \text{ Tr}\left(\mathcal{D}\,Z\, \bar{\Psi }_{    }\,
   Z\, \Psi _{    }\, Z\right)-1.89 \text{ Tr}\left(\mathcal{D}\,Z\, \bar{\Psi }_{    }\, Z\, Z\, \Psi _{    }\right)-1.79
   \text{ Tr}\left(\mathcal{D}\,Z\, \bar{\Psi }_{    }\, \Psi _{    }\, Z\, Z\right)+1.00 \text{ Tr}\left(\mathcal{D}\,Z\, \bar{\Psi }_{    }\,
   Z\, \Psi _{    }\, Z\right)+1.89 \text{ Tr}\left(\mathcal{D}\,Z\, \bar{\Psi }_{    }\, Z\, Z\, \Psi _{    }\right)+0.553
   \text{ Tr}\left(\mathcal{D}\,X\, Z\, Z\, Z\, \mathcal{D}\,\bar{X}\right)+1.45 \text{ Tr}\left(\mathcal{D}\,X\, Z\, Z\, \mathcal{D}\,\bar{X}\, Z\right)+1.45
   \text{ Tr}\left(\mathcal{D}\,X\, Z\, \mathcal{D}\,\bar{X}\, Z\, Z\right)+0.553 \text{ Tr}\left(\mathcal{D}\,X\, \mathcal{D}\,\bar{X}\, Z\, Z\, Z\right)-0.553
   \text{ Tr}\left(\mathcal{D}\,Y\, Z\, Z\, Z\, \mathcal{D}\,\bar{Y}\right)-1.45 \text{ Tr}\left(\mathcal{D}\,Y\, Z\, Z\, \mathcal{D}\,\bar{Y}\, Z\right)-1.45
   \text{ Tr}\left(\mathcal{D}\,Y\, Z\, \mathcal{D}\,\bar{Y}\, Z\, Z\right)-0.553 \text{ Tr}\left(\mathcal{D}\,Y\, \mathcal{D}\,\bar{Y}\, Z\, Z\, Z\right)-1.89
   \text{ Tr}\left(\bar{\Psi }_{    }\, \mathcal{F}_{    }\, Z\, Z\, \bar{\Psi }_{    }\right)-1.00 \text{ Tr}\left(\bar{\Psi }_{   
   }\, \mathcal{F}_{    }\, Z\, \bar{\Psi }_{    }\, Z\right)+1.79 \text{ Tr}\left(\bar{\Psi }_{    }\, \mathcal{F}_{    }\,
   \bar{\Psi }_{    }\, Z\, Z\right)-0.500 \text{ Tr}\left(\bar{\Psi }_{    }\, \Psi _{    }\, \Psi _{    }\, Z\, \bar{\Psi
   }_{    }\right)-3.34 \text{ Tr}\left(\bar{\Psi }_{    }\, \Psi _{    }\, \Psi _{    }\, \bar{\Psi }_{    }\,
   Z\right)-3.34 \text{ Tr}\left(\bar{\Psi }_{    }\, \Psi _{    }\, Z\, \Psi _{    }\, \bar{\Psi }_{    }\right)+0.947
   \text{ Tr}\left(\bar{\Psi }_{    }\, \Psi _{    }\, Z\, Z\, \mathcal{D}\,\bar{X}\right)+1.00 \text{ Tr}\left(\bar{\Psi }_{    }\, \Psi
   _{    }\, Z\, \bar{X}\, \mathcal{D}\,Z\right)+1.00 \text{ Tr}\left(\bar{\Psi }_{    }\, \Psi _{    }\, Z\, \mathcal{D}\,Z\,
   \bar{X}\right)-0.842 \text{ Tr}\left(\bar{\Psi }_{    }\, \Psi _{    }\, Z\, \bar{\Psi }_{    }\, \Psi _{    }\right)-1.84
   \text{ Tr}\left(\bar{\Psi }_{    }\, \Psi _{    }\, \bar{X}\, Z\, \mathcal{D}\,Z\right)+1.00 \text{ Tr}\left(\bar{\Psi }_{    }\, \Psi
   _{    }\, \bar{X}\, \mathcal{D}\,Z\, Z\right)-1.84 \text{ Tr}\left(\bar{\Psi }_{    }\, \Psi _{    }\, \mathcal{D}\,Z\, Z\,
   \bar{X}\right)+1.00 \text{ Tr}\left(\bar{\Psi }_{    }\, \Psi _{    }\, \mathcal{D}\,Z\, \bar{X}\, Z\right)+0.947 \text{ Tr}\left(\bar{\Psi
   }_{    }\, \Psi _{    }\, \mathcal{D}\,\bar{X}\, Z\, Z\right)-0.842 \text{ Tr}\left(\bar{\Psi }_{    }\, \Psi _{    }\, \bar{\Psi
   }_{    }\, \Psi _{    }\, Z\right)-0.842 \text{ Tr}\left(\bar{\Psi }_{    }\, \Psi _{    }\, \bar{\Psi }_{    }\, Z\,
   \Psi _{    }\right)+0.500 \text{ Tr}\left(\bar{\Psi }_{    }\, \Psi _{    }\, \Psi _{    }\, Z\, \bar{\Psi }_{   
   }\right)+3.34 \text{ Tr}\left(\bar{\Psi }_{    }\, \Psi _{    }\, \Psi _{    }\, \bar{\Psi }_{    }\, Z\right)+3.34
   \text{ Tr}\left(\bar{\Psi }_{    }\, \Psi _{    }\, Z\, \Psi _{    }\, \bar{\Psi }_{    }\right)-0.947 \text{ Tr}\left(\bar{\Psi
   }_{    }\, \Psi _{    }\, Z\, Z\, \mathcal{D}\,Y\right)-1.00 \text{ Tr}\left(\bar{\Psi }_{    }\, \Psi _{    }\, Z\, Y\,
   \mathcal{D}\,Z\right)-1.00 \text{ Tr}\left(\bar{\Psi }_{    }\, \Psi _{    }\, Z\, \mathcal{D}\,Z\, Y\right)+0.842 \text{ Tr}\left(\bar{\Psi }_{ 
     }\, \Psi _{    }\, Z\, \bar{\Psi }_{    }\, \Psi _{    }\right)+1.84 \text{ Tr}\left(\bar{\Psi }_{    }\, \Psi _{ 
     }\, Y\, Z\, \mathcal{D}\,Z\right)-1.00 \text{ Tr}\left(\bar{\Psi }_{    }\, \Psi _{    }\, Y\, \mathcal{D}\,Z\, Z\right)+1.84
   \text{ Tr}\left(\bar{\Psi }_{    }\, \Psi _{    }\, \mathcal{D}\,Z\, Z\, Y\right)-1.00 \text{ Tr}\left(\bar{\Psi }_{    }\, \Psi _{ 
     }\, \mathcal{D}\,Z\, Y\, Z\right)-0.947 \text{ Tr}\left(\bar{\Psi }_{    }\, \Psi _{    }\, \mathcal{D}\,Y\, Z\, Z\right)+0.842
   \text{ Tr}\left(\bar{\Psi }_{    }\, \Psi _{    }\, \bar{\Psi }_{    }\, \Psi _{    }\, Z\right)+0.842 \text{ Tr}\left(\bar{\Psi
   }_{    }\, \Psi _{    }\, \bar{\Psi }_{    }\, Z\, \Psi _{    }\right)+1.89 \text{ Tr}\left(\bar{\Psi }_{    }\, \Psi
   _{    }\, Z\, Z\, \mathcal{D}\,Z\right)+1.00 \text{ Tr}\left(\bar{\Psi }_{    }\, \Psi _{    }\, Z\, \mathcal{D}\,Z\, Z\right)-1.79
   \text{ Tr}\left(\bar{\Psi }_{    }\, \Psi _{    }\, \mathcal{D}\,Z\, Z\, Z\right)-1.00 \text{ Tr}\left(\bar{\Psi }_{    }\, Z\,
   \mathcal{F}_{    }\, \bar{\Psi }_{    }\, Z\right)-0.500 \text{ Tr}\left(\bar{\Psi }_{    }\, Z\, \Psi _{    }\, \Psi _{ 
     }\, \bar{\Psi }_{    }\right)+0.500 \text{ Tr}\left(\bar{\Psi }_{    }\, Z\, \Psi _{    }\, Z\, \mathcal{D}\,\bar{X}\right)-1.84
   \text{ Tr}\left(\bar{\Psi }_{    }\, Z\, \Psi _{    }\, \bar{X}\, \mathcal{D}\,Z\right)-1.84 \text{ Tr}\left(\bar{\Psi }_{    }\, Z\, \Psi
   _{    }\, \mathcal{D}\,Z\, \bar{X}\right)+0.500 \text{ Tr}\left(\bar{\Psi }_{    }\, Z\, \Psi _{    }\, \mathcal{D}\,\bar{X}\,
   Z\right)-0.842 \text{ Tr}\left(\bar{\Psi }_{    }\, Z\, \Psi _{    }\, \bar{\Psi }_{    }\, \Psi _{    }\right)+0.500
   \text{ Tr}\left(\bar{\Psi }_{    }\, Z\, \Psi _{    }\, \Psi _{    }\, \bar{\Psi }_{    }\right)-0.500 \text{ Tr}\left(\bar{\Psi
   }_{    }\, Z\, \Psi _{    }\, Z\, \mathcal{D}\,Y\right)+1.84 \text{ Tr}\left(\bar{\Psi }_{    }\, Z\, \Psi _{    }\, Y\,
   \mathcal{D}\,Z\right)+1.84 \text{ Tr}\left(\bar{\Psi }_{    }\, Z\, \Psi _{    }\, \mathcal{D}\,Z\, Y\right)-0.500 \text{ Tr}\left(\bar{\Psi }_{ 
     }\, Z\, \Psi _{    }\, \mathcal{D}\,Y\, Z\right)+0.842 \text{ Tr}\left(\bar{\Psi }_{    }\, Z\, \Psi _{    }\, \bar{\Psi }_{ 
     }\, \Psi _{    }\right)+1.00 \text{ Tr}\left(\bar{\Psi }_{    }\, Z\, \Psi _{    }\, \mathcal{D}\,Z\, Z\right)-1.89
   \text{ Tr}\left(\bar{\Psi }_{    }\, Z\, Z\, \mathcal{F}_{    }\, \bar{\Psi }_{    }\right)-0.894 \text{ Tr}\left(\bar{\Psi }_{ 
     }\, Z\, Z\, \Psi _{    }\, \mathcal{D}\,\bar{X}\right)+0.894 \text{ Tr}\left(\bar{\Psi }_{    }\, Z\, Z\, \Psi _{    }\,
   \mathcal{D}\,Y\right)+1.89 \text{ Tr}\left(\bar{\Psi }_{    }\, Z\, Z\, \Psi _{    }\, \mathcal{D}\,Z\right)-0.894 \text{ Tr}\left(\bar{\Psi }_{ 
     }\, Z\, Z\, Y\, \mathcal{D}\,\Psi _{    }\right)+0.894 \text{ Tr}\left(\bar{\Psi }_{    }\, Z\, Z\, \bar{X}\, \mathcal{D}\,\Psi _{   
   }\right)-0.947 \text{ Tr}\left(\bar{\Psi }_{    }\, Z\, Z\, \mathcal{D}\,\Psi _{    }\, \bar{X}\right)+0.947 \text{ Tr}\left(\bar{\Psi }_{ 
     }\, Z\, Z\, \mathcal{D}\,\Psi _{    }\, Y\right)-1.79 \text{ Tr}\left(\bar{\Psi }_{    }\, Z\, Z\, \mathcal{D}\,Z\, \Psi _{   
   }\right)-0.947 \text{ Tr}\left(\bar{\Psi }_{    }\, Z\, Z\, \mathcal{D}\,Y\, \Psi _{    }\right)+0.947 \text{ Tr}\left(\bar{\Psi }_{    }\,
   Z\, Z\, \mathcal{D}\,\bar{X}\, \Psi _{    }\right)+1.79 \text{ Tr}\left(\bar{\Psi }_{    }\, Z\, Z\, \bar{\Psi }_{    }\,
   \mathcal{F}_{    }\right)-2.34 \text{ Tr}\left(\bar{\Psi }_{    }\, Z\, Y\, \Psi _{    }\, \mathcal{D}\,Z\right)+0.500
   \text{ Tr}\left(\bar{\Psi }_{    }\, Z\, Y\, Z\, \mathcal{D}\,\Psi _{    }\right)+0.500 \text{ Tr}\left(\bar{\Psi }_{    }\, Z\, Y\,
   \mathcal{D}\,\Psi _{    }\, Z\right)-1.00 \text{ Tr}\left(\bar{\Psi }_{    }\, Z\, Y\, \mathcal{D}\,Z\, \Psi _{    }\right)+2.34
   \text{ Tr}\left(\bar{\Psi }_{    }\, Z\, \bar{X}\, \Psi _{    }\, \mathcal{D}\,Z\right)-0.500 \text{ Tr}\left(\bar{\Psi }_{    }\, Z\,
   \bar{X}\, Z\, \mathcal{D}\,\Psi _{    }\right)-0.500 \text{ Tr}\left(\bar{\Psi }_{    }\, Z\, \bar{X}\, \mathcal{D}\,\Psi _{    }\,
   Z\right)+1.00 \text{ Tr}\left(\bar{\Psi }_{    }\, Z\, \bar{X}\, \mathcal{D}\,Z\, \Psi _{    }\right)-0.500 \text{ Tr}\left(\bar{\Psi }_{ 
     }\, Z\, \mathcal{D}\,\Psi _{    }\, \bar{X}\, Z\right)+0.500 \text{ Tr}\left(\bar{\Psi }_{    }\, Z\, \mathcal{D}\,\Psi _{    }\, Y\,
   Z\right)+2.34 \text{ Tr}\left(\bar{\Psi }_{    }\, Z\, \mathcal{D}\,Z\, \Psi _{    }\, \bar{X}\right)-2.34 \text{ Tr}\left(\bar{\Psi }_{   
   }\, Z\, \mathcal{D}\,Z\, \Psi _{    }\, Y\right)+1.00 \text{ Tr}\left(\bar{\Psi }_{    }\, Z\, \mathcal{D}\,Z\, \Psi _{    }\, Z\right)+1.00
   \text{ Tr}\left(\bar{\Psi }_{    }\, Z\, \mathcal{D}\,Z\, Z\, \Psi _{    }\right)-1.00 \text{ Tr}\left(\bar{\Psi }_{    }\, Z\,
   \mathcal{D}\,Z\, Y\, \Psi _{    }\right)+1.00 \text{ Tr}\left(\bar{\Psi }_{    }\, Z\, \mathcal{D}\,Z\, \bar{X}\, \Psi _{    }\right)-0.500
   \text{ Tr}\left(\bar{\Psi }_{    }\, Z\, \mathcal{D}\,Y\, \Psi _{    }\, Z\right)+0.500 \text{ Tr}\left(\bar{\Psi }_{    }\, Z\,
   \mathcal{D}\,\bar{X}\, \Psi _{    }\, Z\right)-1.00 \text{ Tr}\left(\bar{\Psi }_{    }\, Z\, \bar{\Psi }_{    }\, \mathcal{F}_{   
   }\, Z\right)-3.34 \text{ Tr}\left(\bar{\Psi }_{    }\, Z\, \bar{\Psi }_{    }\, \Psi _{    }\, \Psi _{    }\right)+3.34
   \text{ Tr}\left(\bar{\Psi }_{    }\, Z\, \bar{\Psi }_{    }\, \Psi _{    }\, \Psi _{    }\right)-1.00 \text{ Tr}\left(\bar{\Psi
   }_{    }\, Z\, \bar{\Psi }_{    }\, Z\, \mathcal{F}_{    }\right)-2.34 \text{ Tr}\left(\bar{\Psi }_{    }\, Y\, \Psi _{ 
     }\, Z\, \mathcal{D}\,Z\right)-2.34 \text{ Tr}\left(\bar{\Psi }_{    }\, Y\, \Psi _{    }\, \mathcal{D}\,Z\, Z\right)-2.34
   \text{ Tr}\left(\bar{\Psi }_{    }\, Y\, Z\, \Psi _{    }\, \mathcal{D}\,Z\right)+0.947 \text{ Tr}\left(\bar{\Psi }_{    }\, Y\, Z\, Z\,
   \mathcal{D}\,\Psi _{    }\right)+1.84 \text{ Tr}\left(\bar{\Psi }_{    }\, Y\, Z\, \mathcal{D}\,Z\, \Psi _{    }\right)+0.947
   \text{ Tr}\left(\bar{\Psi }_{    }\, Y\, \mathcal{D}\,\Psi _{    }\, Z\, Z\right)+1.84 \text{ Tr}\left(\bar{\Psi }_{    }\, Y\,
   \mathcal{D}\,Z\, \Psi _{    }\, Z\right)-1.00 \text{ Tr}\left(\bar{\Psi }_{    }\, Y\, \mathcal{D}\,Z\, Z\, \Psi _{    }\right)+2.34
   \text{ Tr}\left(\bar{\Psi }_{    }\, \bar{X}\, \Psi _{    }\, Z\, \mathcal{D}\,Z\right)+2.34 \text{ Tr}\left(\bar{\Psi }_{    }\, \bar{X}\,
   \Psi _{    }\, \mathcal{D}\,Z\, Z\right)+2.34 \text{ Tr}\left(\bar{\Psi }_{    }\, \bar{X}\, Z\, \Psi _{    }\, \mathcal{D}\,Z\right)-0.947
   \text{ Tr}\left(\bar{\Psi }_{    }\, \bar{X}\, Z\, Z\, \mathcal{D}\,\Psi _{    }\right)-1.84 \text{ Tr}\left(\bar{\Psi }_{    }\, \bar{X}\,
   Z\, \mathcal{D}\,Z\, \Psi _{    }\right)-0.947 \text{ Tr}\left(\bar{\Psi }_{    }\, \bar{X}\, \mathcal{D}\,\Psi _{    }\, Z\, Z\right)-1.84
   \text{ Tr}\left(\bar{\Psi }_{    }\, \bar{X}\, \mathcal{D}\,Z\, \Psi _{    }\, Z\right)+1.00 \text{ Tr}\left(\bar{\Psi }_{    }\, \bar{X}\,
   \mathcal{D}\,Z\, Z\, \Psi _{    }\right)-0.947 \text{ Tr}\left(\bar{\Psi }_{    }\, \mathcal{D}\,\Psi _{    }\, Z\, Z\, \bar{X}\right)-0.500
   \text{ Tr}\left(\bar{\Psi }_{    }\, \mathcal{D}\,\Psi _{    }\, Z\, \bar{X}\, Z\right)+0.894 \text{ Tr}\left(\bar{\Psi }_{    }\,
   \mathcal{D}\,\Psi _{    }\, \bar{X}\, Z\, Z\right)+0.947 \text{ Tr}\left(\bar{\Psi }_{    }\, \mathcal{D}\,\Psi _{    }\, Z\, Z\,
   Y\right)+0.500 \text{ Tr}\left(\bar{\Psi }_{    }\, \mathcal{D}\,\Psi _{    }\, Z\, Y\, Z\right)-0.894 \text{ Tr}\left(\bar{\Psi }_{    }\,
   \mathcal{D}\,\Psi _{    }\, Y\, Z\, Z\right)+2.34 \text{ Tr}\left(\bar{\Psi }_{    }\, \mathcal{D}\,Z\, \Psi _{    }\, Z\,
   \bar{X}\right)+2.34 \text{ Tr}\left(\bar{\Psi }_{    }\, \mathcal{D}\,Z\, \Psi _{    }\, \bar{X}\, Z\right)-2.34 \text{ Tr}\left(\bar{\Psi }_{ 
     }\, \mathcal{D}\,Z\, \Psi _{    }\, Z\, Y\right)-2.34 \text{ Tr}\left(\bar{\Psi }_{    }\, \mathcal{D}\,Z\, \Psi _{    }\, Y\,
   Z\right)+1.89 \text{ Tr}\left(\bar{\Psi }_{    }\, \mathcal{D}\,Z\, \Psi _{    }\, Z\, Z\right)+2.34 \text{ Tr}\left(\bar{\Psi }_{    }\,
   \mathcal{D}\,Z\, Z\, \Psi _{    }\, \bar{X}\right)-2.34 \text{ Tr}\left(\bar{\Psi }_{    }\, \mathcal{D}\,Z\, Z\, \Psi _{    }\,
   Y\right)+1.89 \text{ Tr}\left(\bar{\Psi }_{    }\, \mathcal{D}\,Z\, Z\, Z\, \Psi _{    }\right)+1.84 \text{ Tr}\left(\bar{\Psi }_{    }\,
   \mathcal{D}\,Z\, Z\, Y\, \Psi _{    }\right)-1.84 \text{ Tr}\left(\bar{\Psi }_{    }\, \mathcal{D}\,Z\, Z\, \bar{X}\, \Psi _{   
   }\right)+1.84 \text{ Tr}\left(\bar{\Psi }_{    }\, \mathcal{D}\,Z\, Y\, \Psi _{    }\, Z\right)-1.00 \text{ Tr}\left(\bar{\Psi }_{    }\,
   \mathcal{D}\,Z\, Y\, Z\, \Psi _{    }\right)-1.84 \text{ Tr}\left(\bar{\Psi }_{    }\, \mathcal{D}\,Z\, \bar{X}\, \Psi _{    }\,
   Z\right)+1.00 \text{ Tr}\left(\bar{\Psi }_{    }\, \mathcal{D}\,Z\, \bar{X}\, Z\, \Psi _{    }\right)+0.894 \text{ Tr}\left(\bar{\Psi }_{ 
     }\, \mathcal{D}\,Y\, \Psi _{    }\, Z\, Z\right)-0.500 \text{ Tr}\left(\bar{\Psi }_{    }\, \mathcal{D}\,Y\, Z\, \Psi _{    }\,
   Z\right)-0.947 \text{ Tr}\left(\bar{\Psi }_{    }\, \mathcal{D}\,Y\, Z\, Z\, \Psi _{    }\right)-0.894 \text{ Tr}\left(\bar{\Psi }_{    }\,
   \mathcal{D}\,\bar{X}\, \Psi _{    }\, Z\, Z\right)+0.500 \text{ Tr}\left(\bar{\Psi }_{    }\, \mathcal{D}\,\bar{X}\, Z\, \Psi _{    }\,
   Z\right)+0.947 \text{ Tr}\left(\bar{\Psi }_{    }\, \mathcal{D}\,\bar{X}\, Z\, Z\, \Psi _{    }\right)-1.89 \text{ Tr}\left(\bar{\Psi }_{ 
     }\, \bar{\Psi }_{    }\, \mathcal{F}_{    }\, Z\, Z\right)-0.500 \text{ Tr}\left(\bar{\Psi }_{    }\, \bar{\Psi }_{    }\,
   \Psi _{    }\, \Psi _{    }\, Z\right)-3.34 \text{ Tr}\left(\bar{\Psi }_{    }\, \bar{\Psi }_{    }\, \Psi _{    }\, Z\,
   \Psi _{    }\right)+0.500 \text{ Tr}\left(\bar{\Psi }_{    }\, \bar{\Psi }_{    }\, \Psi _{    }\, \Psi _{    }\,
   Z\right)+3.34 \text{ Tr}\left(\bar{\Psi }_{    }\, \bar{\Psi }_{    }\, \Psi _{    }\, Z\, \Psi _{    }\right)-0.500
   \text{ Tr}\left(\bar{\Psi }_{    }\, \bar{\Psi }_{    }\, Z\, \Psi _{    }\, \Psi _{    }\right)+0.500 \text{ Tr}\left(\bar{\Psi
   }_{    }\, \bar{\Psi }_{    }\, Z\, \Psi _{    }\, \Psi _{    }\right)-1.89 \text{ Tr}\left(\bar{\Psi }_{    }\,
   \bar{\Psi }_{    }\, Z\, Z\, \mathcal{F}_{    }\right)-0.947 \text{ Tr}\left(\bar{\Psi }_{    }\, \Psi _{    }\, Z\, Z\,
   \mathcal{D}\,\bar{Y}\right)-1.00 \text{ Tr}\left(\bar{\Psi }_{    }\, \Psi _{    }\, Z\, \bar{Y}\, \mathcal{D}\,Z\right)-1.00
   \text{ Tr}\left(\bar{\Psi }_{    }\, \Psi _{    }\, Z\, \mathcal{D}\,Z\, \bar{Y}\right)+1.84 \text{ Tr}\left(\bar{\Psi }_{    }\, \Psi
   _{    }\, \bar{Y}\, Z\, \mathcal{D}\,Z\right)-1.00 \text{ Tr}\left(\bar{\Psi }_{    }\, \Psi _{    }\, \bar{Y}\, \mathcal{D}\,Z\,
   Z\right)+1.84 \text{ Tr}\left(\bar{\Psi }_{    }\, \Psi _{    }\, \mathcal{D}\,Z\, Z\, \bar{Y}\right)-1.00 \text{ Tr}\left(\bar{\Psi }_{   
   }\, \Psi _{    }\, \mathcal{D}\,Z\, \bar{Y}\, Z\right)-0.947 \text{ Tr}\left(\bar{\Psi }_{    }\, \Psi _{    }\, \mathcal{D}\,\bar{Y}\, Z\,
   Z\right)+0.947 \text{ Tr}\left(\bar{\Psi }_{    }\, \Psi _{    }\, Z\, Z\, \mathcal{D}\,X\right)+1.00 \text{ Tr}\left(\bar{\Psi }_{    }\,
   \Psi _{    }\, Z\, X\, \mathcal{D}\,Z\right)+1.00 \text{ Tr}\left(\bar{\Psi }_{    }\, \Psi _{    }\, Z\, \mathcal{D}\,Z\, X\right)-1.84
   \text{ Tr}\left(\bar{\Psi }_{    }\, \Psi _{    }\, X\, Z\, \mathcal{D}\,Z\right)+1.00 \text{ Tr}\left(\bar{\Psi }_{    }\, \Psi _{ 
     }\, X\, \mathcal{D}\,Z\, Z\right)-1.84 \text{ Tr}\left(\bar{\Psi }_{    }\, \Psi _{    }\, \mathcal{D}\,Z\, Z\, X\right)+1.00
   \text{ Tr}\left(\bar{\Psi }_{    }\, \Psi _{    }\, \mathcal{D}\,Z\, X\, Z\right)+0.947 \text{ Tr}\left(\bar{\Psi }_{    }\, \Psi _{ 
     }\, \mathcal{D}\,X\, Z\, Z\right)-1.89 \text{ Tr}\left(\bar{\Psi }_{    }\, \Psi _{    }\, Z\, Z\, \mathcal{D}\,Z\right)-1.00
   \text{ Tr}\left(\bar{\Psi }_{    }\, \Psi _{    }\, Z\, \mathcal{D}\,Z\, Z\right)+1.79 \text{ Tr}\left(\bar{\Psi }_{    }\, \Psi _{ 
     }\, \mathcal{D}\,Z\, Z\, Z\right)-0.500 \text{ Tr}\left(\bar{\Psi }_{    }\, Z\, \Psi _{    }\, Z\, \mathcal{D}\,\bar{Y}\right)+1.84
   \text{ Tr}\left(\bar{\Psi }_{    }\, Z\, \Psi _{    }\, \bar{Y}\, \mathcal{D}\,Z\right)+1.84 \text{ Tr}\left(\bar{\Psi }_{    }\, Z\, \Psi
   _{    }\, \mathcal{D}\,Z\, \bar{Y}\right)-0.500 \text{ Tr}\left(\bar{\Psi }_{    }\, Z\, \Psi _{    }\, \mathcal{D}\,\bar{Y}\,
   Z\right)+0.500 \text{ Tr}\left(\bar{\Psi }_{    }\, Z\, \Psi _{    }\, Z\, \mathcal{D}\,X\right)-1.84 \text{ Tr}\left(\bar{\Psi }_{    }\,
   Z\, \Psi _{    }\, X\, \mathcal{D}\,Z\right)-1.84 \text{ Tr}\left(\bar{\Psi }_{    }\, Z\, \Psi _{    }\, \mathcal{D}\,Z\, X\right)+0.500
   \text{ Tr}\left(\bar{\Psi }_{    }\, Z\, \Psi _{    }\, \mathcal{D}\,X\, Z\right)-1.00 \text{ Tr}\left(\bar{\Psi }_{    }\, Z\, \Psi
   _{    }\, \mathcal{D}\,Z\, Z\right)+0.894 \text{ Tr}\left(\bar{\Psi }_{    }\, Z\, Z\, \Psi _{    }\, \mathcal{D}\,\bar{Y}\right)-0.894
   \text{ Tr}\left(\bar{\Psi }_{    }\, Z\, Z\, \Psi _{    }\, \mathcal{D}\,X\right)-1.89 \text{ Tr}\left(\bar{\Psi }_{    }\, Z\, Z\, \Psi
   _{    }\, \mathcal{D}\,Z\right)+0.894 \text{ Tr}\left(\bar{\Psi }_{    }\, Z\, Z\, X\, \mathcal{D}\,\Psi _{    }\right)-0.894
   \text{ Tr}\left(\bar{\Psi }_{    }\, Z\, Z\, \bar{Y}\, \mathcal{D}\,\Psi _{    }\right)+0.947 \text{ Tr}\left(\bar{\Psi }_{    }\, Z\, Z\,
   \mathcal{D}\,\Psi _{    }\, \bar{Y}\right)-0.947 \text{ Tr}\left(\bar{\Psi }_{    }\, Z\, Z\, \mathcal{D}\,\Psi _{    }\, X\right)+1.79
   \text{ Tr}\left(\bar{\Psi }_{    }\, Z\, Z\, \mathcal{D}\,Z\, \Psi _{    }\right)+0.947 \text{ Tr}\left(\bar{\Psi }_{    }\, Z\, Z\,
   \mathcal{D}\,X\, \Psi _{    }\right)-0.947 \text{ Tr}\left(\bar{\Psi }_{    }\, Z\, Z\, \mathcal{D}\,\bar{Y}\, \Psi _{    }\right)+2.34
   \text{ Tr}\left(\bar{\Psi }_{    }\, Z\, X\, \Psi _{    }\, \mathcal{D}\,Z\right)-0.500 \text{ Tr}\left(\bar{\Psi }_{    }\, Z\, X\, Z\,
   \mathcal{D}\,\Psi _{    }\right)-0.500 \text{ Tr}\left(\bar{\Psi }_{    }\, Z\, X\, \mathcal{D}\,\Psi _{    }\, Z\right)+1.00
   \text{ Tr}\left(\bar{\Psi }_{    }\, Z\, X\, \mathcal{D}\,Z\, \Psi _{    }\right)-2.34 \text{ Tr}\left(\bar{\Psi }_{    }\, Z\, \bar{Y}\,
   \Psi _{    }\, \mathcal{D}\,Z\right)+0.500 \text{ Tr}\left(\bar{\Psi }_{    }\, Z\, \bar{Y}\, Z\, \mathcal{D}\,\Psi _{    }\right)+0.500
   \text{ Tr}\left(\bar{\Psi }_{    }\, Z\, \bar{Y}\, \mathcal{D}\,\Psi _{    }\, Z\right)-1.00 \text{ Tr}\left(\bar{\Psi }_{    }\, Z\,
   \bar{Y}\, \mathcal{D}\,Z\, \Psi _{    }\right)+0.500 \text{ Tr}\left(\bar{\Psi }_{    }\, Z\, \mathcal{D}\,\Psi _{    }\, \bar{Y}\,
   Z\right)-0.500 \text{ Tr}\left(\bar{\Psi }_{    }\, Z\, \mathcal{D}\,\Psi _{    }\, X\, Z\right)-2.34 \text{ Tr}\left(\bar{\Psi }_{    }\,
   Z\, \mathcal{D}\,Z\, \Psi _{    }\, \bar{Y}\right)+2.34 \text{ Tr}\left(\bar{\Psi }_{    }\, Z\, \mathcal{D}\,Z\, \Psi _{    }\,
   X\right)-1.00 \text{ Tr}\left(\bar{\Psi }_{    }\, Z\, \mathcal{D}\,Z\, \Psi _{    }\, Z\right)-1.00 \text{ Tr}\left(\bar{\Psi }_{    }\,
   Z\, \mathcal{D}\,Z\, Z\, \Psi _{    }\right)+1.00 \text{ Tr}\left(\bar{\Psi }_{    }\, Z\, \mathcal{D}\,Z\, X\, \Psi _{    }\right)-1.00
   \text{ Tr}\left(\bar{\Psi }_{    }\, Z\, \mathcal{D}\,Z\, \bar{Y}\, \Psi _{    }\right)+0.500 \text{ Tr}\left(\bar{\Psi }_{    }\, Z\,
   \mathcal{D}\,X\, \Psi _{    }\, Z\right)-0.500 \text{ Tr}\left(\bar{\Psi }_{    }\, Z\, \mathcal{D}\,\bar{Y}\, \Psi _{    }\, Z\right)+2.34
   \text{ Tr}\left(\bar{\Psi }_{    }\, X\, \Psi _{    }\, Z\, \mathcal{D}\,Z\right)+2.34 \text{ Tr}\left(\bar{\Psi }_{    }\, X\, \Psi
   _{    }\, \mathcal{D}\,Z\, Z\right)+2.34 \text{ Tr}\left(\bar{\Psi }_{    }\, X\, Z\, \Psi _{    }\, \mathcal{D}\,Z\right)-0.947
   \text{ Tr}\left(\bar{\Psi }_{    }\, X\, Z\, Z\, \mathcal{D}\,\Psi _{    }\right)-1.84 \text{ Tr}\left(\bar{\Psi }_{    }\, X\, Z\,
   \mathcal{D}\,Z\, \Psi _{    }\right)-0.947 \text{ Tr}\left(\bar{\Psi }_{    }\, X\, \mathcal{D}\,\Psi _{    }\, Z\, Z\right)-1.84
   \text{ Tr}\left(\bar{\Psi }_{    }\, X\, \mathcal{D}\,Z\, \Psi _{    }\, Z\right)+1.00 \text{ Tr}\left(\bar{\Psi }_{    }\, X\,
   \mathcal{D}\,Z\, Z\, \Psi _{    }\right)-2.34 \text{ Tr}\left(\bar{\Psi }_{    }\, \bar{Y}\, \Psi _{    }\, Z\, \mathcal{D}\,Z\right)-2.34
   \text{ Tr}\left(\bar{\Psi }_{    }\, \bar{Y}\, \Psi _{    }\, \mathcal{D}\,Z\, Z\right)-2.34 \text{ Tr}\left(\bar{\Psi }_{    }\, \bar{Y}\,
   Z\, \Psi _{    }\, \mathcal{D}\,Z\right)+0.947 \text{ Tr}\left(\bar{\Psi }_{    }\, \bar{Y}\, Z\, Z\, \mathcal{D}\,\Psi _{    }\right)+1.84
   \text{ Tr}\left(\bar{\Psi }_{    }\, \bar{Y}\, Z\, \mathcal{D}\,Z\, \Psi _{    }\right)+0.947 \text{ Tr}\left(\bar{\Psi }_{    }\,
   \bar{Y}\, \mathcal{D}\,\Psi _{    }\, Z\, Z\right)+1.84 \text{ Tr}\left(\bar{\Psi }_{    }\, \bar{Y}\, \mathcal{D}\,Z\, \Psi _{    }\,
   Z\right)-1.00 \text{ Tr}\left(\bar{\Psi }_{    }\, \bar{Y}\, \mathcal{D}\,Z\, Z\, \Psi _{    }\right)+0.947 \text{ Tr}\left(\bar{\Psi }_{ 
     }\, \mathcal{D}\,\Psi _{    }\, Z\, Z\, \bar{Y}\right)+0.500 \text{ Tr}\left(\bar{\Psi }_{    }\, \mathcal{D}\,\Psi _{    }\, Z\,
   \bar{Y}\, Z\right)-0.894 \text{ Tr}\left(\bar{\Psi }_{    }\, \mathcal{D}\,\Psi _{    }\, \bar{Y}\, Z\, Z\right)-0.947 \text{ Tr}\left(\bar{\Psi
   }_{    }\, \mathcal{D}\,\Psi _{    }\, Z\, Z\, X\right)-0.500 \text{ Tr}\left(\bar{\Psi }_{    }\, \mathcal{D}\,\Psi _{    }\, Z\,
   X\, Z\right)+0.894 \text{ Tr}\left(\bar{\Psi }_{    }\, \mathcal{D}\,\Psi _{    }\, X\, Z\, Z\right)-2.34 \text{ Tr}\left(\bar{\Psi }_{   
   }\, \mathcal{D}\,Z\, \Psi _{    }\, Z\, \bar{Y}\right)-2.34 \text{ Tr}\left(\bar{\Psi }_{    }\, \mathcal{D}\,Z\, \Psi _{    }\, \bar{Y}\,
   Z\right)+2.34 \text{ Tr}\left(\bar{\Psi }_{    }\, \mathcal{D}\,Z\, \Psi _{    }\, Z\, X\right)+2.34 \text{ Tr}\left(\bar{\Psi }_{    }\,
   \mathcal{D}\,Z\, \Psi _{    }\, X\, Z\right)-1.89 \text{ Tr}\left(\bar{\Psi }_{    }\, \mathcal{D}\,Z\, \Psi _{    }\, Z\, Z\right)-2.34
   \text{ Tr}\left(\bar{\Psi }_{    }\, \mathcal{D}\,Z\, Z\, \Psi _{    }\, \bar{Y}\right)+2.34 \text{ Tr}\left(\bar{\Psi }_{    }\,
   \mathcal{D}\,Z\, Z\, \Psi _{    }\, X\right)-1.89 \text{ Tr}\left(\bar{\Psi }_{    }\, \mathcal{D}\,Z\, Z\, Z\, \Psi _{    }\right)-1.84
   \text{ Tr}\left(\bar{\Psi }_{    }\, \mathcal{D}\,Z\, Z\, X\, \Psi _{    }\right)+1.84 \text{ Tr}\left(\bar{\Psi }_{    }\, \mathcal{D}\,Z\,
   Z\, \bar{Y}\, \Psi _{    }\right)-1.84 \text{ Tr}\left(\bar{\Psi }_{    }\, \mathcal{D}\,Z\, X\, \Psi _{    }\, Z\right)+1.00
   \text{ Tr}\left(\bar{\Psi }_{    }\, \mathcal{D}\,Z\, X\, Z\, \Psi _{    }\right)+1.84 \text{ Tr}\left(\bar{\Psi }_{    }\, \mathcal{D}\,Z\,
   \bar{Y}\, \Psi _{    }\, Z\right)-1.00 \text{ Tr}\left(\bar{\Psi }_{    }\, \mathcal{D}\,Z\, \bar{Y}\, Z\, \Psi _{    }\right)-0.894
   \text{ Tr}\left(\bar{\Psi }_{    }\, \mathcal{D}\,X\, \Psi _{    }\, Z\, Z\right)+0.500 \text{ Tr}\left(\bar{\Psi }_{    }\, \mathcal{D}\,X\,
   Z\, \Psi _{    }\, Z\right)+0.947 \text{ Tr}\left(\bar{\Psi }_{    }\, \mathcal{D}\,X\, Z\, Z\, \Psi _{    }\right)+0.894
   \text{ Tr}\left(\bar{\Psi }_{    }\, \mathcal{D}\,\bar{Y}\, \Psi _{    }\, Z\, Z\right)-0.500 \text{ Tr}\left(\bar{\Psi }_{    }\,
   \mathcal{D}\,\bar{Y}\, Z\, \Psi _{    }\, Z\right)-0.947 \text{ Tr}\left(\bar{\Psi }_{    }\, \mathcal{D}\,\bar{Y}\, Z\, Z\, \Psi _{   
   }\right)+0.553 \text{ Tr}\left(\bar{\Psi }_{    }\, Z\, Z\, Z\, \mathcal{D}\,\Psi _{    }\right)+1.45 \text{ Tr}\left(\bar{\Psi }_{    }\,
   Z\, Z\, \mathcal{D}\,\Psi _{    }\, Z\right)+1.45 \text{ Tr}\left(\bar{\Psi }_{    }\, Z\, \mathcal{D}\,\Psi _{    }\, Z\, Z\right)+0.553
   \text{ Tr}\left(\bar{\Psi }_{    }\, \mathcal{D}\,\Psi _{    }\, Z\, Z\, Z\right)+0.553 \text{ Tr}\left(\mathcal{D}^2\,Z\, Z\, Z\, Z\,
   \bar{Z}\right)-0.947 \text{ Tr}\left(\mathcal{D}^2\,Z\, Z\, Z\, X\, \bar{X}\right)+0.947 \text{ Tr}\left(\mathcal{D}^2\,Z\, Z\, Z\, Y\, \bar{Y}\right)+0.947
   \text{ Tr}\left(\mathcal{D}^2\,Z\, Z\, Z\, \bar{Y}\, Y\right)-0.947 \text{ Tr}\left(\mathcal{D}^2\,Z\, Z\, Z\, \bar{X}\, X\right)+1.45
   \text{ Tr}\left(\mathcal{D}^2\,Z\, Z\, Z\, \bar{Z}\, Z\right)-0.500 \text{ Tr}\left(\mathcal{D}^2\,Z\, Z\, X\, Z\, \bar{X}\right)+0.500
   \text{ Tr}\left(\mathcal{D}^2\,Z\, Z\, Y\, Z\, \bar{Y}\right)+0.500 \text{ Tr}\left(\mathcal{D}^2\,Z\, Z\, \bar{Y}\, Z\, Y\right)-0.500
   \text{ Tr}\left(\mathcal{D}^2\,Z\, Z\, \bar{X}\, Z\, X\right)+1.45 \text{ Tr}\left(\mathcal{D}^2\,Z\, Z\, \bar{Z}\, Z\, Z\right)+0.894
   \text{ Tr}\left(\mathcal{D}^2\,Z\, X\, Z\, Z\, \bar{X}\right)-0.500 \text{ Tr}\left(\mathcal{D}^2\,Z\, X\, Z\, \bar{X}\, Z\right)-0.947
   \text{ Tr}\left(\mathcal{D}^2\,Z\, X\, \bar{X}\, Z\, Z\right)-0.894 \text{ Tr}\left(\mathcal{D}^2\,Z\, Y\, Z\, Z\, \bar{Y}\right)+0.500
   \text{ Tr}\left(\mathcal{D}^2\,Z\, Y\, Z\, \bar{Y}\, Z\right)+0.947 \text{ Tr}\left(\mathcal{D}^2\,Z\, Y\, \bar{Y}\, Z\, Z\right)-0.894
   \text{ Tr}\left(\mathcal{D}^2\,Z\, \bar{Y}\, Z\, Z\, Y\right)+0.500 \text{ Tr}\left(\mathcal{D}^2\,Z\, \bar{Y}\, Z\, Y\, Z\right)+0.947
   \text{ Tr}\left(\mathcal{D}^2\,Z\, \bar{Y}\, Y\, Z\, Z\right)+0.894 \text{ Tr}\left(\mathcal{D}^2\,Z\, \bar{X}\, Z\, Z\, X\right)-0.500
   \text{ Tr}\left(\mathcal{D}^2\,Z\, \bar{X}\, Z\, X\, Z\right)-0.947 \text{ Tr}\left(\mathcal{D}^2\,Z\, \bar{X}\, X\, Z\, Z\right)+0.553
   \text{ Tr}\left(\mathcal{D}^2\,Z\, \bar{Z}\, Z\, Z\, Z\right)+0.894 \text{ Tr}\left(\mathcal{D}\,\bar{\Psi }_{    }\, \Psi _{    }\, Z\, Z\,
   \bar{X}\right)-0.500 \text{ Tr}\left(\mathcal{D}\,\bar{\Psi }_{    }\, \Psi _{    }\, Z\, \bar{X}\, Z\right)-0.947
   \text{ Tr}\left(\mathcal{D}\,\bar{\Psi }_{    }\, \Psi _{    }\, \bar{X}\, Z\, Z\right)-0.894 \text{ Tr}\left(\mathcal{D}\,\bar{\Psi }_{   
   }\, \Psi _{    }\, Z\, Z\, Y\right)+0.500 \text{ Tr}\left(\mathcal{D}\,\bar{\Psi }_{    }\, \Psi _{    }\, Z\, Y\, Z\right)+0.947
   \text{ Tr}\left(\mathcal{D}\,\bar{\Psi }_{    }\, \Psi _{    }\, Y\, Z\, Z\right)-0.553 \text{ Tr}\left(\mathcal{D}\,\bar{\Psi }_{    }\,
   \Psi _{    }\, Z\, Z\, Z\right)-0.500 \text{ Tr}\left(\mathcal{D}\,\bar{\Psi }_{    }\, Z\, \Psi _{    }\, Z\, \bar{X}\right)+0.500
   \text{ Tr}\left(\mathcal{D}\,\bar{\Psi }_{    }\, Z\, \Psi _{    }\, Z\, Y\right)-1.45 \text{ Tr}\left(\mathcal{D}\,\bar{\Psi }_{    }\, Z\,
   \Psi _{    }\, Z\, Z\right)-0.947 \text{ Tr}\left(\mathcal{D}\,\bar{\Psi }_{    }\, Z\, Z\, \Psi _{    }\, \bar{X}\right)+0.947
   \text{ Tr}\left(\mathcal{D}\,\bar{\Psi }_{    }\, Z\, Z\, \Psi _{    }\, Y\right)-1.45 \text{ Tr}\left(\mathcal{D}\,\bar{\Psi }_{    }\, Z\,
   Z\, \Psi _{    }\, Z\right)-0.553 \text{ Tr}\left(\mathcal{D}\,\bar{\Psi }_{    }\, Z\, Z\, Z\, \Psi _{    }\right)+0.947
   \text{ Tr}\left(\mathcal{D}\,\bar{\Psi }_{    }\, Z\, Z\, Y\, \Psi _{    }\right)-0.947 \text{ Tr}\left(\mathcal{D}\,\bar{\Psi }_{    }\, Z\,
   Z\, \bar{X}\, \Psi _{    }\right)+0.500 \text{ Tr}\left(\mathcal{D}\,\bar{\Psi }_{    }\, Z\, Y\, Z\, \Psi _{    }\right)-0.500
   \text{ Tr}\left(\mathcal{D}\,\bar{\Psi }_{    }\, Z\, \bar{X}\, Z\, \Psi _{    }\right)+0.947 \text{ Tr}\left(\mathcal{D}\,\bar{\Psi }_{   
   }\, Y\, \Psi _{    }\, Z\, Z\right)+0.500 \text{ Tr}\left(\mathcal{D}\,\bar{\Psi }_{    }\, Y\, Z\, \Psi _{    }\, Z\right)-0.894
   \text{ Tr}\left(\mathcal{D}\,\bar{\Psi }_{    }\, Y\, Z\, Z\, \Psi _{    }\right)-0.947 \text{ Tr}\left(\mathcal{D}\,\bar{\Psi }_{    }\,
   \bar{X}\, \Psi _{    }\, Z\, Z\right)-0.500 \text{ Tr}\left(\mathcal{D}\,\bar{\Psi }_{    }\, \bar{X}\, Z\, \Psi _{    }\, Z\right)+0.894
   \text{ Tr}\left(\mathcal{D}\,\bar{\Psi }_{    }\, \bar{X}\, Z\, Z\, \Psi _{    }\right)-0.894 \text{ Tr}\left(\mathcal{D}\,\bar{\Psi }_{   
   }\, \Psi _{    }\, Z\, Z\, \bar{Y}\right)+0.500 \text{ Tr}\left(\mathcal{D}\,\bar{\Psi }_{    }\, \Psi _{    }\, Z\, \bar{Y}\,
   Z\right)+0.947 \text{ Tr}\left(\mathcal{D}\,\bar{\Psi }_{    }\, \Psi _{    }\, \bar{Y}\, Z\, Z\right)+0.894 \text{ Tr}\left(\mathcal{D}\,\bar{\Psi
   }_{    }\, \Psi _{    }\, Z\, Z\, X\right)-0.500 \text{ Tr}\left(\mathcal{D}\,\bar{\Psi }_{    }\, \Psi _{    }\, Z\, X\,
   Z\right)-0.947 \text{ Tr}\left(\mathcal{D}\,\bar{\Psi }_{    }\, \Psi _{    }\, X\, Z\, Z\right)+0.553 \text{ Tr}\left(\mathcal{D}\,\bar{\Psi
   }_{    }\, \Psi _{    }\, Z\, Z\, Z\right)+0.500 \text{ Tr}\left(\mathcal{D}\,\bar{\Psi }_{    }\, Z\, \Psi _{    }\, Z\,
   \bar{Y}\right)-0.500 \text{ Tr}\left(\mathcal{D}\,\bar{\Psi }_{    }\, Z\, \Psi _{    }\, Z\, X\right)+1.45 \text{ Tr}\left(\mathcal{D}\,\bar{\Psi
   }_{    }\, Z\, \Psi _{    }\, Z\, Z\right)+0.947 \text{ Tr}\left(\mathcal{D}\,\bar{\Psi }_{    }\, Z\, Z\, \Psi _{    }\,
   \bar{Y}\right)-0.947 \text{ Tr}\left(\mathcal{D}\,\bar{\Psi }_{    }\, Z\, Z\, \Psi _{    }\, X\right)+1.45 \text{ Tr}\left(\mathcal{D}\,\bar{\Psi
   }_{    }\, Z\, Z\, \Psi _{    }\, Z\right)+0.553 \text{ Tr}\left(\mathcal{D}\,\bar{\Psi }_{    }\, Z\, Z\, Z\, \Psi _{   
   }\right)-0.947 \text{ Tr}\left(\mathcal{D}\,\bar{\Psi }_{    }\, Z\, Z\, X\, \Psi _{    }\right)+0.947 \text{ Tr}\left(\mathcal{D}\,\bar{\Psi
   }_{    }\, Z\, Z\, \bar{Y}\, \Psi _{    }\right)-0.500 \text{ Tr}\left(\mathcal{D}\,\bar{\Psi }_{    }\, Z\, X\, Z\, \Psi _{   
   }\right)+0.500 \text{ Tr}\left(\mathcal{D}\,\bar{\Psi }_{    }\, Z\, \bar{Y}\, Z\, \Psi _{    }\right)-0.947 \text{ Tr}\left(\mathcal{D}\,\bar{\Psi
   }_{    }\, X\, \Psi _{    }\, Z\, Z\right)-0.500 \text{ Tr}\left(\mathcal{D}\,\bar{\Psi }_{    }\, X\, Z\, \Psi _{    }\,
   Z\right)+0.894 \text{ Tr}\left(\mathcal{D}\,\bar{\Psi }_{    }\, X\, Z\, Z\, \Psi _{    }\right)+0.947 \text{ Tr}\left(\mathcal{D}\,\bar{\Psi
   }_{    }\, \bar{Y}\, \Psi _{    }\, Z\, Z\right)+0.500 \text{ Tr}\left(\mathcal{D}\,\bar{\Psi }_{    }\, \bar{Y}\, Z\, \Psi _{   
   }\, Z\right)-0.894 \text{ Tr}\left(\mathcal{D}\,\bar{\Psi }_{    }\, \bar{Y}\, Z\, Z\, \Psi _{    }\right)+0.553
   \text{ Tr}\left(\bar{\mathcal{F}}_{    }\, \mathcal{F}_{    }\, Z\, Z\, Z\right)+1.89 \text{ Tr}\left(\bar{\mathcal{F}}_{    }\, \Psi
   _{    }\, \Psi _{    }\, Z\, Z\right)+1.00 \text{ Tr}\left(\bar{\mathcal{F}}_{    }\, \Psi _{    }\, Z\, \Psi _{    }\,
   Z\right)-1.79 \text{ Tr}\left(\bar{\mathcal{F}}_{    }\, \Psi _{    }\, Z\, Z\, \Psi _{    }\right)-1.89
   \text{ Tr}\left(\bar{\mathcal{F}}_{    }\, \Psi _{    }\, \Psi _{    }\, Z\, Z\right)-1.00 \text{ Tr}\left(\bar{\mathcal{F}}_{   
   }\, \Psi _{    }\, Z\, \Psi _{    }\, Z\right)+1.79 \text{ Tr}\left(\bar{\mathcal{F}}_{    }\, \Psi _{    }\, Z\, Z\, \Psi
   _{    }\right)+1.45 \text{ Tr}\left(\bar{\mathcal{F}}_{    }\, Z\, \mathcal{F}_{    }\, Z\, Z\right)+1.00
   \text{ Tr}\left(\bar{\mathcal{F}}_{    }\, Z\, \Psi _{    }\, Z\, \Psi _{    }\right)-1.00 \text{ Tr}\left(\bar{\mathcal{F}}_{   
   }\, Z\, \Psi _{    }\, Z\, \Psi _{    }\right)+1.45 \text{ Tr}\left(\bar{\mathcal{F}}_{    }\, Z\, Z\, \mathcal{F}_{    }\,
   Z\right)+1.89 \text{ Tr}\left(\bar{\mathcal{F}}_{    }\, Z\, Z\, \Psi _{    }\, \Psi _{    }\right)-1.89
   \text{ Tr}\left(\bar{\mathcal{F}}_{    }\, Z\, Z\, \Psi _{    }\, \Psi _{    }\right)+0.553 \text{ Tr}\left(\bar{\mathcal{F}}_{ 
     }\, Z\, Z\, Z\, \mathcal{F}_{    }\right)$
\endgroup

In the previous formulas, $\Psi$ are Weyl fermions, $\mathcal{D}$ the covariant derivative, $X,Y,Z$ the complex scalars and $\mathcal{F}$ the field strength. We omitted indices for simplicity. The coefficients are truncated at the 3rd significant digit.

The scaling dimension of those operators can be computed using the QSC numerical algorithm introduced in Section \ref{SM:QSC}. In the following Tables, we collect some numerical data for their non-perturbative dimensions $\Delta_{\pm}$. The number of significant digits obtained by numerical QSC ranges between 15 and 30. In the following, we include the first 10 digits.

\begin{table}[!h]%[H]
    \centering
    \begin{tabular}{||cc|cc|cc|cc||}
    \hline
    $g$ & $\Delta_-$ & $g$ & $\Delta_-$ & $g$ & $\Delta_-$ & $g$ & $\Delta_-$ \\
    \hline\hline
0.01 & 7.0005527733 & 0.045 & 7.0111884907 & 0.35 & 7.6499415977 & 0.70 & 9.2018914868 \\ 0.015 & 7.0012437029 & 0.05 & 7.0138113604 & 0.40 & 7.8358141677 & 0.75 & 9.4464029289 \\ 0.02 & 7.0022109351 & 0.10 & 7.0551415766 & 0.45 & 8.0384744343 & 0.80 & 9.6903172850 \\ 0.025 & 7.0034544006 & 0.15 & 7.1236494312 & 0.50 & 8.2549906792 & 0.85 & 9.9324630913 \\ 0.03 & 7.0049740097 & 0.20 & 7.2186792998 & 0.55 & 8.4823723018 & 0.90 & 10.172007082 \\ 0.035 & 7.0067696520 & 0.25 & 7.3391690994 & 0.60 & 8.7177161225 & 0.95 & 10.408392845 \\ 0.04 & 7.0088411965 & 0.30 & 7.4835892287 & 0.65 & 8.9583407840 & 1.00 & 10.641278141\\
    \hline
    \end{tabular}
    \caption{Non-perturbative scaling dimension of the operator $\mathcal{O}_{\Delta_-}$.}
    \label{tab:Deltaminus}
\end{table}

\begin{table}[!h]%[H]
    \centering
    \begin{tabular}{||cc|cc|cc|cc||}
    \hline
    $g$ & $\Delta_+$ & $g$ & $\Delta_+$ & $g$ & $\Delta_+$ & $g$ & $\Delta_+$ \\
    \hline\hline
0.01 & 7.0014465972 & 0.045 & 7.0290569733 & 0.35 & 8.2588166049 & 0.70 & 10.140935196 \\ 0.015 & 7.0032531131 & 0.05 & 7.0358020135 & 0.40 & 8.5328052140 & 0.75 & 10.391411331 \\ 0.02 & 7.0057790146 & 0.10 & 7.1389837552 & 0.45 & 8.8088584099 & 0.80 & 10.636223609 \\ 0.025 & 7.0090211034 & 0.15 & 7.2988221912 & 0.50 & 9.0838311026 & 0.85 & 10.875558214 \\ 0.03 & 7.0129752979 & 0.20 & 7.5021227152 & 0.55 & 9.3555960723 & 0.90 & 11.109651881 \\ 0.035 & 7.0176366572 & 0.25 & 7.7364118674 & 0.60 & 9.6228093196 & 0.95 & 11.338761993 \\ 0.04 & 7.0229994074 & 0.30 & 7.9913310732 & 0.65 & 9.8847071830 & 1.00 & 11.563148729\\
    \hline
    \end{tabular}
    \caption{Non-perturbative scaling dimension of the operator $\mathcal{O}_{\Delta_+}$.}
    \label{tab:Deltaplus}
\end{table}

\end{document}